\documentclass{emulateapj}
\usepackage{natbib}
\newcommand{\nhintr}{\ifmmode{ N_{H}^{intr}} \else N$_{H}^{intr}$\fi}
\newcommand{\nhgal}{\ifmmode{ N_{H}^{Gal}} \else N$_{H}^{Gal}$\fi}
\newcommand{\Lbol}{\ifmmode{L_{\rm Bol}} \else $L_{\rm Bol}$\fi}
\newcommand{\LEdd}{\ifmmode{L_{\rm Edd}} \else $L_{\rm Edd}$\fi}
\newcommand{\REdd}{\ifmmode{L_{\rm Bol}/L_{\rm Edd}} \else $L_{\rm Bol}/L_{\rm Edd}$\fi}
\begin{document}

\shorttitle{ChaMP Optical Spectroscopy} 
\shortauthors{Trichas et al.}
\received{2012 March}
\title{The Chandra Multi-Wavelength Project: Optical Spectroscopy and the Broadband Spectral Energy Distributions of X-ray Selected AGN}
\author{Markos Trichas\altaffilmark{1}}
\email{mtrichas@cfa.harvard.edu}
\author{Paul J. Green\altaffilmark{1},
  John D. Silverman\altaffilmark{2},
  Tom Aldcroft\altaffilmark{1},
  Wayne Barkhouse\altaffilmark{3},
  Robert A. Cameron \altaffilmark{4},
  Anca Constantin\altaffilmark{5},
  Sara L. Ellison \altaffilmark{6},
  Craig Foltz \altaffilmark{7},
  Daryl Haggard\altaffilmark{8},
  Buell T. Jannuzi\altaffilmark{9},
  Dong-Woo Kim\altaffilmark{1},
  Herman L. Marshall\altaffilmark{10},
  Amy Mossman\altaffilmark{1},
  Laura M. P\'erez\altaffilmark{11},
  Encarni Romero-Colmenero\altaffilmark{12},
  Angel Ruiz\altaffilmark{13},
  Malcolm G.  Smith\altaffilmark{14},
  Paul S. Smith \altaffilmark{15},
  Guillermo Torres \altaffilmark{1},
  Daniel R. Wik \altaffilmark{16},
  Belinda~J.~Wilkes\altaffilmark{1},
  Angie Wolfgang \altaffilmark{17}}
\altaffiltext{1}{Harvard-Smithsonian Center for Astrophysics,
  Cambridge, MA 02138}
\altaffiltext{2}{Institute for the Physics and Mathematics of the
  Universe (IPMU), University of Tokyo, Kashiwanoha 5-1-5,
  Kashiwa-shi, Chiba 277-8568, Japan}
\altaffiltext{3}{Department of Physics and Astrophysics, University of
  North Dakota, Grand Forks, ND 58202, USA}
\altaffiltext{4}{W. W. Hansen Experimental Physics Laboratory, Kavli Institute for Particle Astrophysics and Cosmology, Department of Physics and SLAC National Accelerator Laboratory, Stanford University, Stanford, CA 94305, USA}
\altaffiltext{5}{Department of Physics and Astronomy, James Madison
  University, PHCH, Harrisonburg, VA 22807}
\altaffiltext{6}{Department of Physics and Astronomy, University of Victoria,
  Victoria, British Columbia V8P 1A1, Canada}
\altaffiltext{7}{Division of Astronomical Sciences, National Science Foundation, 4201 Wilson Blvd., Arlington, VA 22230}
\altaffiltext{8}{Center for Interdisciplinary Exploration and Research
  in Astrophysics, Northwestern University, 2145 Sheridan Road,
  Evanston, IL 60208, USA}
\altaffiltext{9}{NOAO, Kitt Peak National Observatory, Tucson, AZ 85726 USA}
\altaffiltext{10}{Kavli Institute for Astrophysics and Space Research,
  Massachusetts Institute of Technology, 77 Massachusetts Ave.,
  Cambridge, MA 02139, USA}
\altaffiltext{11}{Department of Astronomy, California Institute of Technology, 1200 East California Blvd, Pasadena, CA 91125, USA}
\altaffiltext{12}{South African Astronomical Observatory, P.O. Box 9,
  Observatory, 7935, South Africa}
\altaffiltext{13}{Osservatorio Astronomico di Brera - INAF, Milan, Italy}
\altaffiltext{14}{Cerro Tololo Interamerican Observatory, La Serena, Chile}
\altaffiltext{15}{Steward Observatory, University of Arizona, 933 Cherry Ave., Tucson, AZ 85721, USA}
\altaffiltext{16}{NASA Goddard Space Flight Center, Greenbelt, MD 20771}
\altaffiltext{17}{Department of Astronomy and Astrophysics, University
of California, Santa Cruz, CA 95064, USA}
\begin{abstract}
\noindent 
From optical spectroscopy of X-ray sources observed as part of the
Chandra Multiwavelength Project (ChaMP), we present redshifts and
classifications for a total of 1569 Chandra sources from our targeted
spectroscopic follow up using the FLWO/1.5m, SAAO/1.9m, WIYN 3.5m,
CTIO/4m, KPNO/4m, Magellan/6.5m, MMT/6.5m and Gemini/8m telescopes,
and from archival SDSS spectroscopy. We classify the optical
counterparts as 50$\%$ broad line AGN, 16$\%$ emission line galaxies,
14$\%$ absorption line galaxies, and 20$\%$ stars. We detect QSOs out
to z$\sim$5.5 and galaxies out to z$\sim$3. We have compiled extensive
photometry, including X-ray (ChaMP), ultra-violet (GALEX), optical
(SDSS, ChaMP-NOAO/MOSAIC follow-up), near-infrared (UKIDSS, 2MASS,
ChaMP-CTIO/ISPI follow-up), mid-infrared (WISE) and radio (FIRST,
NVSS) bands.  Together with our spectroscopic information, this
enables us to derive detailed spectral energy distributions (SEDs) for
our extragalactic sources. We fit a variety of template SEDs to
determine bolometric luminosities, and to constrain AGN and starburst
components where both are present. While $\sim$58$\%$ of X-ray
Seyferts (10$^{42}$$<$$L_{2-10~keV}$$<$10$^{44}$ erg s$^{-1}$) require
a starburst event ($>$5\% starburst contribution to bolometric
luminosity) to fit observed photometry only 26$\%$ of the X-ray QSO
($L_{2-10~keV}$$>$10$^{44}$ erg s$^{-1}$) population appear to have
some kind of star formation contribution. This is significantly lower
than for the Seyferts, especially if we take into account torus
contamination at z$>$1 where the majority of our X-ray QSOs lie. In
addition, we observe a rapid drop of the percentage of starburst
contribution as X-ray luminosity increases. This is consistent with
the quenching of star formation by powerful QSOs, as predicted by the
merger model, or with a time lag between the peak of star formation
and QSO activity.  We have tested the hypothesis that there should be
a strong connection between X-ray obscuration and star-formation but
we do not find any association between X-ray column density and star
formation rate both in the general population or the star-forming
X-ray Seyferts. Our large compilation also allows us to report here
the identification of 81 X-ray Bright Optically inactive Galaxies
(XBONG), 78 z$>$3 X-ray sources and 8 Type-2 QSO candidates.  Also we
have identified the highest redshift (z$=$5.4135) X-ray selected QSO
with optical spectroscopy.
\end{abstract}


\section{Introduction}
\noindent Our understanding of how galaxies form and evolve has 
significantly advanced in the last few years. Large spectroscopic 
programs (e.g. 2dFGRS, SDSS, DEEP2) have shown that the 
evolution of galaxies strongly depends on their position in the 
cosmic web. A striking manifestation of this link is the suppression of the star formation in increasingly dense environments 
(e.g. Kauffmann et al. 2004; Cooper et al. 2006; Netzer 2009; Schawinski et al. 2009). In addition, there  is now  strong evidence  that powerful  active  galactic nuclei
(AGN) play a key role  in the evolution of galaxies. The correlation
of central  black hole  and stellar bulge  mass ($M_{BH}$-$\sigma$; e.g. Magorrian et al. 1998), and  the similarity between  the cosmic star  formation history
(e.g. Hopkins  $\&$ Beacom 2006)  and cosmic black hole  mass assembly
history  (e.g.  Aird  et  al 2010)  in massive galaxies, both  suggest that  the growth  of
supermassive  black holes  (SMBH) is  related  to the  growth of  host
galaxies.  The apparently independent observational trends above are believed to hold the key to galaxy assembly, but the detailed physical mechanism(s) behind them remain 
poorly understood. For example, although processes like galaxy 
suffocation, harassment and ram-pressure stripping are proposed 
to explain the star formation/density relation (e.g. Haines et al. 
2006), they usually operate in rich and hence rare environments 
(i.e. massive clusters). The vast majority of galaxies inhabit less 
dense regions (poor clusters, groups, field), where alternative 
mechanisms should dominate. Understanding what  drives the  formation and  evolution of
galaxies and their  central SMBHs remains one of  the most significant
challenges   in  extragalactic   astrophysics. \\ \\
 \noindent Recent attention has focused on models where AGN feedback 
regulates the star formation in the host galaxy. These scenarios 
are consistent with the $M_{BH}$ - $\sigma$ relation and make various predictions for AGN properties, including the environmental dependence of the AGN/galaxy interplay and the relative timing of periods of peak star formation and nuclear accretion activity. The key feature of these 
models is that they can potentially link the apparently independent observed relations between star formation, AGN activity and large 
scale structure to the same underlying physical process. For example, in the ``radio-mode" model of Croton et al. (2006), accretion of gas from cooling flows in dense environments (e.g. group, 
cluster) may produce relatively low-luminosity AGN, which in 
turn heat the bulk of the cooling gas and prevent it from falling 
into the galaxy center to form stars. Alternatively, Hopkins et al. 
(2006) propose that mergers trigger luminous QSOs and circumnuclear starbursts, which both feed and obscure the central engine for most of its active lifetime. In this scenario, AGN outflows eventually sweep away the dust and gas clouds, thereby quenching the star formation. This ``QSO-mode" likely dominates in poor environments (e.g. field, group), as the high-velocity encounters, common in dense regions, do not favour mergers. These proposed models make clear, testable predictions about the 
properties of AGN, while observational constraints provide first-order confirmation of this theoretical picture (e.g. Lehmer et al. 2009). Merger-driven scenarios for example, predict an association between optical morphological disturbances, 
star formation and an intense obscured AGN phase in low density regions. The ``radio mode" model,  in contrast, invokes 
milder AGN activity in early-type hosts and relatively dense environments with little or no star formation.\\ \\
\noindent While it  is now recognized that
black holes play a fundamental  role in shaping the galaxy population,
the sequence in  which galaxies build up their  stellar and black hole
mass  and the  relationship between  the two  components are,  as yet,
poorly  understood.  In  some  models  (e.g. Di  Matteo  et al.  2005,
Hopkins et al.  2006) the stellar population and the SMBHs form almost
simultaneously and therefore predict a correlation between
star formation  and AGN activity.   Contrary to  that scenario,  it is
also proposed  that the stellar population of  galaxies is accumulated
first, followed by  the main epoch of SMBH  growth (e.g.  Archibald et
al.  2002, Cen 2011).  In the latter class of models, one expects
star formation and  AGN activity to  be unrelated, or  even negatively
correlated.  A major complexity in addressing this question observationally, is assessing the level of star formation in
AGN.  Particularly  at  $z>1$,  the  main epoch  of  galaxy  and  SMBH
formation, it is difficult to decompose the stellar  from the AGN
emission, especially in the case of dust enshrouded systems. When they  are  accreting rapidly,  AGN  can dominate  the
radiation from stars over  almost the entire electromagnetic spectrum.
There are, however, two key energy ranges which allow the most effective separation between the
radiation from  accretion and from star  formation.  The  X-ray band  provides  a  clean  window in  which  the
radiation from luminous AGN ($L_X\rm (2-10keV)>10^{42}\,erg\,s^{-1}$),
can be observed with  minimal contamination from star formation, while
star-forming galaxies emit a  large fraction of their energy  in the infrared
band where the AGN contribution is minimal.  Combining X-ray with longer wavelength data and optical spectroscopy can provide a handle on this issue (e.g. Trichas et al. 2009; 2010, Kalfountzou et al. 2011).  X-ray surveys have proved to be by far the most efficient way of finding AGN, and in relatively shallow surveys, AGN will completely dominate the source population. When a hard bandpass is available ($>$2 keV), as with Chandra, one can detect X-ray AGN that might otherwise be completely missed in other surveys due to obscuration. However, lack of observational data limits the information on the interplay between AGN, star formation and local density, particularly at z$>$1, close to the peak of the AGN and star formation activity of the Universe (e.g. Barger et al. 2005). Addressing these key questions requires the identification of  large numbers of AGN at z$>$1 over a broad range of environments for which a precise estimation of their bolometric luminosity will allow us to determine the relative contribution of AGN and star formation to the bolometric emission.\\\\
\begin{figure} 
\begin{center}
\begin{minipage}[c]{8.8 cm}
        \includegraphics[width=1.0\textwidth]{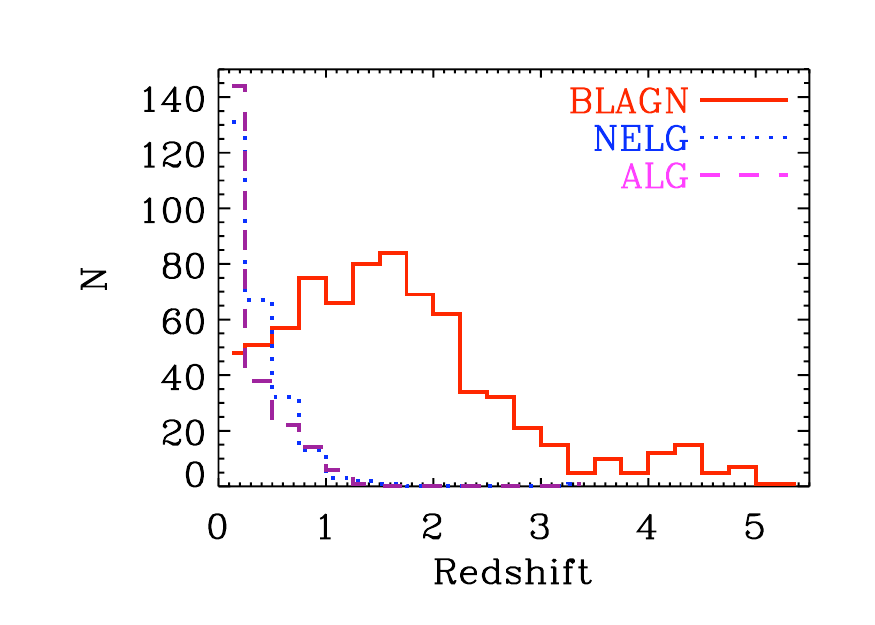}
\end{minipage}
\end{center}
\caption
           {\scriptsize Redshift  distribution for all 1242 spectroscopically identified ChaMP extragalactic sources. }
\end{figure}
\noindent The above questions have motivated efforts to study the multi-wavelength properties of AGN combining data from various space and ground-based telescopes. The Chandra and XMM-Newton Observatories are, for the 
first time, resolving the hard (2-10 keV) Cosmic X-ray Background (CXRB) into individual sources. Based on the spectral shape of the CXRB, 
the majority of emission from accretion powered sources 
has been obscured from our view. While unabsorbed AGN dominate the soft (0.1-2 keV) CXRB 
(Hasinger et al. 1998), the high energy CXRB spectrum 
(2-30 keV) is harder than that of known AGN (Gilli et al 
2001). With recent Chandra observations of the Chandra 
Deep Field North (CDF-N; Brandt et al. 2001), South (CDF- 
S; Rosati et al. 2002) and XMM observations of the Lockman 
Hole (Hasinger et al. 2001) , $\sim$75$\%$ of the hard (2-8 keV) 
XRB has been resolved into point sources. The resolved fraction
decreases with increasing X-ray energy (Worsley et al. 2006),
but approaches 80\% e.g., when X-rays at the position of faint optical
sources are stacked (Hickox \& Markevitch 2006). Many of the hardest serendipitous sources found so far arise in otherwise unremarkable bright galaxies (Hornschemeier et al. 2001; Tozzi 
et al. 2001), which may contain very heavily obscured AGN. 
In addition, Chandra has begun to detect the extremely rare, 
heavily obscured, dust enshrouded quasars (Norman et al., 
2001; Stern et al., 2002). Wider area surveys are needed to 
study these X-ray emitting populations with significant statistics. 
The Chandra Multiwavelength Project (ChaMP) is a medium-depth, wide area sample of serendipitous X-ray sources from archival 
Chandra fields. The 
ChaMP effectively bridges the gap between flux limits attained by past large-area X-ray surveys and recent Chandra Deep 
fields. The complete project has detected a total of $>$19,000 X-ray sources (Green et al. 2009) over 33 deg$^{2}$ with $>$9000 X-ray sources positionally matched to SDSS optical sources (Kim et al. 2007).  However, to fully study the properties of X-ray detected AGN, good quality spectra are needed for redshifts, luminosities and source classification. \\ \\
\noindent In this paper we describe the large sample of optical spectroscopy available for our ChaMP sources with a suite of available multi-wavelength data ranging from X-rays to radio. These data are used to test the predictions of the different feedback models proposed to explain the apparent relationship between AGN and star formation activity.  In $\S$2 we describe our optical spectroscopy and ancillary multi-wavelength data. $\S$3 summarizes the template fitting method used to produce spectral energy distributions for all our extragalactic objects while $\S$4 and $\S$5 describe the X-ray spectral fitting, star formation rates and black hole mass estimates. $\S$6, $\S$7 and  $\S$8 discuss what our observations suggest regarding the different feedback models while $\S$9 and $\S$10 briefly summarize the interesting populations of XBONG and high redshift objects found within our sample. $\S$11 is a summary of our findings. A cosmological model with $\Omega_{o}~=~0.3$, $\lambda_{o}~=~0.7$ and a Hubble constant of 72 $km~s^{-1}~Mpc^{-1}$ is used throughout. 
\section{ChaMP Observations}
\subsection{Imaging}
\noindent The Chandra Multiwavelength Project (ChaMP) 
is a wide-area serendipitous X-ray survey based on 
archival X-ray images of the ($|$b$|$ $>$ 20 deg) sky observed with the AXAF CCD Imaging Spectrometer 
(ACIS) onboard Chandra. The ChaMP covers a total of 392 fields, omitting pointings from dedicated 
serendipitous surveys like the Chandra Deep Fields as well as fields with large bright optical or X-ray sources. The list of Chandra pointings  avoids any overlapping observations by eliminating the observation with the shorter exposure time. As described in Green et al. (2004), we also 
avoid fields with extended sources ($>$3$'$) in either 
optical or X-rays. Spurious X-ray sources have been flagged 
and removed as described in Kim et al. (2007). Of the 
392 ChaMP obsids, which average 0.1\,deg$^2$ sky area each, at the brightest
fluxes, 323 overlap with the SDSS (DR5) footprint (Covey et al. 2008; Green et al. 2009; Haggard et al. 2010).\\ \\
\noindent Optical imaging provides optical fluxes, preliminary source classiÞcation, and accurate centroiding for 
spectroscopic follow-up.  As a result,  the ChaMP team supplemented observed Chandra imaging with deep optical observations (Green et al. 2004). ChaMP fields were observed with NOAO 4-meter imaging with the Mosaic CCD cameras (Muller et al. 1998) which provided adequate depth, spatial resolution ($\sim$0.6$"$/pixel), and a large field of view (36$'$$\times$36$'$) over the full Chandra field of view. NOAO filters similar to Sloan Digital Sky Survey (SDSS) $g'$, $r'$ and $i'$
passbands were used for 66 such fields, reaching down to AB magnitudes
of 26.1, 25.4 and 24.4 respectively (Green et al. 2004). The
positional uncertainty of ChaMP X-ray source  centroids has been 
analyzed via X-ray simulations by Kim et al. (2007).  An automated  
matching procedure between each optical position 
and the ChaMPÕs X-ray source catalog was first performed with
$\sim$95$\%$ of the matched sample  having an X-ray/optical position
difference of $<$3\arcsec\, yielding a sample of 1376 unique matches.  In
addition to the automated matching procedure, we  also performed
visual inspection of both X-ray and  optical images, overplotting the
centroids and their associated position errors retaining the highest
confidence matches (Green et al. 2009). 
\subsection{Optical Spectroscopy}
\noindent The spectroscopic follow-up of Chandra sources operated 
in three modes based on optical magnitude. Spectra for the 
brightest sources ($r$$<$17) are obtained primarily with the FLWO/1.5m 
FAST spectrograph and the SAAO/1.9m grating spectrograph. For most sources with 17$<$$r$$<$21, we 
used the WIYN and the CTIO/4m with HYDRA, a multi-fiber 
spectrograph. To obtain spectra for the faint source population ($r$$>$21), slit and multi-object spectroscopy with a 4 to 8 m 
class telescope is required (i.e. KPNO/4m, MMT, Magellan, Gemini).  In total, 22 nights of WIYN/Hydra (multi-fiber) time, 32 nights of
Magellan/IMACS and LDSS-2 (multi-slit), 9 nights of MMT with Hectospec
(multi-fiber) or Blue Channel (single slit), 7 nights of CTIO-4m/Hydra
(multi-fiber), 5 nights of KPNO-4m/MARS (multi-slit), and 2 nights of
Gemini-N/GMOS (multi-slit) were used to obtain optical spectroscopy
for our ChaMP sources.  These spectra were supplemented with optical spectra from the SDSS DR6 archive. Among the 1569 spectra, 50$\%$ are broad line AGN (BLAGN; FWHM$>$1000
km/sec), 16$\%$ narrow emission line galaxies (NELG; EW$>$5$\AA$,
FWHM$<$1000 km/sec), 14$\%$ absorption line galaxies (ALG; no emission
lines with  EW$>$5$\AA$) and 20$\%$ stars. Results from the stellar sample are published in Covey et al. (2008).\\ \\
\begin{figure} 
\begin{center}
\begin{minipage}[c]{8.cm}
        \includegraphics[width=1.0\textwidth]{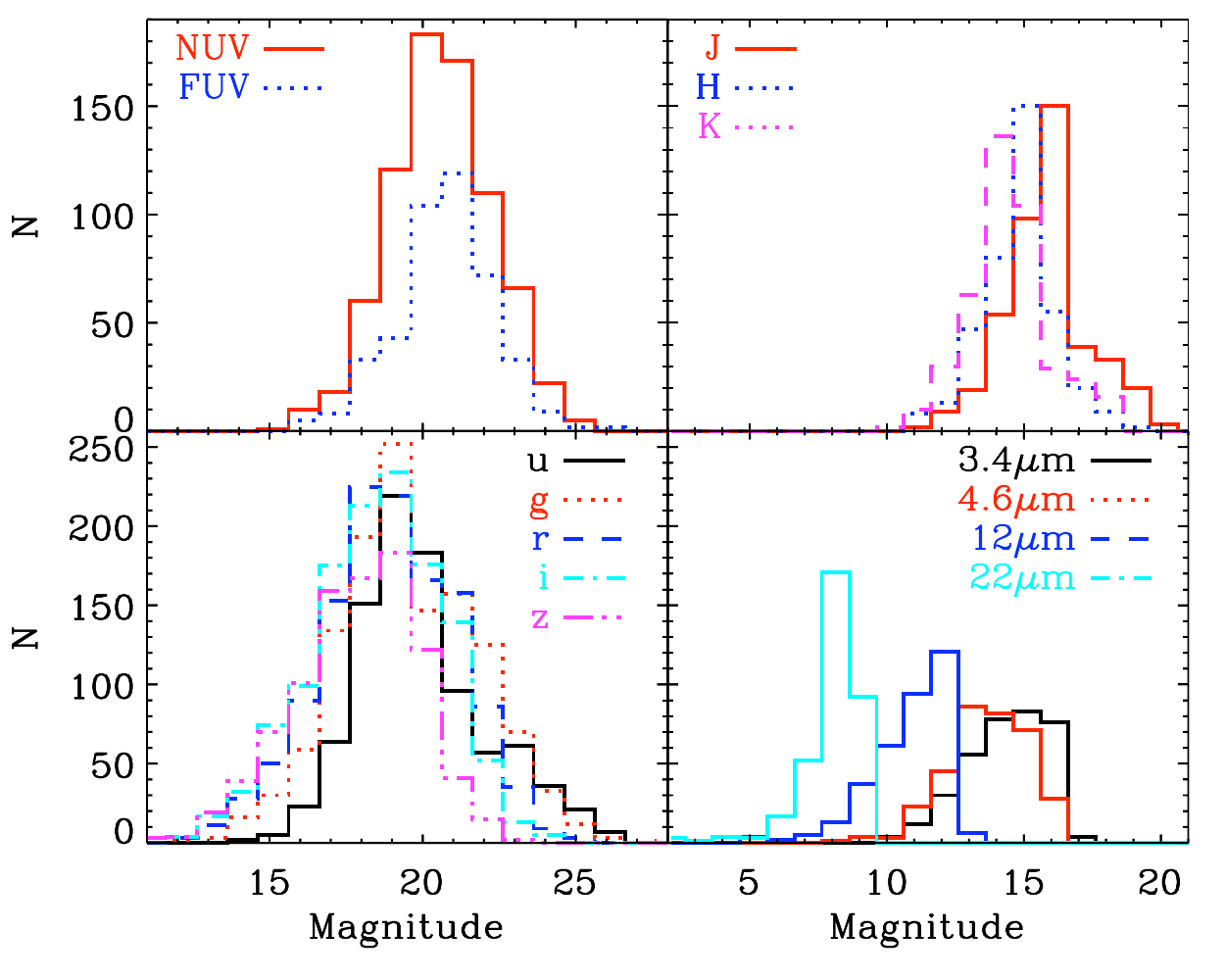}
\end{minipage}
\label{maghistos}
\end{center}
    \caption
           {\scriptsize Multi-wavelength magnitude distributions for all our 1242 spectroscopically identified ChaMP extragalactic sources. UV and optical magnitudes are in the AB system, near- and mid-infrared magnitudes are in the Vega system.}
\end{figure}
\begin{figure*} 
\begin{center}
\begin{minipage}[c]{8.cm}
        \includegraphics[width=1.0\textwidth]{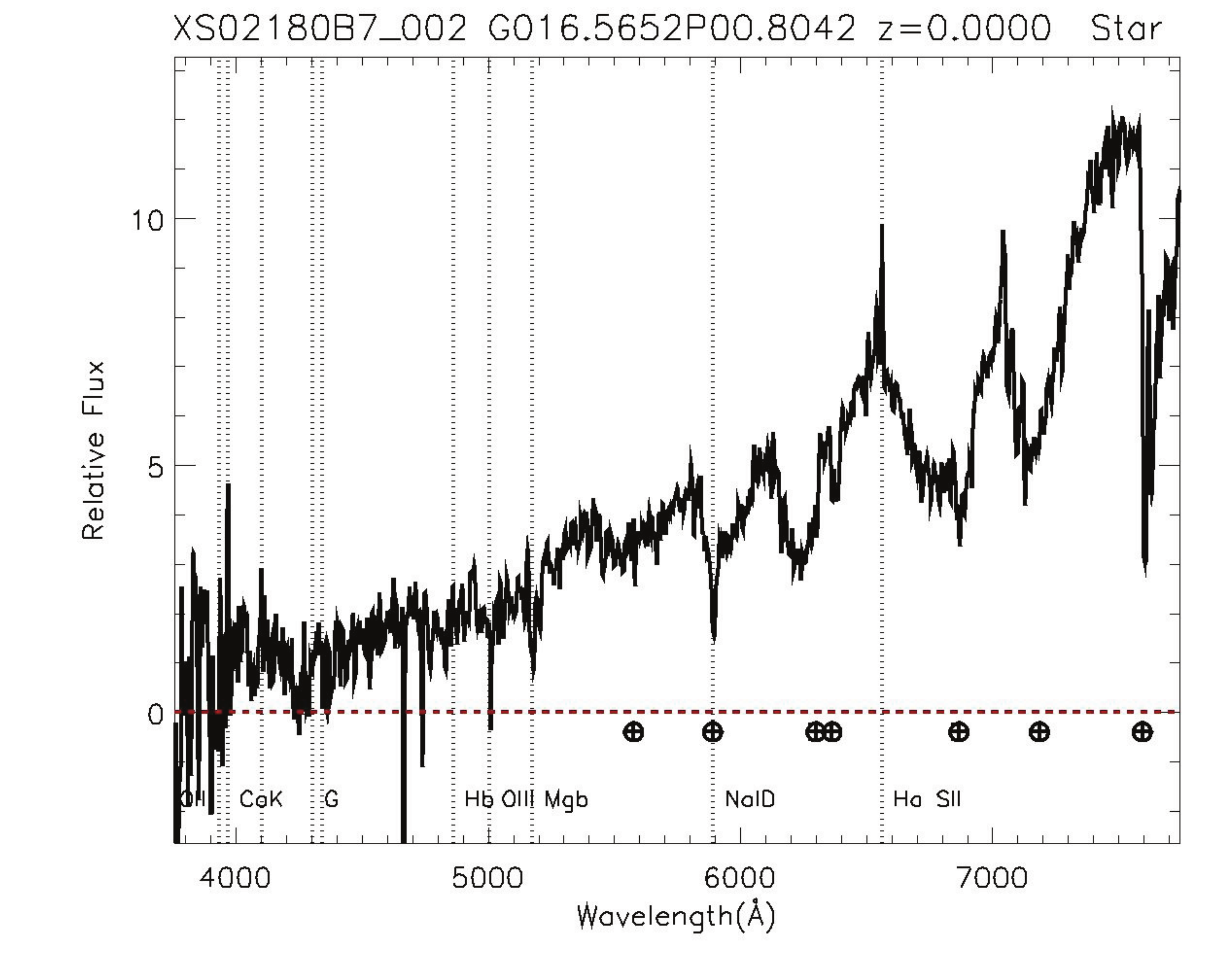}
\end{minipage}
\begin{minipage}[c]{8.cm}
        \includegraphics[width=1.0\textwidth]{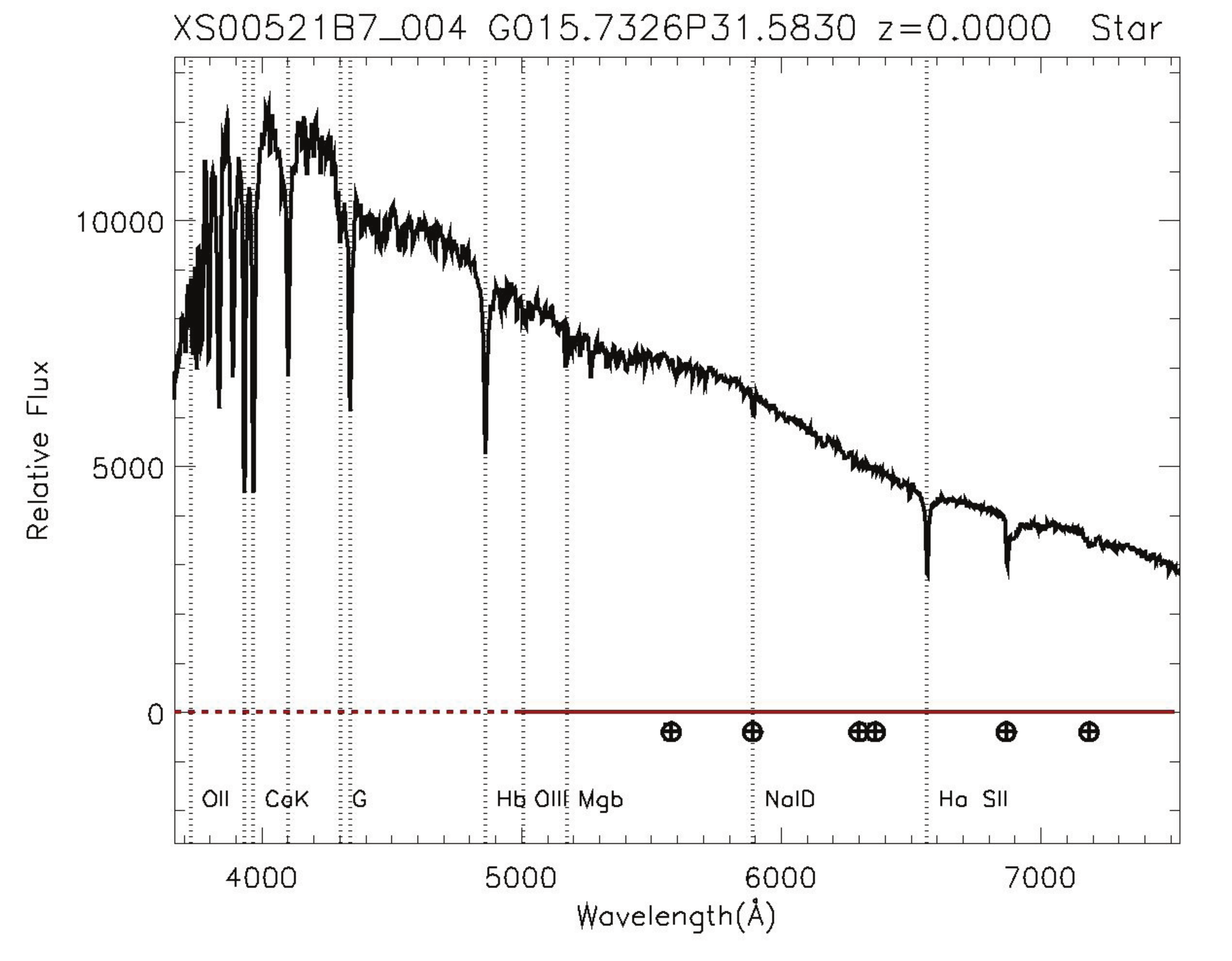}
\end{minipage}\\ 
\begin{minipage}[c]{8cm}
        \includegraphics[width=1.0\textwidth]{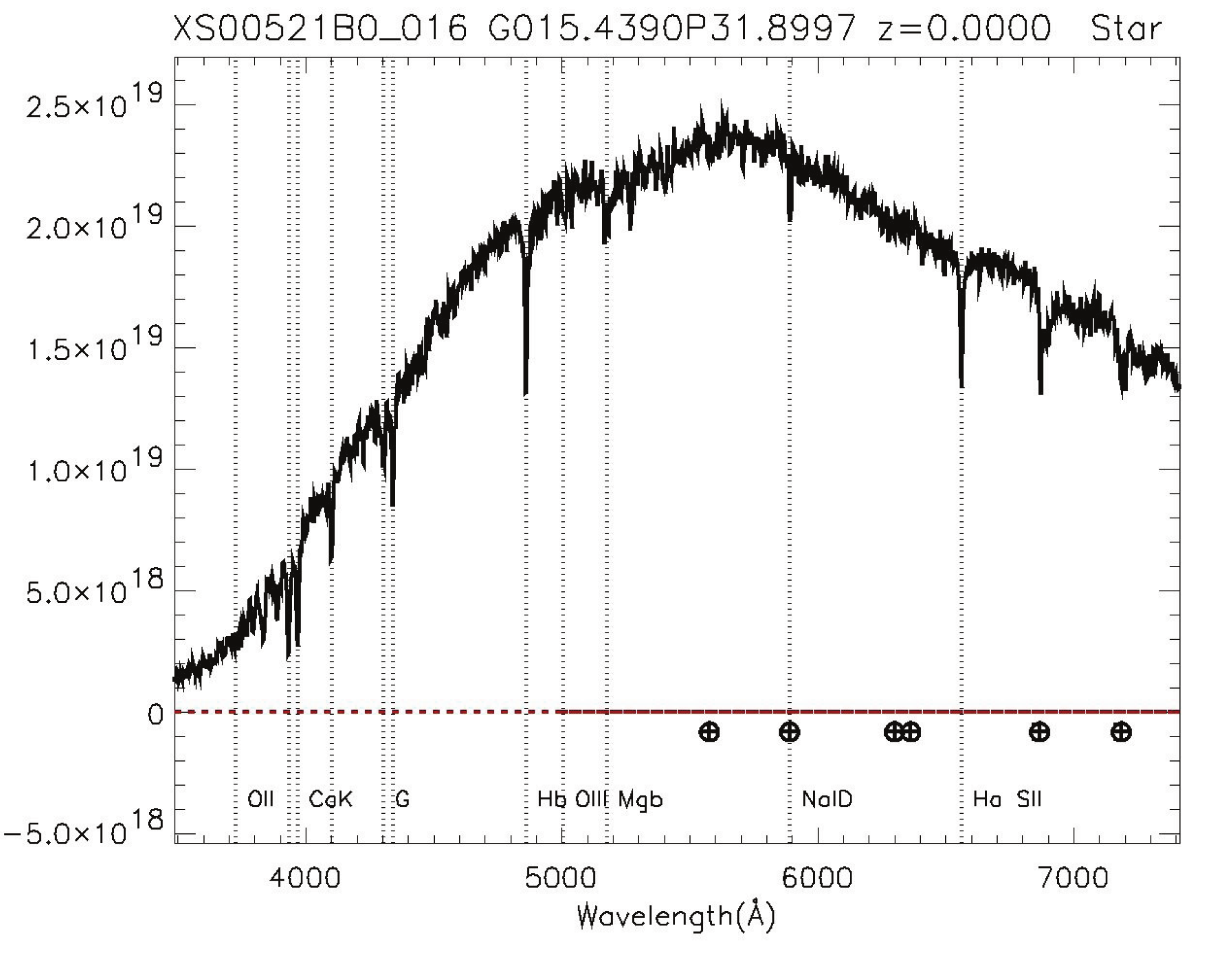}
\end{minipage}
\begin{minipage}[c]{8cm}
        \includegraphics[width=1.0\textwidth]{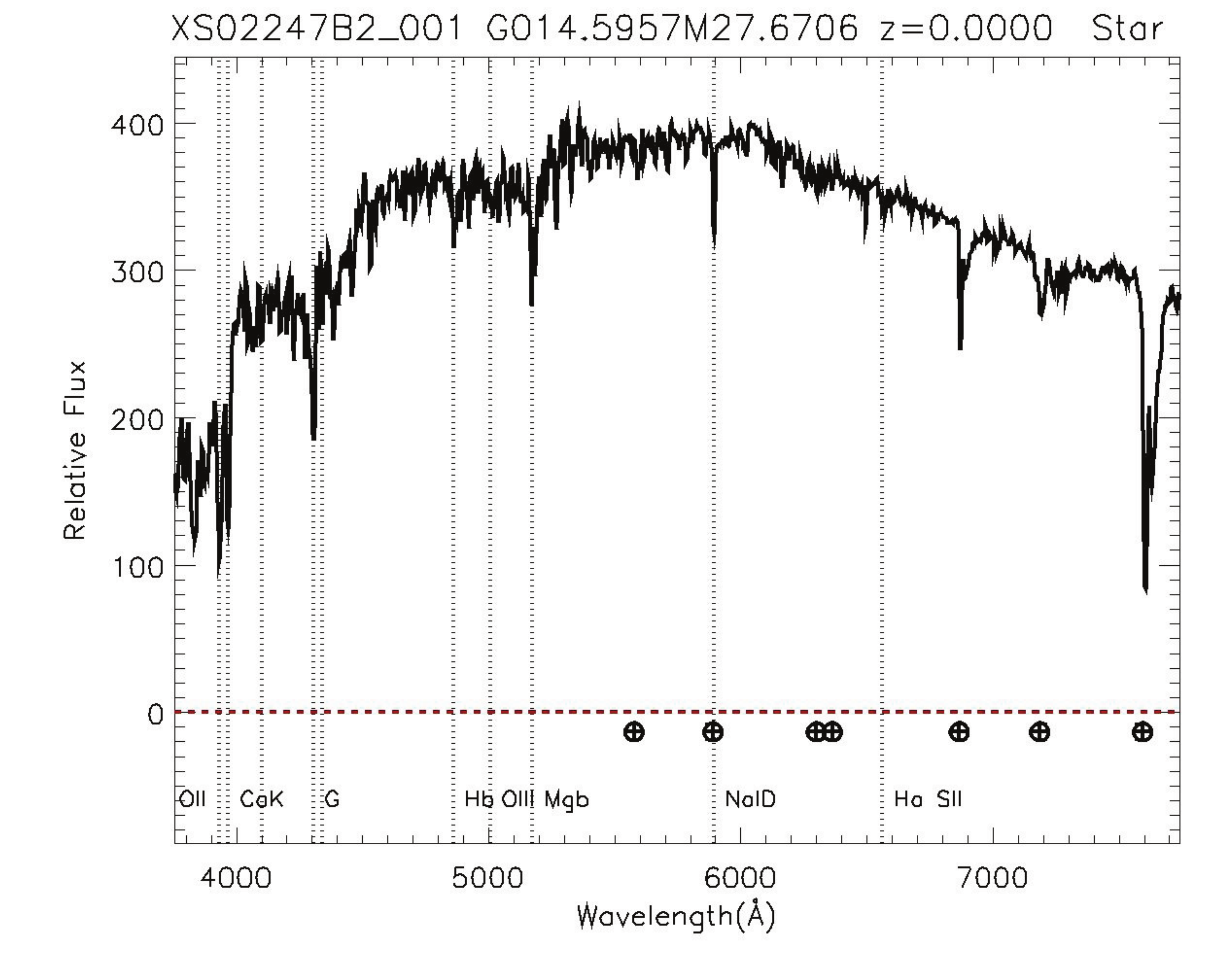}
\end{minipage}\\
\begin{minipage}[c]{8cm}
        \includegraphics[width=1.0\textwidth]{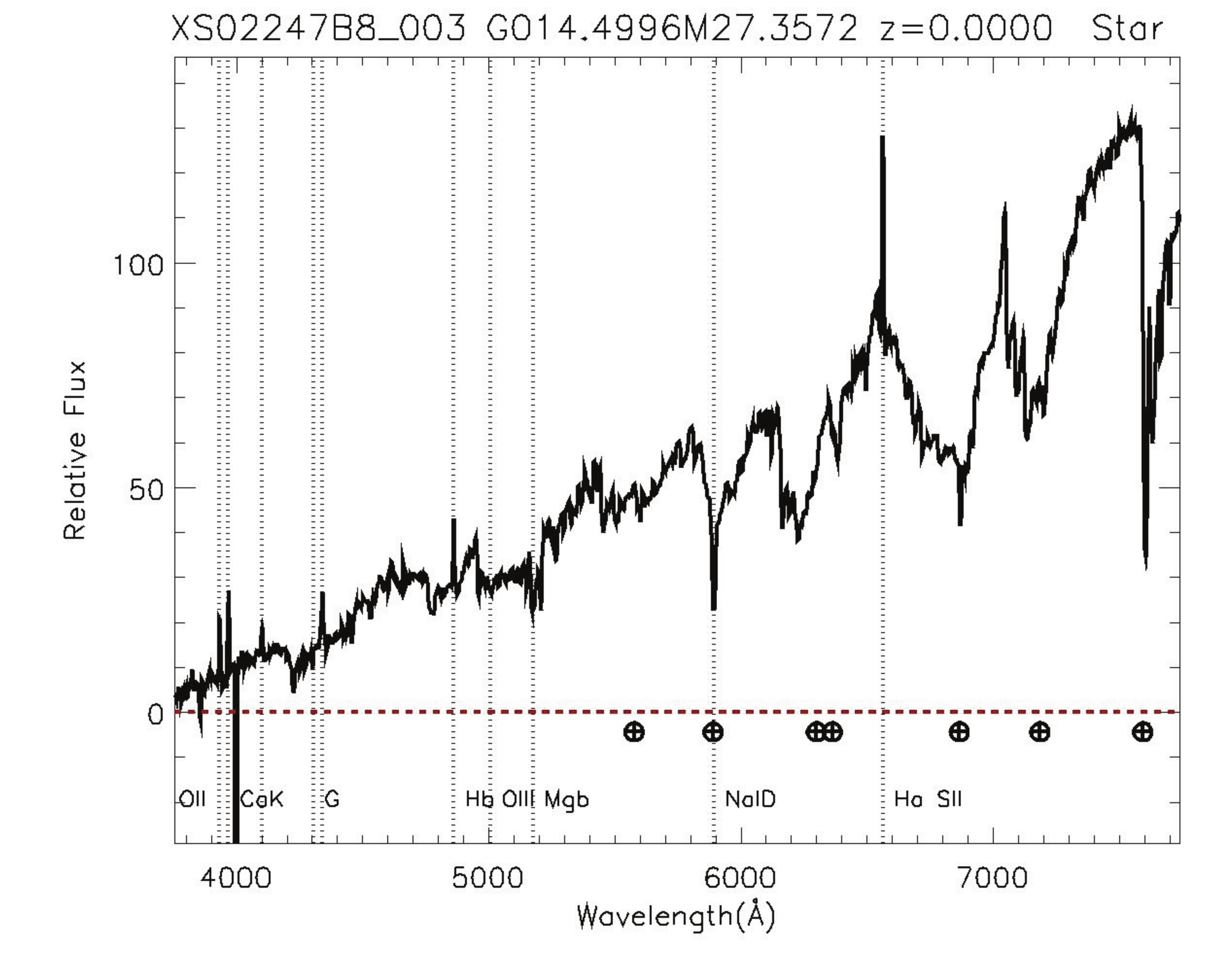}
\end{minipage}
\begin{minipage}[c]{8cm}
        \includegraphics[width=1.0\textwidth]{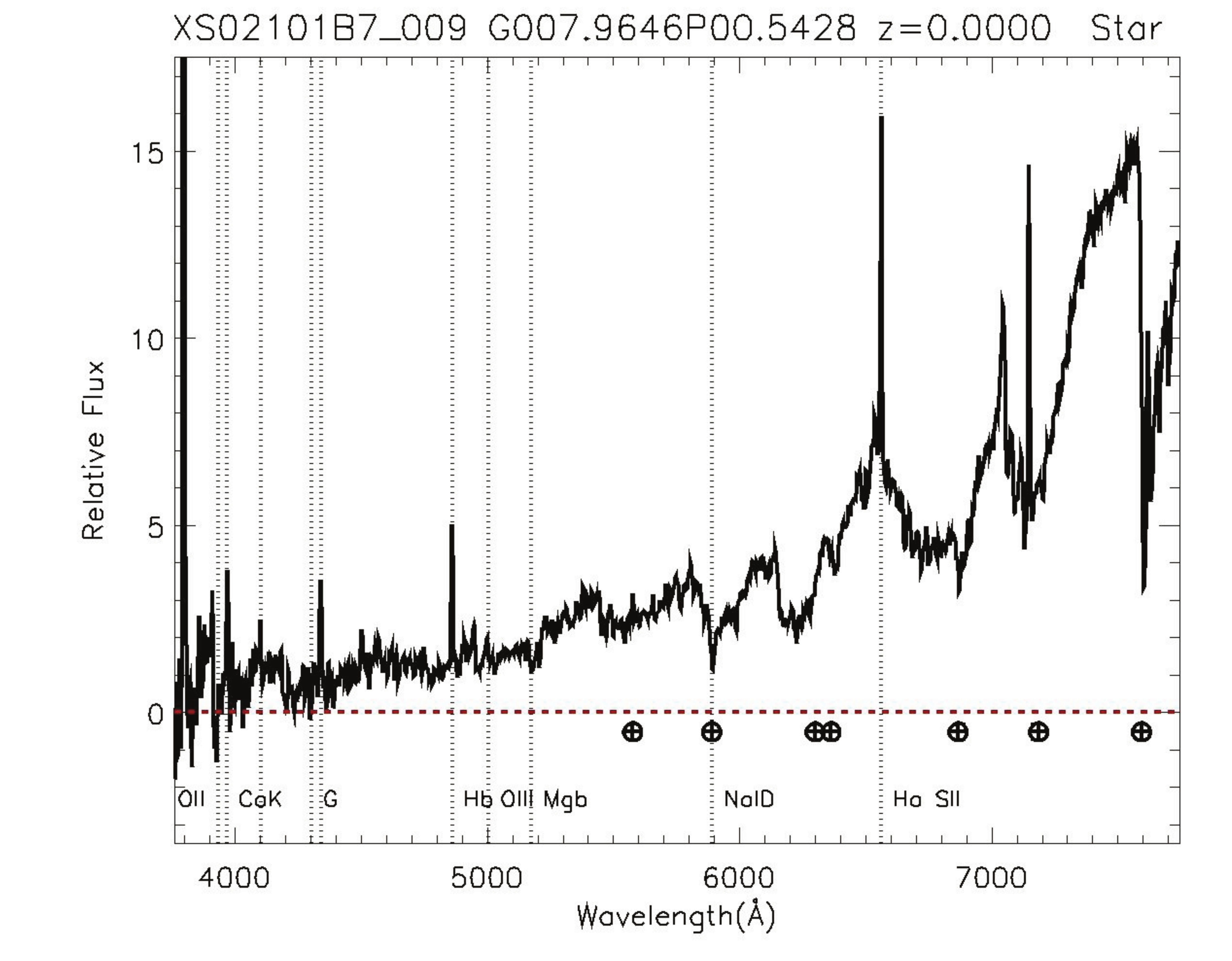}
\end{minipage}
\end{center}
    \caption
           {\scriptsize Example ChaMP spectra of stellar objects observed with FAST. CHANDRAOBSID, SPECOBJID, REDSHIFT and CLASS are given in the top of each plot.}
\end{figure*}
\begin{figure*} 
\begin{center}
\begin{minipage}[c]{8.cm}
        \includegraphics[width=1.0\textwidth]{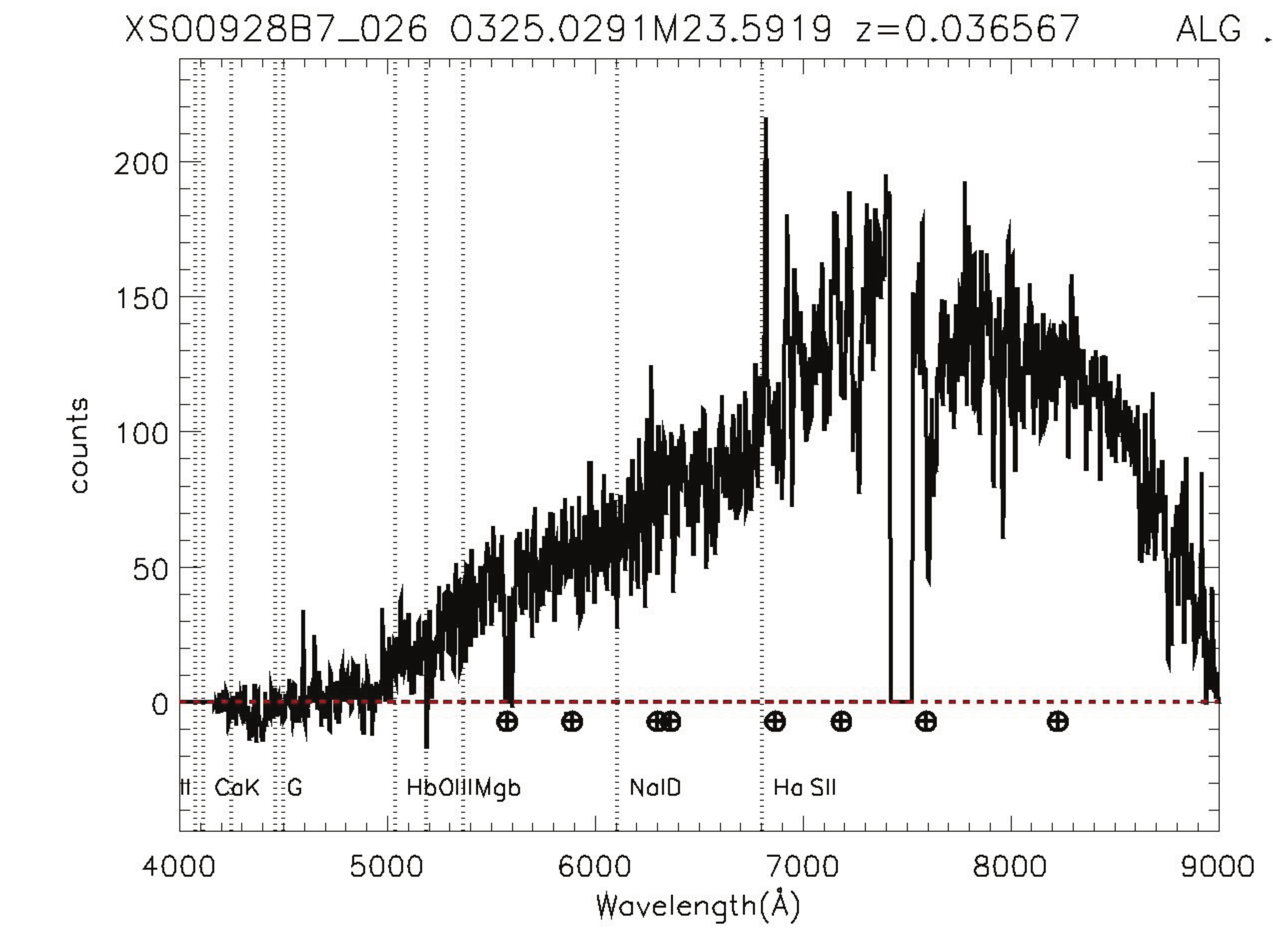}
\end{minipage}
\begin{minipage}[c]{8.cm}
        \includegraphics[width=1.0\textwidth]{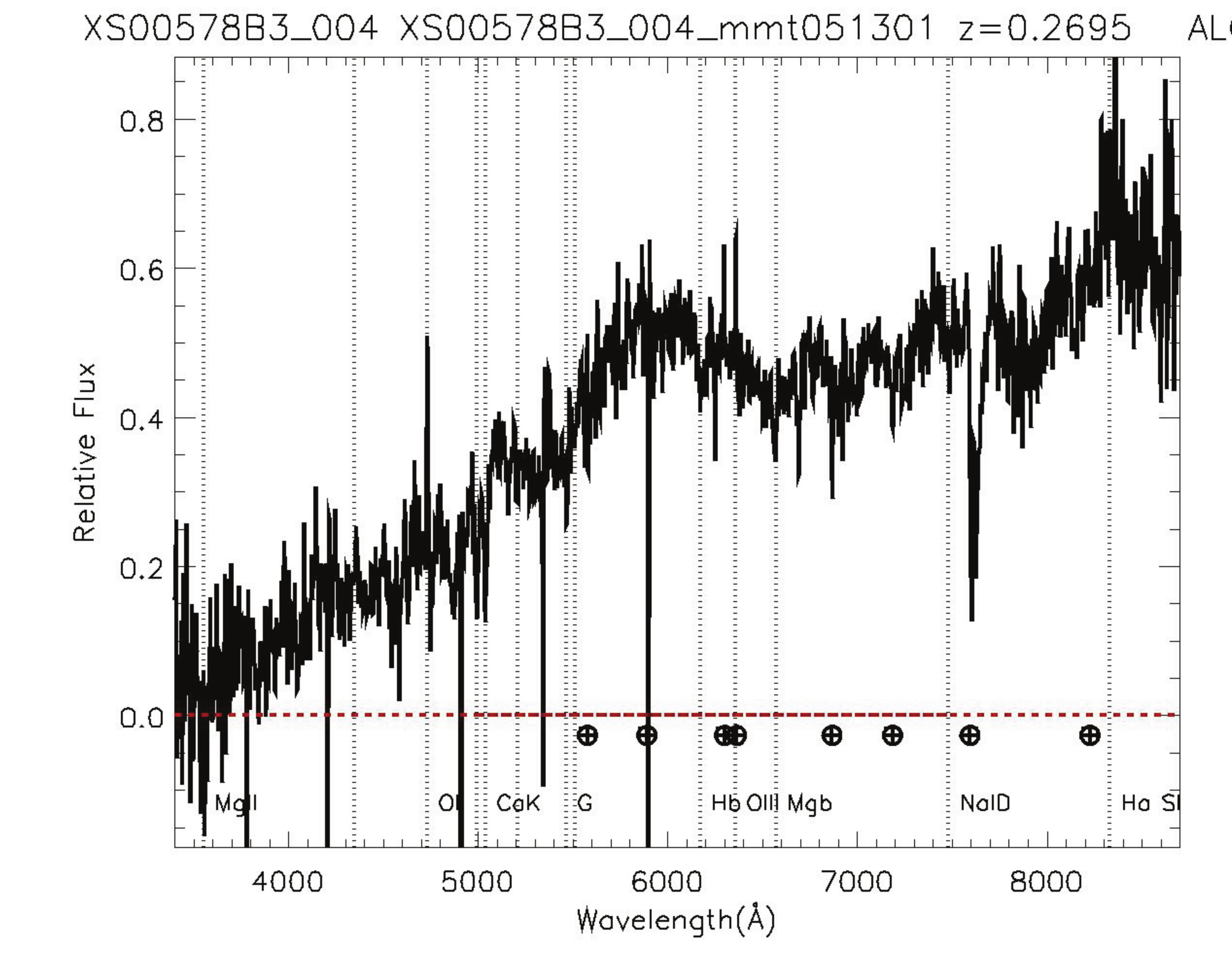}
\end{minipage}\\ 
\begin{minipage}[c]{8cm}
        \includegraphics[width=1.0\textwidth]{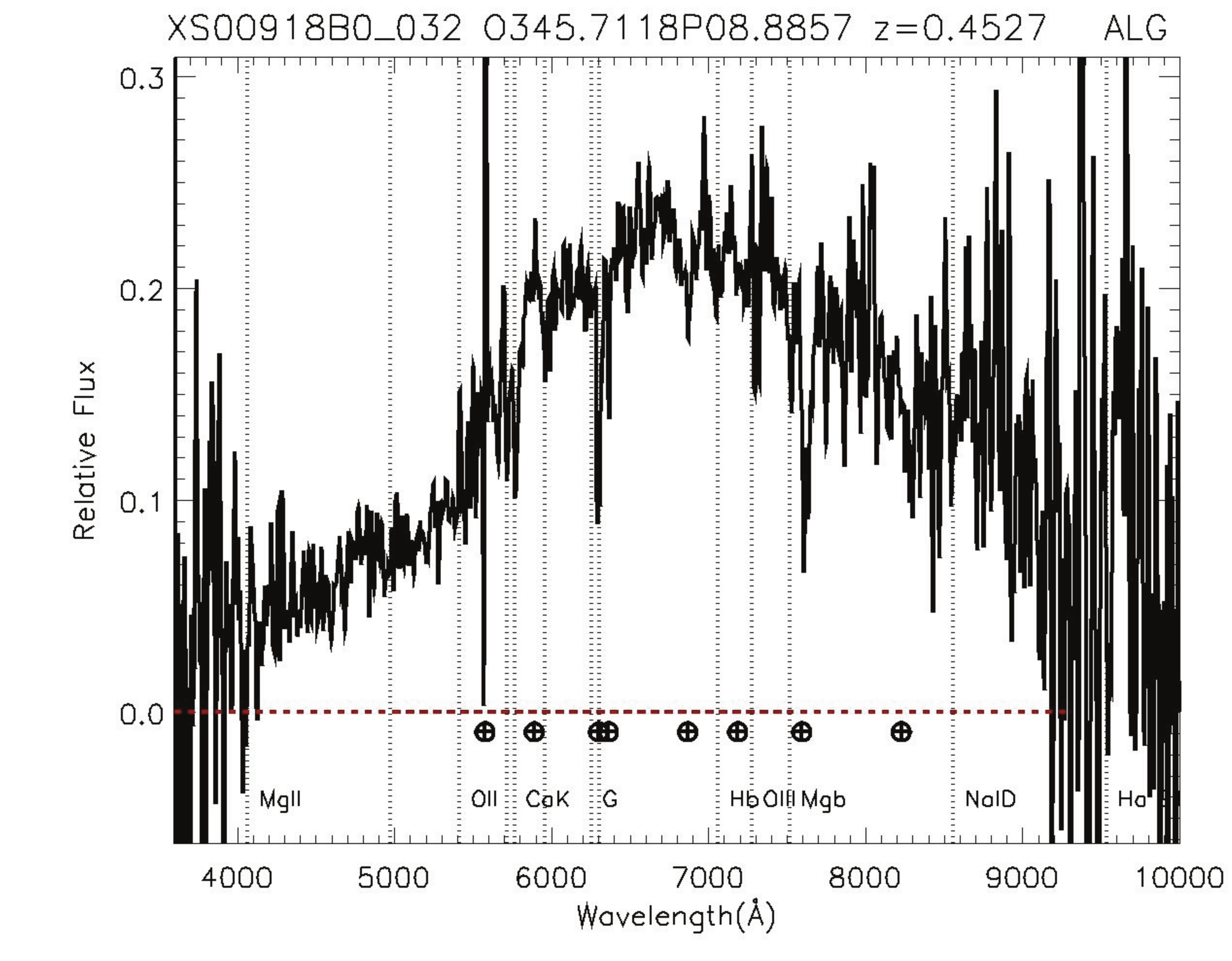}
\end{minipage}
\begin{minipage}[c]{8cm}
        \includegraphics[width=1.0\textwidth]{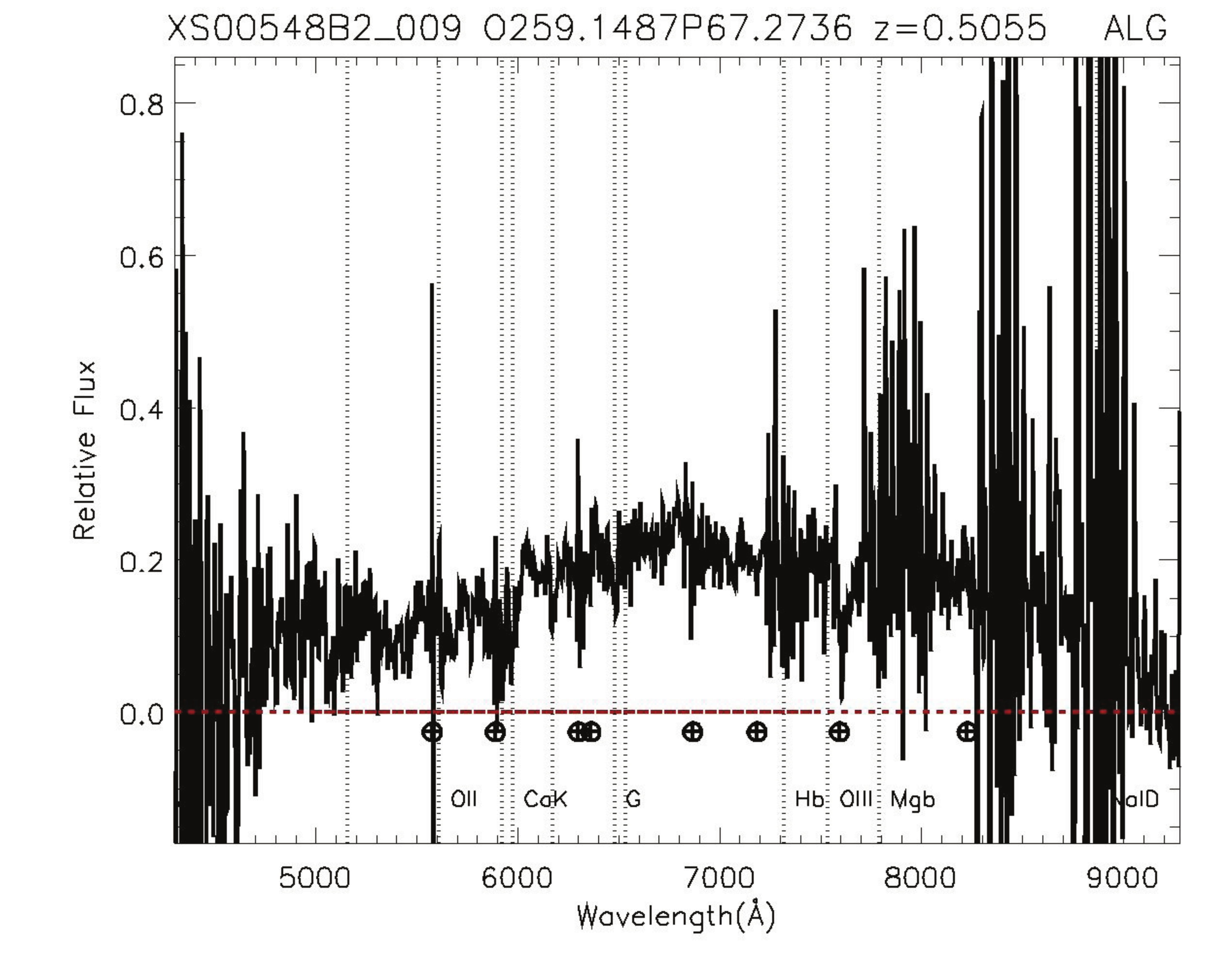}
\end{minipage}\\
\begin{minipage}[c]{8cm}
        \includegraphics[width=1.0\textwidth]{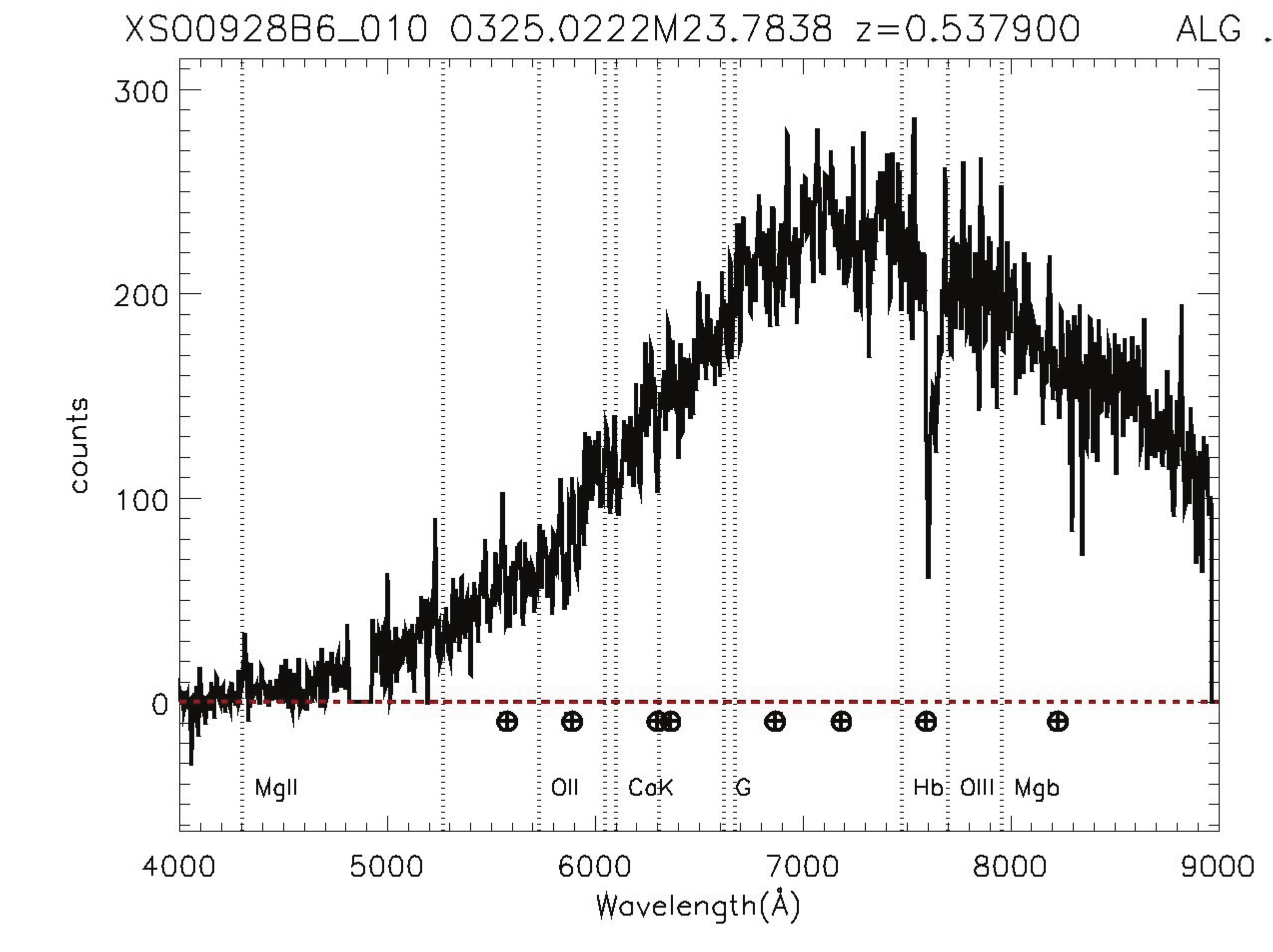}
\end{minipage}
\begin{minipage}[c]{8cm}
        \includegraphics[width=1.0\textwidth]{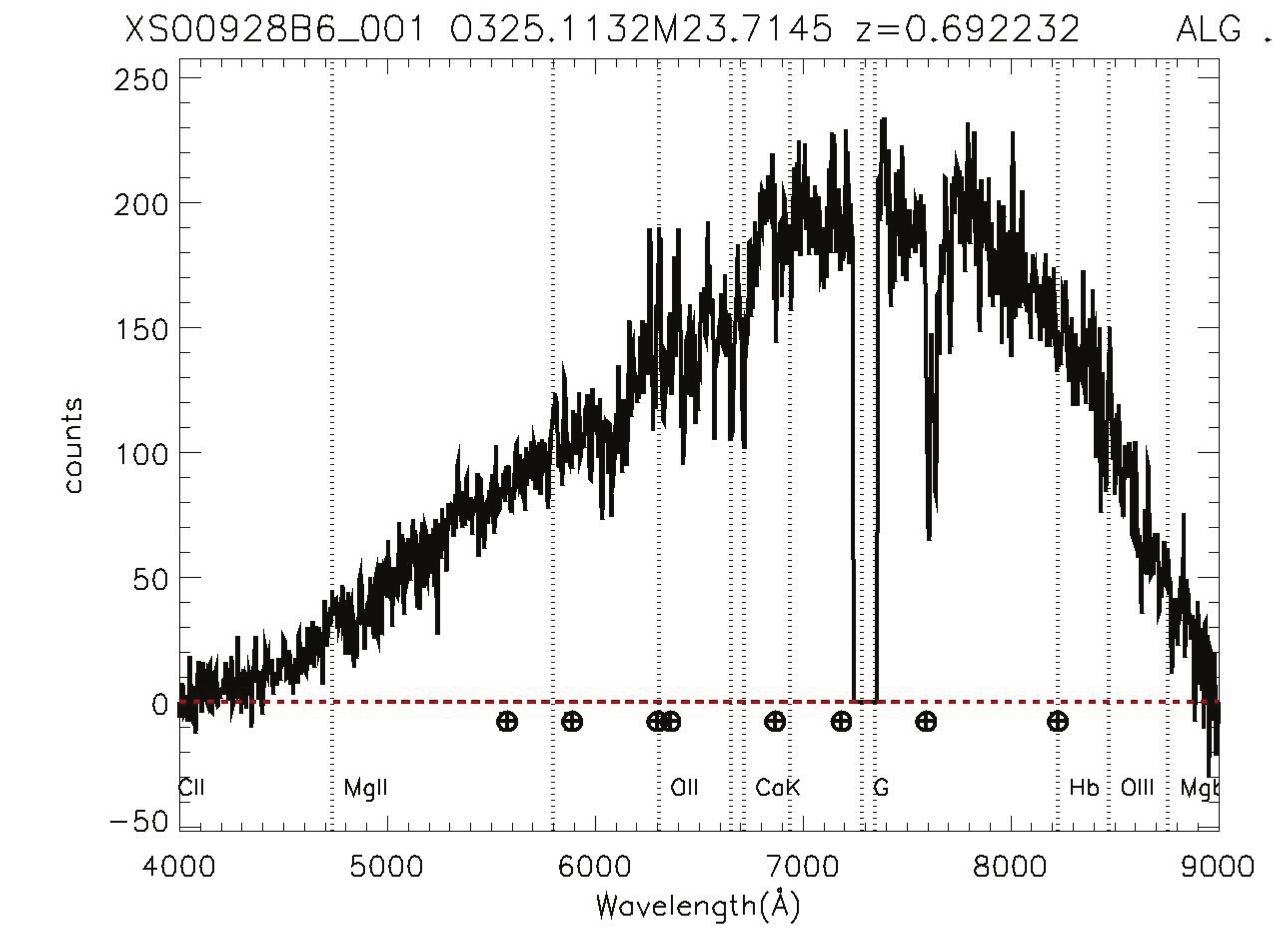}
\end{minipage}
\end{center}
    \caption
           {\scriptsize Example ChaMP spectra of absorption line galaxies  observed with Magellan, MMT and WIYN telescopes. Spectra are not flux calibrated. CHANDRAOBSID, SPECOBJID, REDSHIFT and CLASS are given in the top of each plot.}
\end{figure*}
\begin{figure*} 
\begin{center}
\begin{minipage}[c]{8.cm}
        \includegraphics[width=1.0\textwidth]{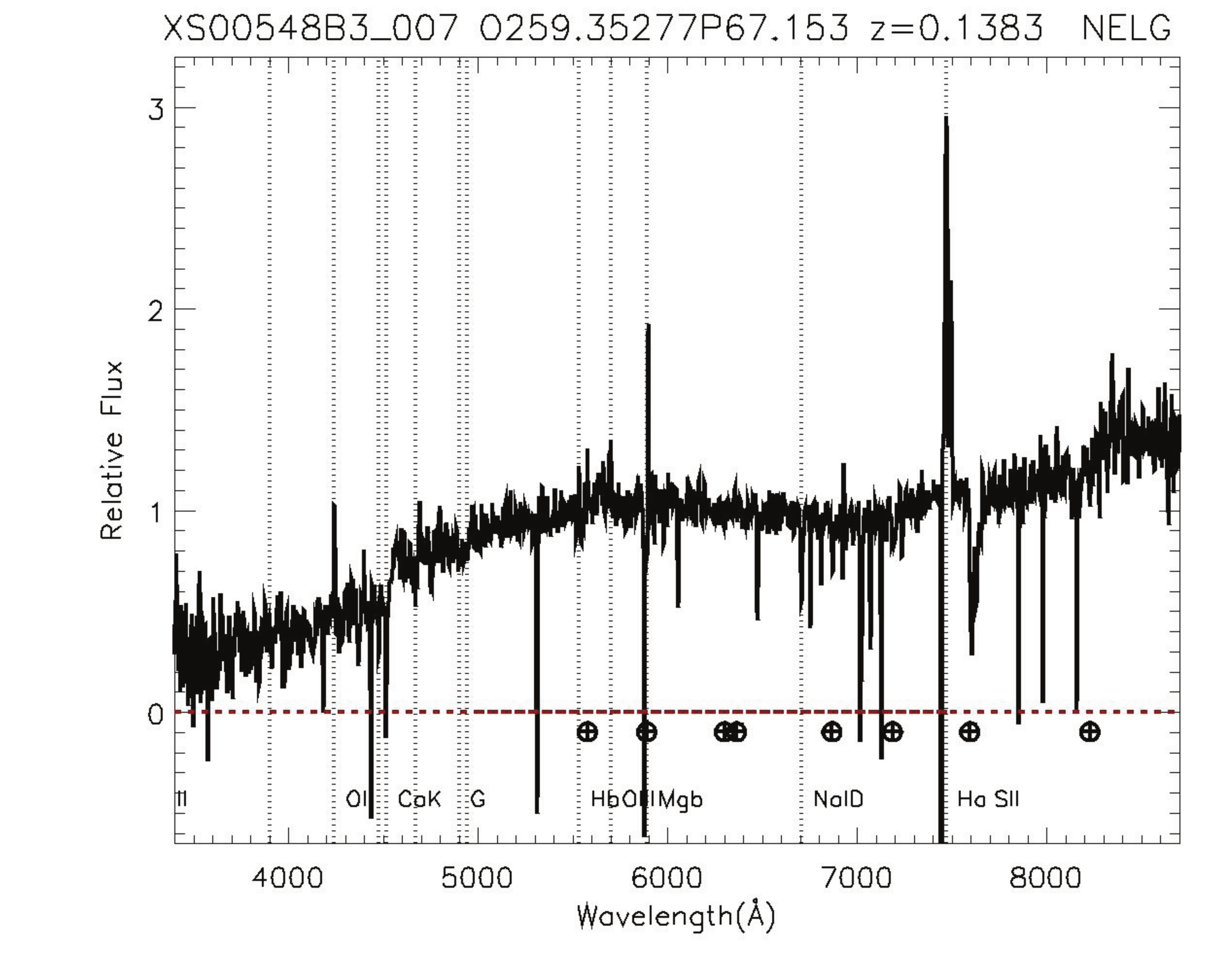}
\end{minipage}
\begin{minipage}[c]{8.cm}
        \includegraphics[width=1.0\textwidth]{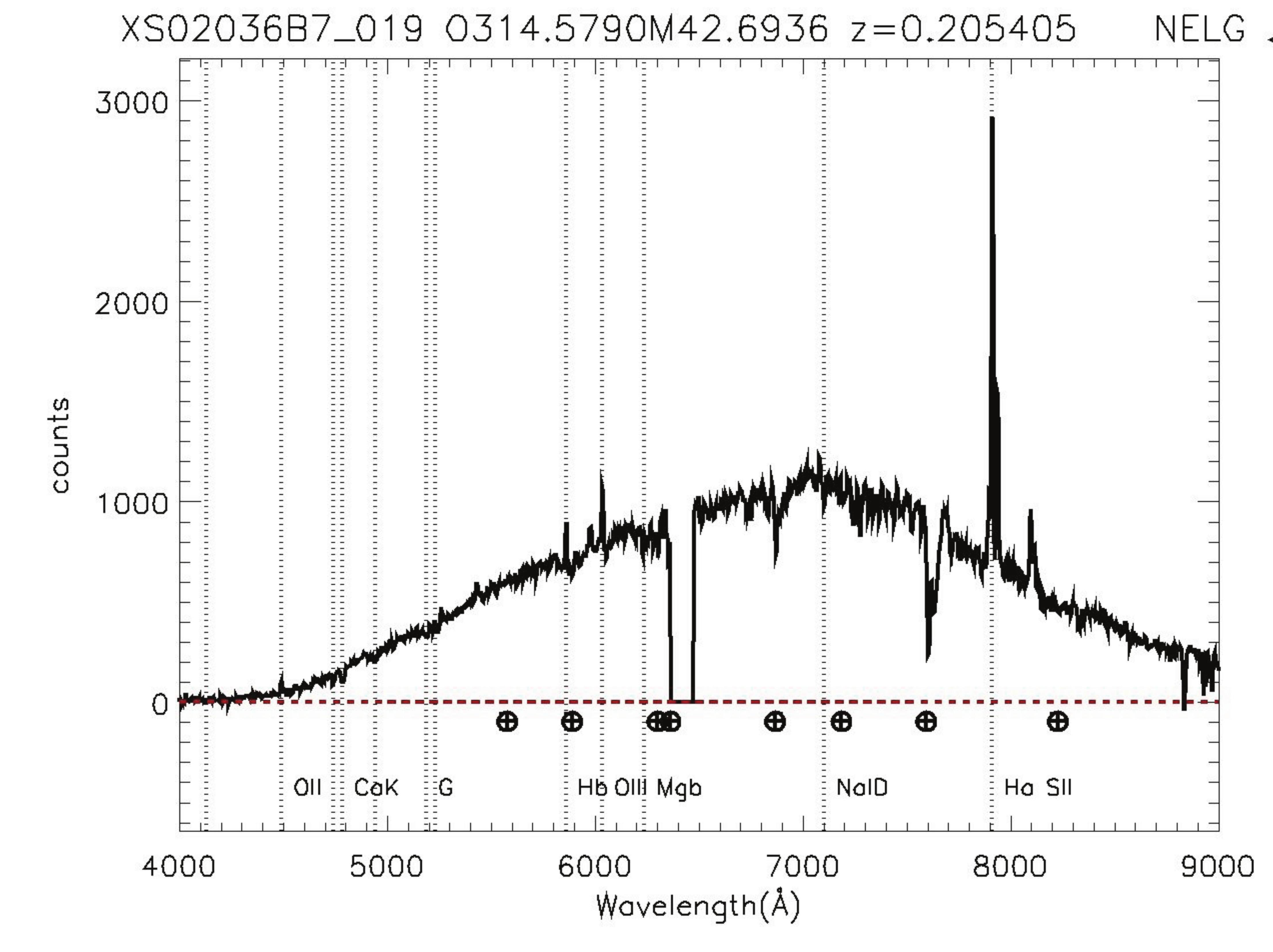}
\end{minipage}\\ 
\begin{minipage}[c]{8cm}
        \includegraphics[width=1.0\textwidth]{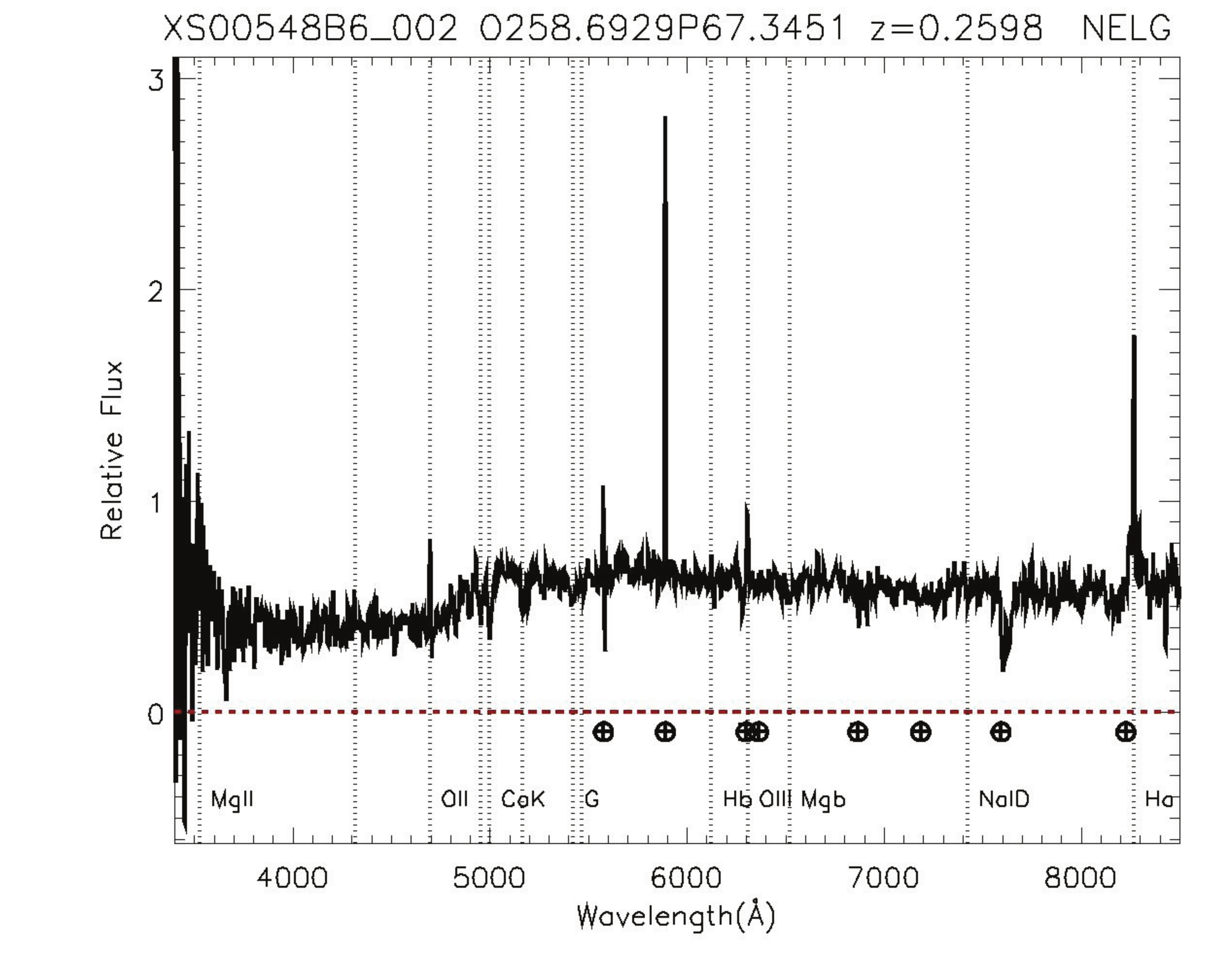}
\end{minipage}
\begin{minipage}[c]{8cm}
        \includegraphics[width=1.0\textwidth]{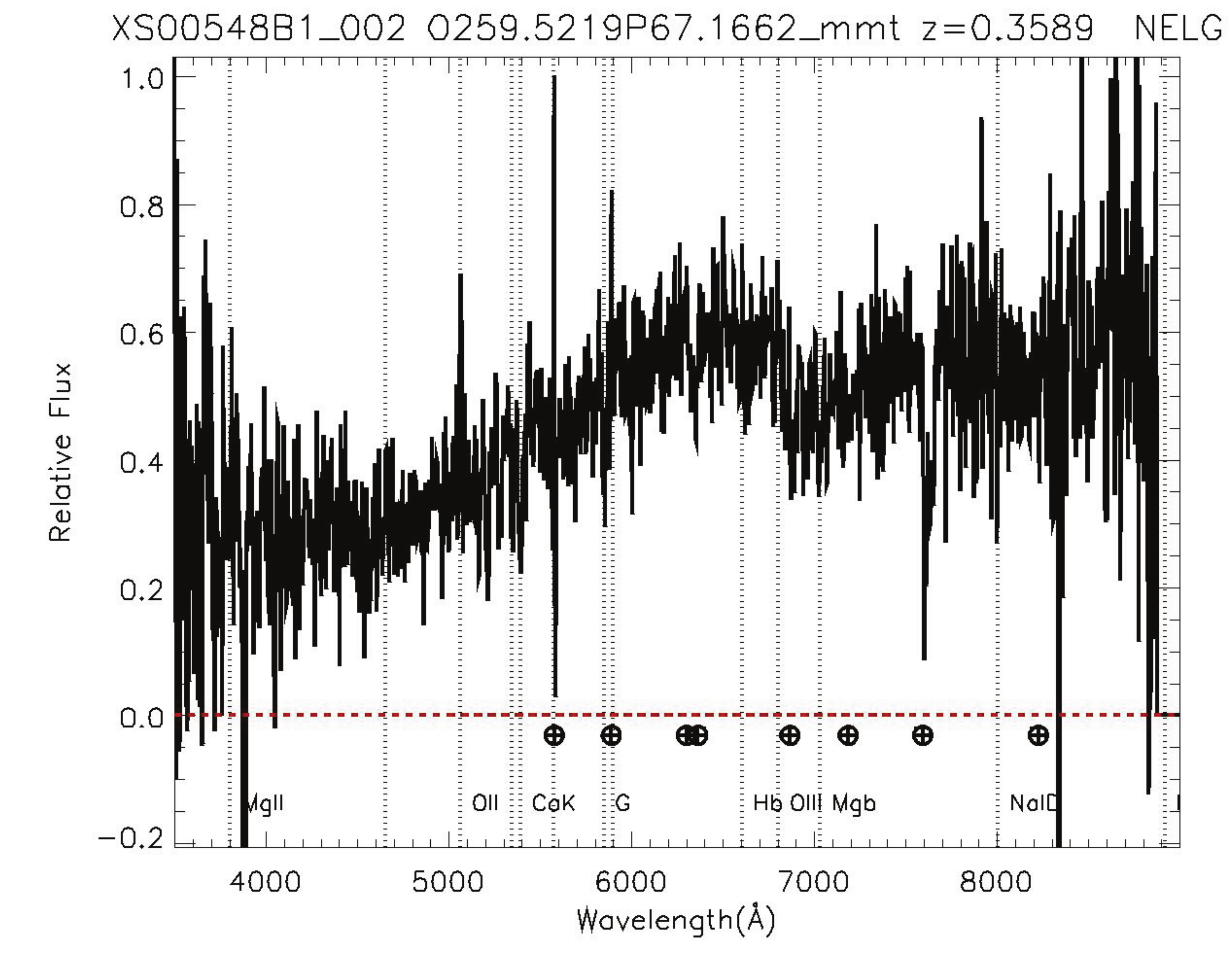}
\end{minipage}\\
\begin{minipage}[c]{8cm}
        \includegraphics[width=1.0\textwidth]{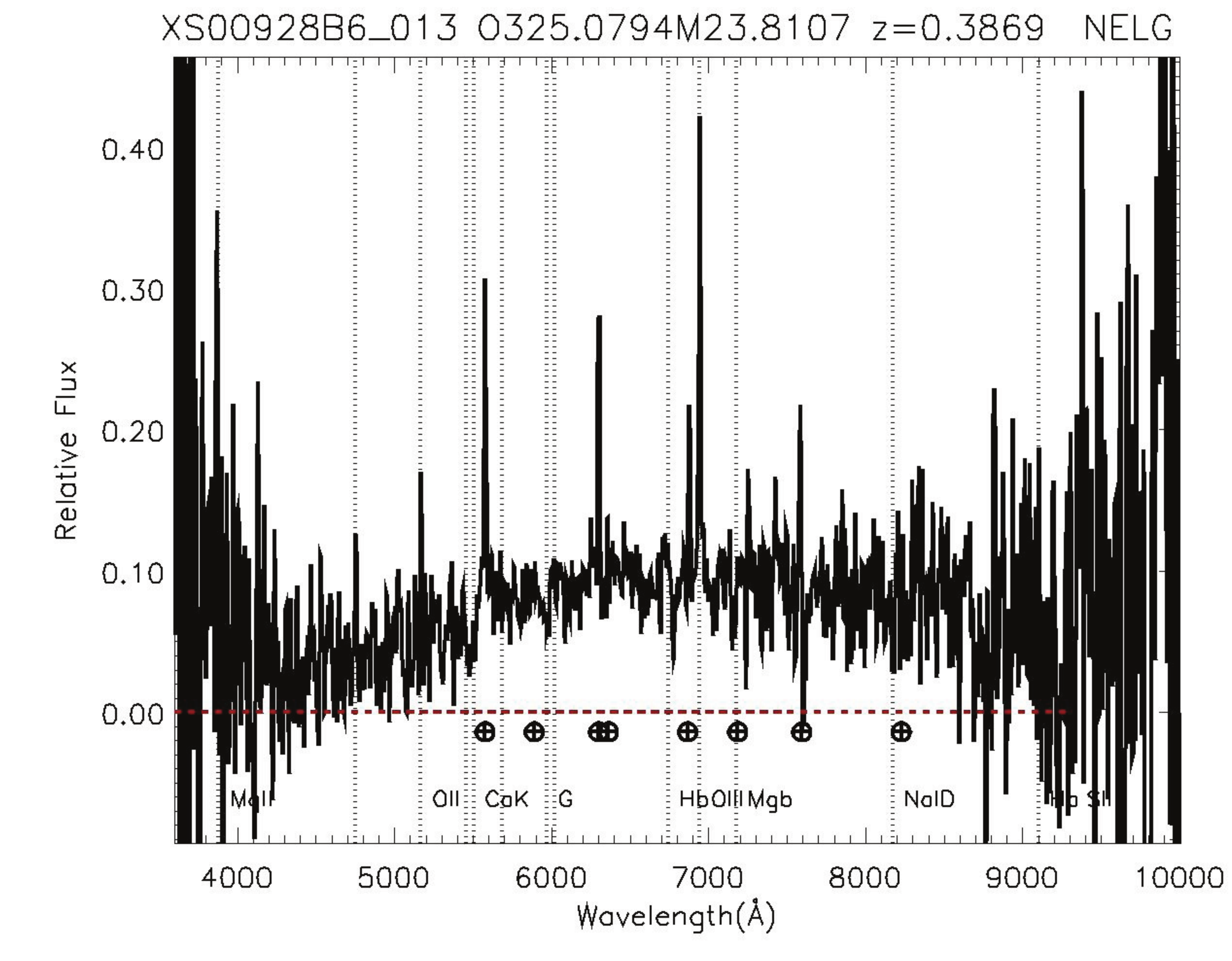}
\end{minipage}
\begin{minipage}[c]{8cm}
        \includegraphics[width=1.0\textwidth]{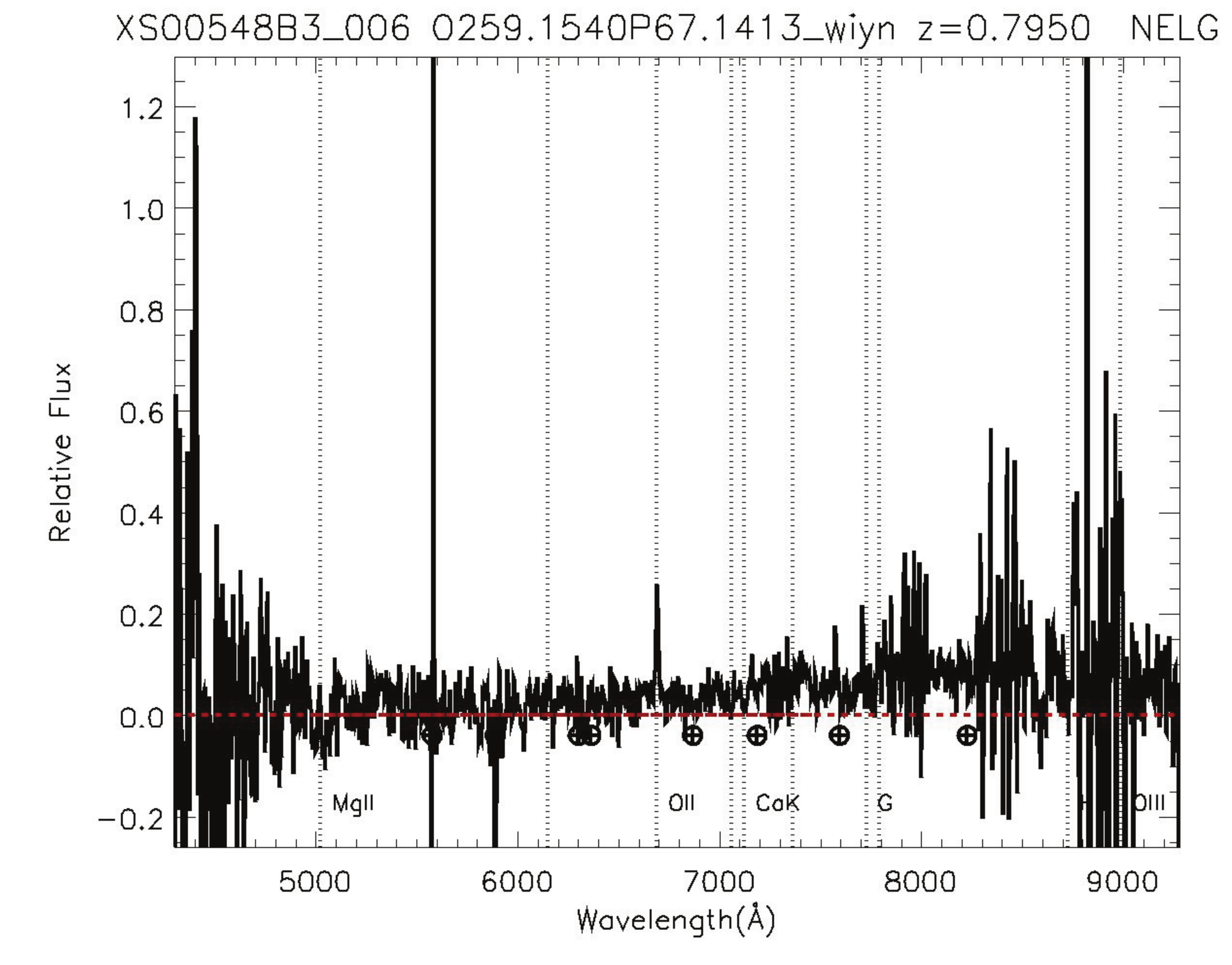}
\end{minipage}
\end{center}
    \caption
           {\scriptsize Example ChaMP spectra of narrow emission line galaxies observed with Magellan, MMT and WIYN telescopes. Spectra are not flux calibrated. CHANDRAOBSID, SPECOBJID, REDSHIFT and CLASS are given in the top of each plot.}
\end{figure*}
\begin{figure*} 
\begin{center}
\begin{minipage}[c]{8.cm}
        \includegraphics[width=1.0\textwidth]{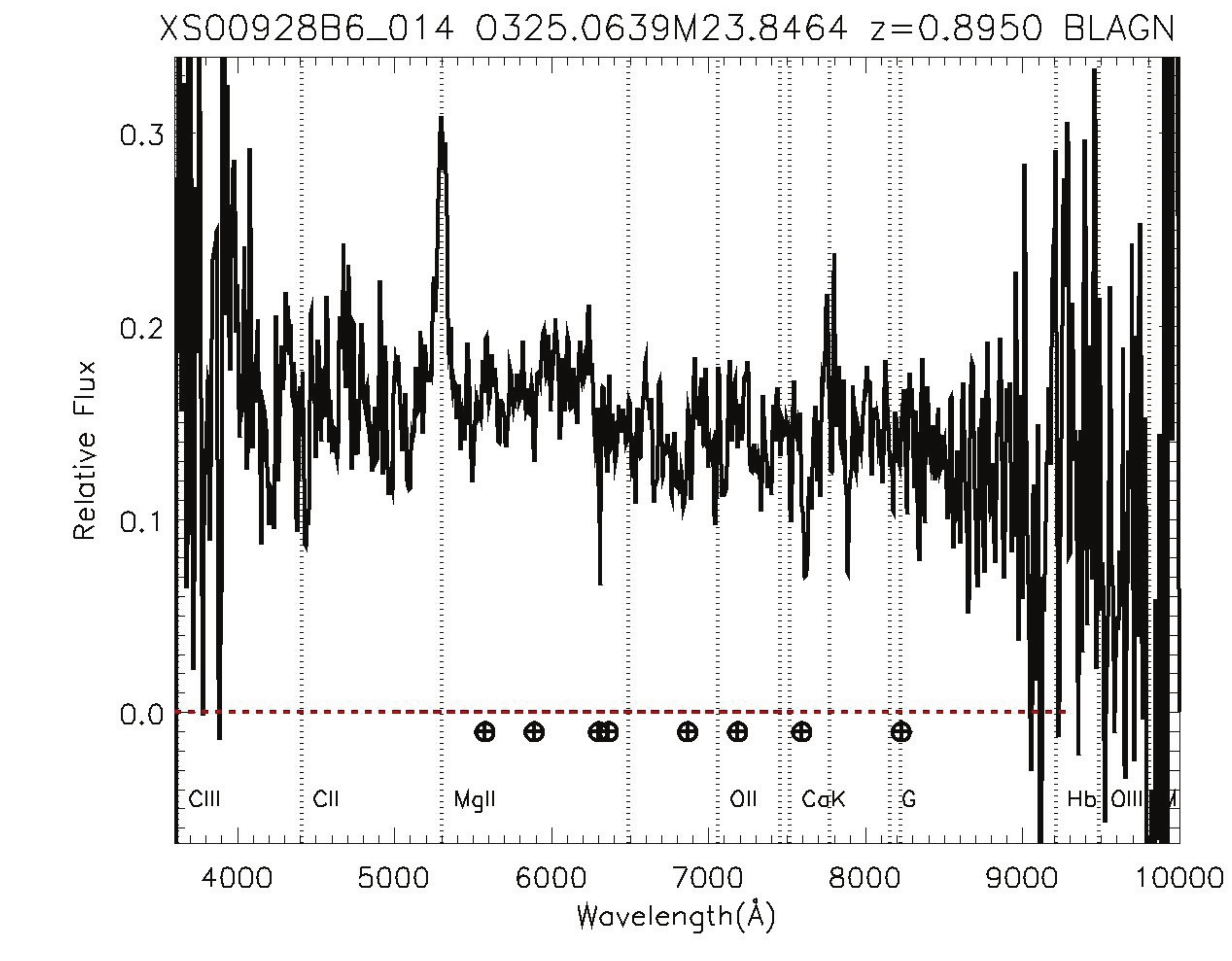}
\end{minipage}
\begin{minipage}[c]{8.cm}
        \includegraphics[width=1.0\textwidth]{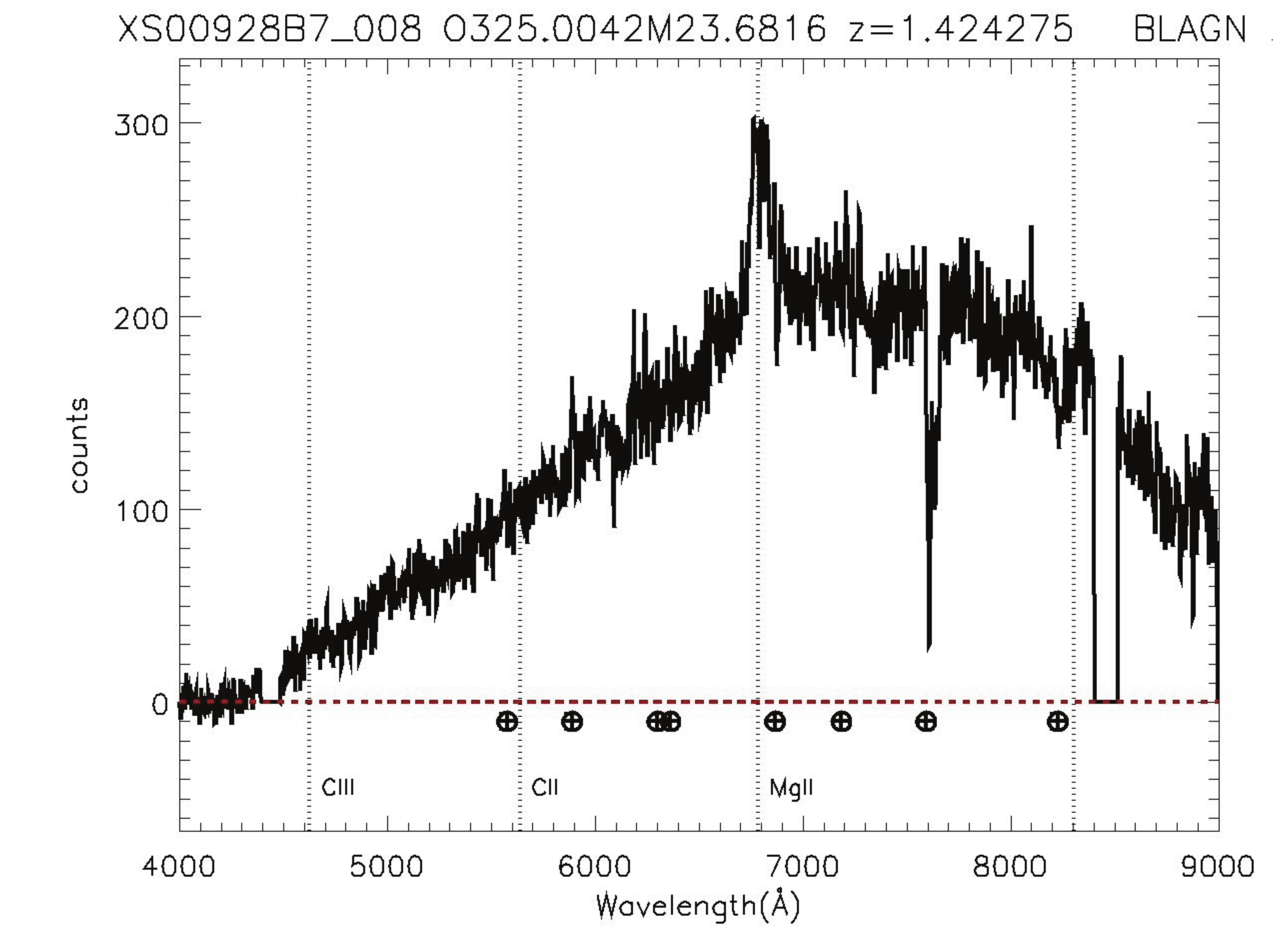}
\end{minipage}\\ 
\begin{minipage}[c]{8cm}
        \includegraphics[width=1.0\textwidth]{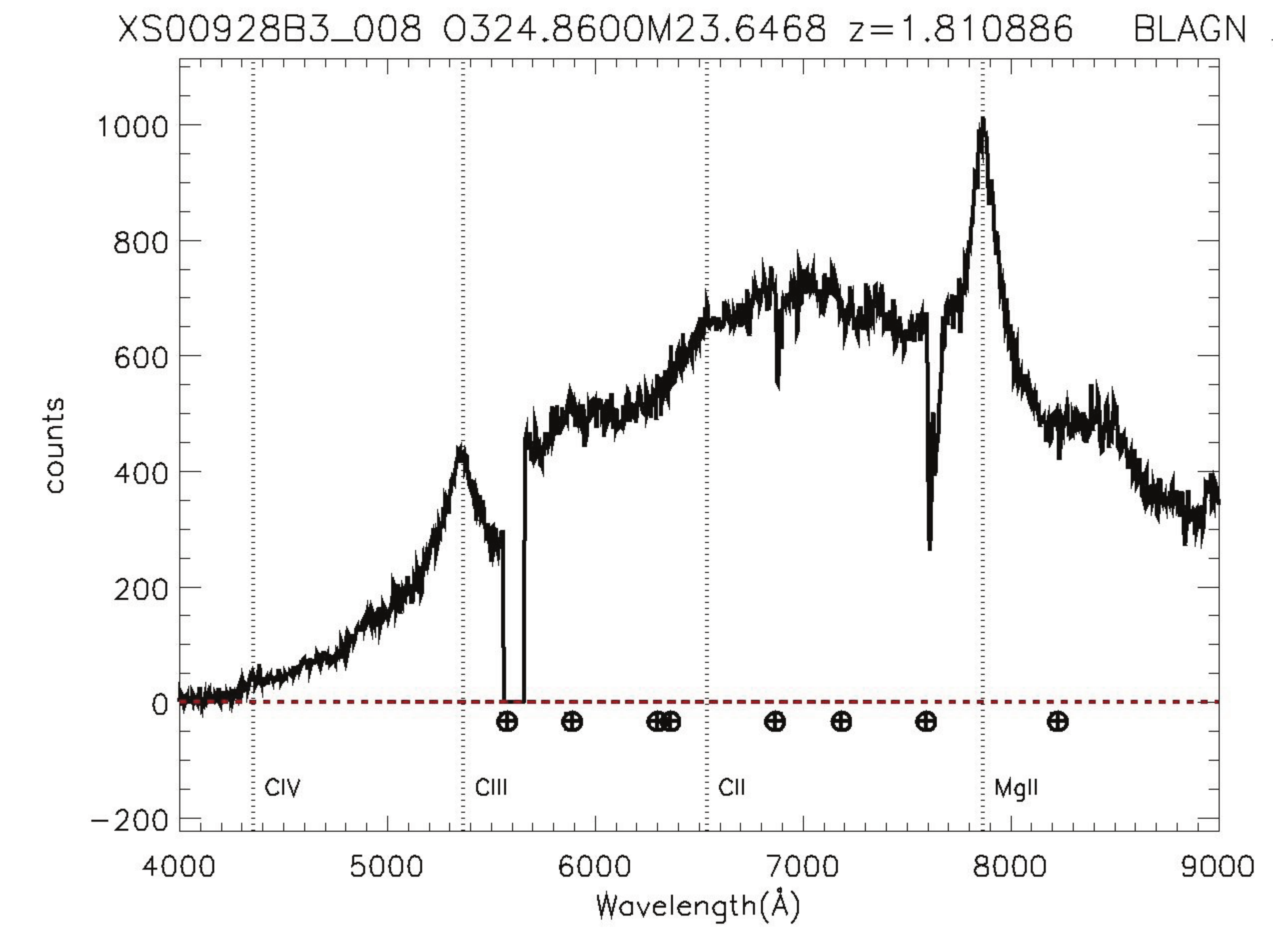}
\end{minipage}
\begin{minipage}[c]{8cm}
        \includegraphics[width=1.0\textwidth]{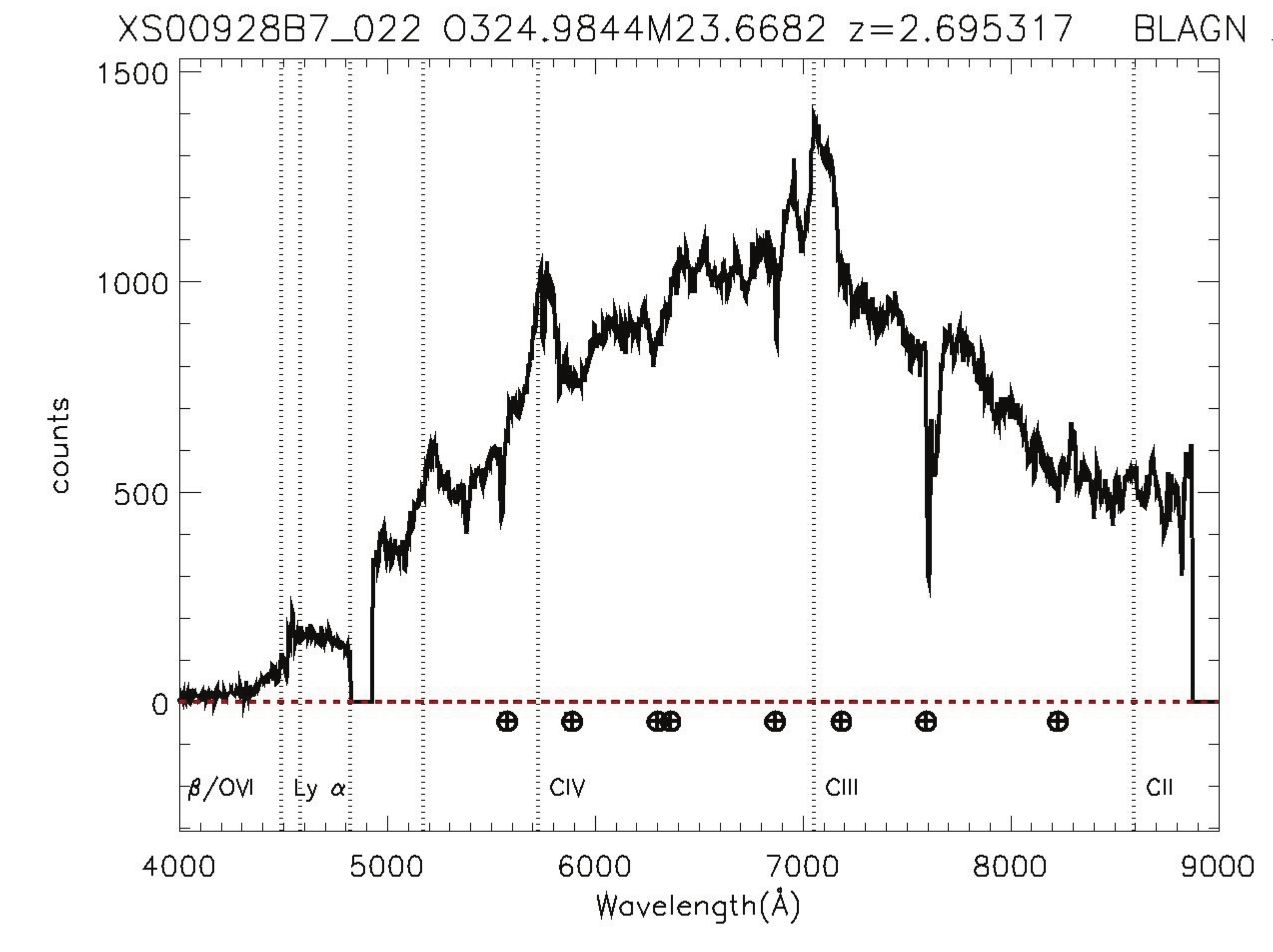}
\end{minipage}\\
\begin{minipage}[c]{8cm}
        \includegraphics[width=1.0\textwidth]{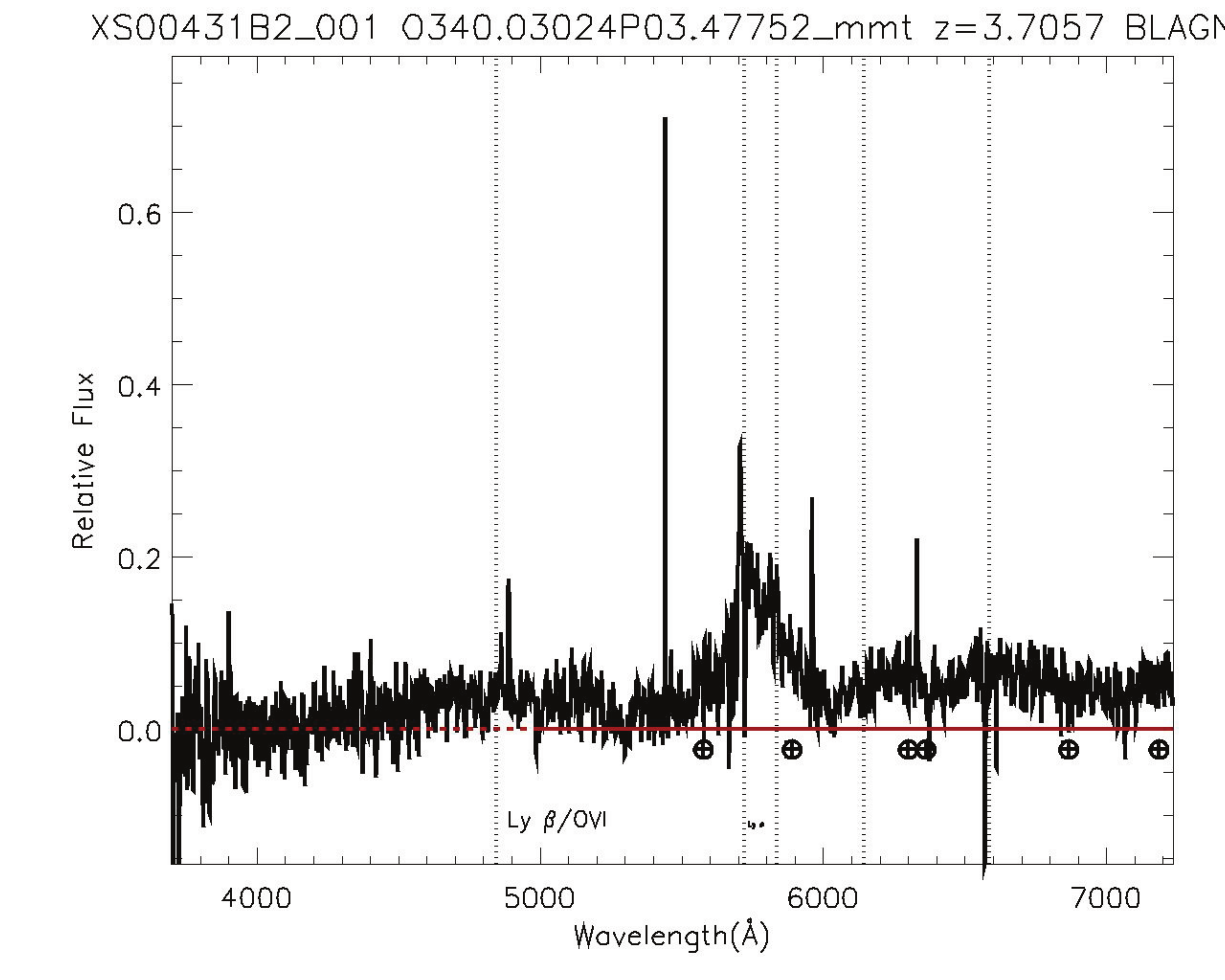}
\end{minipage}
\begin{minipage}[c]{8cm}
        \includegraphics[width=1.0\textwidth]{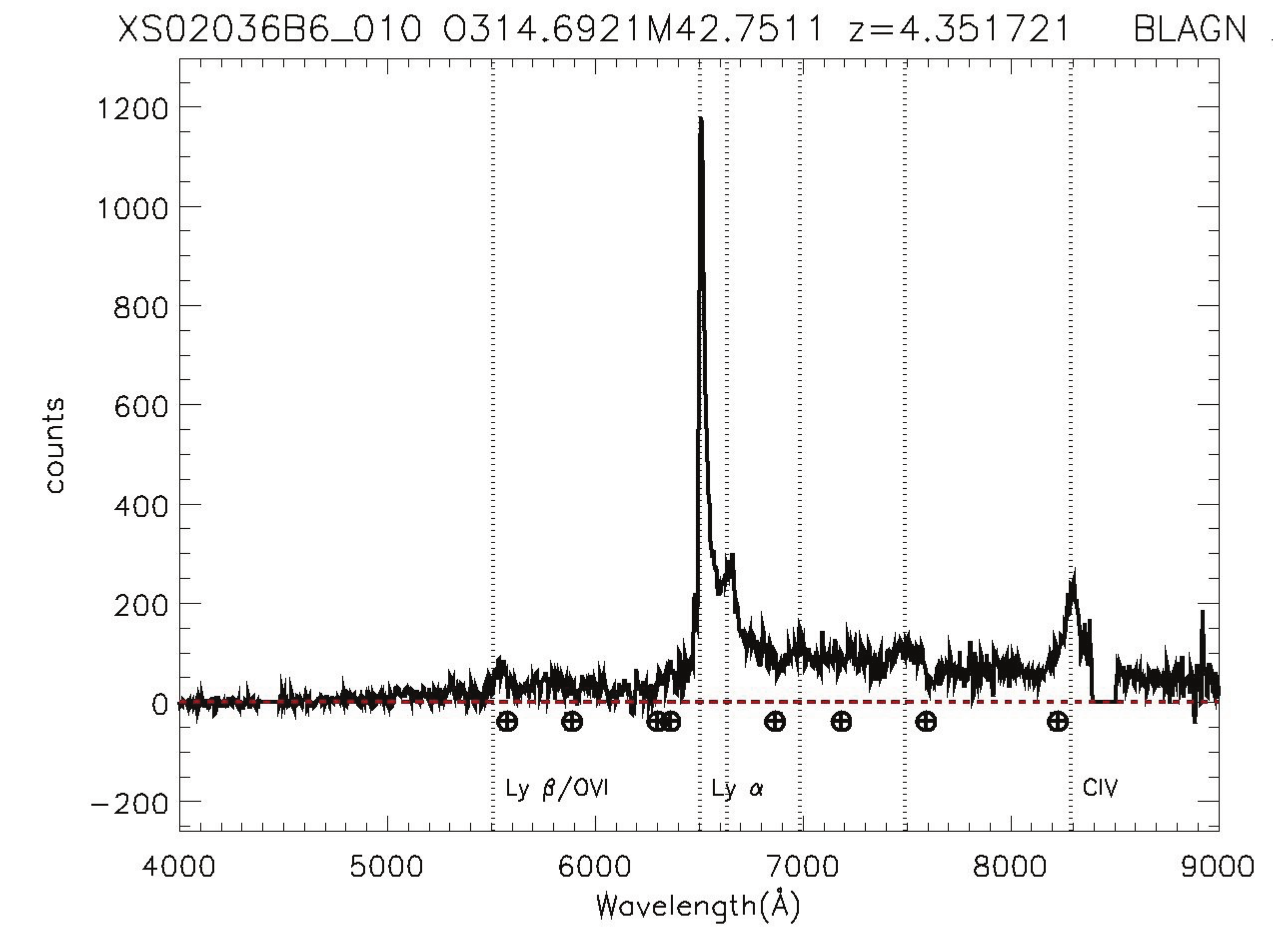}
\end{minipage}
\end{center}
    \caption
           {\scriptsize Example ChaMP spectra of broad line AGN observed with Magellan, MMT and WIYN telescopes. Spectra are not flux calibrated. CHANDRAOBSID, SPECOBJID, REDSHIFT and CLASS are given in the top of each plot.}
\end{figure*}

\noindent A table of our final extragalactic
spectroscopic catalog is 
provided in electronic format. The catalog omits the 327 stars, and so contains a total of 1242 entries. The columns listed are:\\ \\
\noindent CHANDRAOBSID, Chandra observation identifier\\
SPECOBJID, Spectroscopic observation identifier\\
CXOMPNAME, ChaMP identifier\\
RA, optical RA ($J$2000)\\
DEC, optical DEC ($J$2000)\\
TELESCOPE, telescope used for obtaining spectrum\\
SPEC, spectrograph used for obtaining spectrum\\
DATE, date of spectroscopic observation\\
REDSHIFT, spectroscopic redshift\\
CLASS, spectroscopic classification, BLAGN (broad-line AGN), NELG (narrow emission line galaxy), ALG (absorption line galaxy)\\
NETB, number of counts and associated errors\\
FSC, soft (0.5 - 2 $keV$) X-ray flux in units of 10$^{-13}$ erg s$^{-1}$ cm$^{-2}$ and associated errors\\
FHC, hard (2 - 8 $keV$) X-ray flux in units of 10$^{-13}$ erg s$^{-1}$ cm$^{-2}$ and associated errors\\
FUV, far-UV AB magnitude and associated errors\\
NUV, near-UV AB magnitude and associated errors\\
U, u-band AB magnitude and associated errors\\
G, g-band AB magnitude and associated errors\\
R, r-band AB magnitude and associated errors\\
I, i-band AB magnitude and associated errors\\
Z,  z-band AB magnitude and associated errors\\
Y,  Y-band Vega magnitude and associated errors\\
J,  J-band Vega magnitude and associated errors\\
H, H-band Vega magnitude and associated errors\\
K, K-band Vega magnitude and associated errors\\
MAG34, 3.4$\mu$m Vega magnitude and associated errors\\
MAG46, 4.6$\mu$m Vega magnitude and associated errors\\
MAG12, 12$\mu$m Vega magnitude and associated errors\\
MAG22, 22$\mu$m Vega magnitude and associated errors\\
S20, 20cm radio flux in mJy and associated errors\\
LOG LX, logarithmic 2-10keV X-ray luminosity in units of erg s$^{-1}$\\
GAMMA, power-law slope from X-ray spectral fitting and associated 90$\%$ confidence\\
NHINT, \nhintr and associated 90$\%$ confidence\\
NPHOT, number of photometric bands per catalogue used for the SED fitting (X: Chandra, G: GALEX, O: Optical, N: near-infrared, W: WISE, R: radio)\\ 
AGNTEMP: AGN template used for the SED fit (0: No AGN used, 1: Elvis et al. 1994 + Richards et al. 2006, 2: Hopkins et al. 2007, 3-6: Seyfert-2 from Bianchi et al. 2006, 7: NGC1068 by Matt et al. 1997, 8: Mrk231 by Berta et al. 2006, 9: IRAS19254-7245 by Berta et al. 2003, 10: IRAS22491-1808 by Farrah et al. 2003\\
GALTEMP: Galaxy template used for the SED fit (0: No galaxy template used, 1: NGC5253 by Beck et al. 1996, 2: NGC7714 by Brandl et al. 2004, 3: M82 by Strickland et al. 2004, 4: IRAS12112+0305 by Imanishi et al. 2007, 5: Elliptical from Rowan-Robinson et al. 2008, 6: Young elliptical from Maraston 2005)\\
FSB: Fraction of starburst contribution to the bolometric luminosity\\
FA: Fraction of AGN contribution to the bolometric luminosity\\
LBOL: Bolometric luminosity (10$^9$-10$^{19}$ Hz) in erg s$^{-1}$\\
MBH: logarithmic estimate of black hole mass\\

\noindent Figures 1 and 2 show the redshift
and multi-wavelength photometry distributions respectively of all our 1242 extragalactic
sources. Our sample includes a significant population of 78 z$>$3 X-ray QSOs, including two with redshifts greater than five. Figures 3, 4, 5 and 6 show examples of the various types of spectra found in our sample.  Figure 7 shows the distribution of broad-band X-ray flux versus optical magnitude to illustrate the parameter space spanned by the various populations. The majority of BLAGN follow the trend of 0$<$log (f$_{X}$/ f$_{r}$)$<$1. However, there is a significant population of BLAGN that lie at f$_{X}$/ f$_{r}$$>$10. From the latter 30$\%$ of them appear to be X-ray Seyferts (10$^{42}$$<$$L_{2-10~keV}$$<$10$^{44}$ erg s$^{-1}$) with \nhintr$>$10$^{22}$ which is consistent with previous studies that find that this parameter space is occupied by obscured X-ray Seyferts (e.g. Silverman et al. 2010). NELG seem to span a wide range of f$_{X}$/ f$_{r}$ as expected, as among them we can find Seyferts, LINERs and star-forming sources. At faint flux levels, a significant number of ALGs are evident, spanning a wide range of optical magnitudes. \\\\ 
\begin{figure} 
\begin{center}
\begin{minipage}[c]{8.6cm}
        \includegraphics[width=1.0\textwidth]{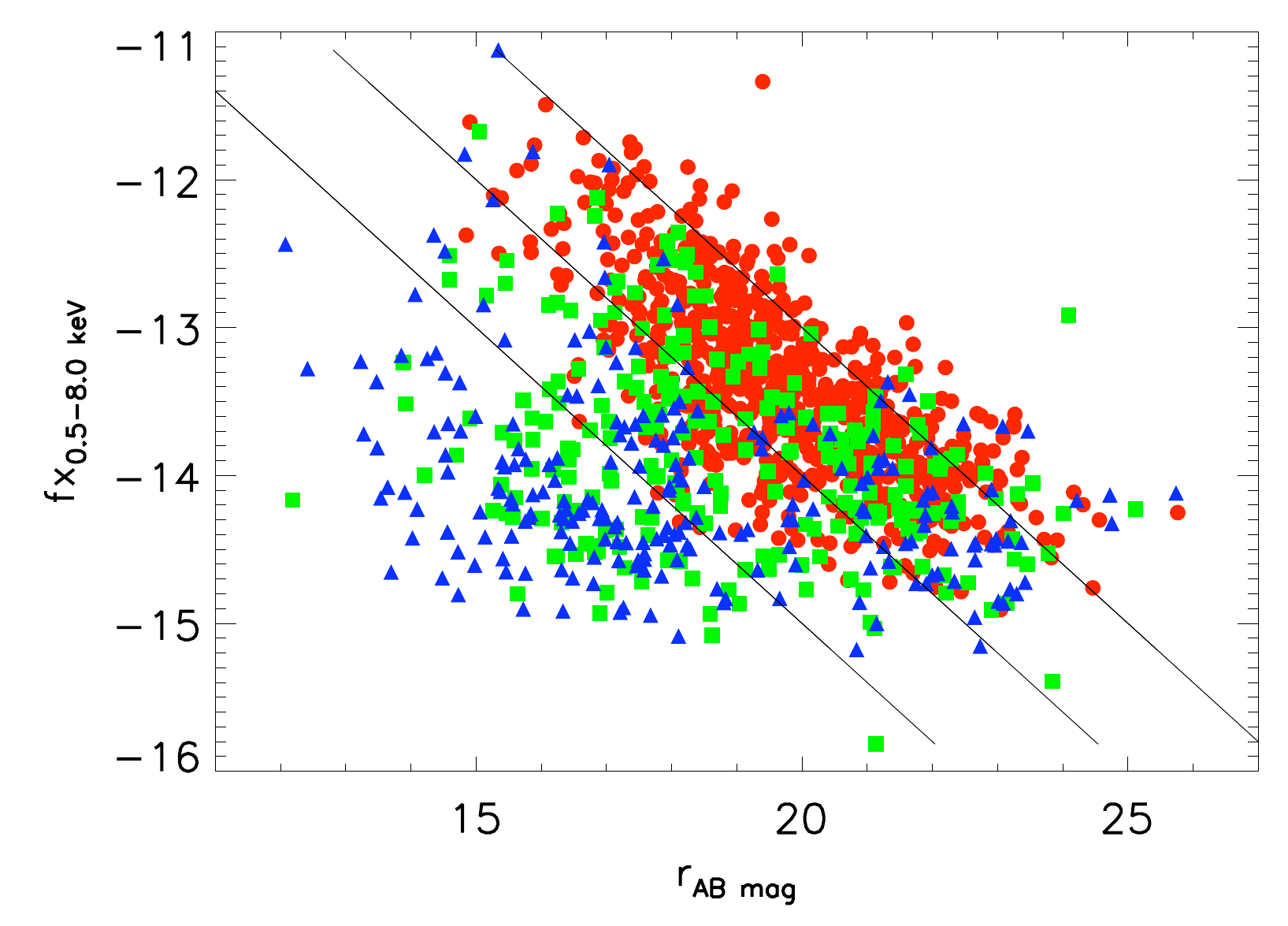}
\end{minipage}
\end{center}
\caption
           {\scriptsize Broad-band X-ray flux in logarithmic scale versus r-band magnitude for all our 1242 extragalactic sources. Red circles, green squares and blue triangles denote broad-line AGN, narrow emission line and absorption line galaxies respectively. Line of constant X-ray-to-optical flux ratio (0.1, 1, 10) are given by Equation 1 of Szokoly et al. (2004).}
\end{figure}

\subsection{Multi-Wavelength Photometry}
\noindent A prime advantage of ChaMP in comparison to deeper pencil-beam X-ray surveys, is its relatively shallow depth that allows for easier source identification in other wavelengths. We have cross-correlated our spectroscopic sample with GALEX (Morrissey et al. 2007), UKIDSS DR4 (Lawrence et al. 2007), 2MASS (Skrutskie et al. 2006), VLA (Condon et al. 1998) and WISE (Wright et al. 2010) catalogs. In the case of GALEX, the list was generated from a search for GALEX counterparts to optical counterpart positions in the ChaMP X-ray catalog. We retrieved the GALEX GR6 catalog using
the Virtual Observatory (VO) TOPCAT tool (Taylor 2005). Using Monte-Carlo
simulations and the Fadda et al (2002) method, we have concluded that
a search radius of 2.5 arcsec provides us with a $P(d)$ $<$ 0.02, where $P(d)$ is the Poisson probability of a GALEX source to have a
random association within a distance $d$, yielding an expected rate of
random associations of less than 5$\%$. The catalog contains only
sources that were detected at S/N $>$ 5 in at least one of the NUV,
FUV filters. All matches were then visually inspected  to
remove any apparent spurious associations. \\\\
\noindent We adopted a similar method 
for near-infrared with the 2MASS and UKIDSS (DR4)
catalogs. The ChaMP team has also obtained deep near-infared
imaging for 35 fields using the ISPI camera on the CTIO 4m Blanco 
telescope.  We shifted the ChaMP source coordinates 6 arcmin in both
RA and Dec in a large number of directions, performing
positional cross-correlation with UKIDSS and 2MASS using a search
radius of 5 arcsec. We thereby conclude that $P(d)$$<$ 0.02,
corresponding here to 4.5$\%$ random associations. In cases of multiple
matches where one of the matches is at separation $<$1.5 arcsec and
the other at separation $>$1.5 arcsec, the nearest match has been
selected. In all other cases where matches are at similar
distances, or both below 1.5 arcsec, visual inspection usually
has broken the counterpart ambiguity. When both 2MASS and UKIDSS imaging are available, we use the deeper UKIDSS photometry for SED fitting ($\S$3). The same method and statistics were used for
WISE data.\\ \\
\noindent In the case of NVSS and VLA-FIRST radio
catalogs, a 5 arcsec match radius yields less than 2$\%$ random
associations. All  
matches were visually inspected to remove any possible spurious
associations. In the cases of FIRST/ChaMP associations where 
the radio/X-ray position is associated with an extended feature of the
radio galaxy (e.g. radio lobe) the NVSS flux is used instead. \\ \\
\noindent From our
1242 ChaMP spectroscopic extragalactic sources, 63\% have detections in UV, 100\% in optical,  33\% in near-infrared, 30\% in mid-infrared and 15\% in radio.
\section{Broad-Band Spectral Energy Distributions}
\noindent To characterize the Spectral Energy Distributions (SEDs) of extragalactic objects, estimate bolometric luminosities and check
for the presence of starburst and/or AGN activity in our sample,
we fit the X-ray-to-radio fluxes with various empirical SEDs of well-observed sources as presented in Ruiz et al. (2010).  We have used a total of 16 such templates: two QSO templates (Elvis et al. 1994 + Richards et al. 2006, Hopkins et al. 2007), four Seyfert-2 galaxies (Bianchi et al. 2006), four starburst galaxies (NGC5253 by Beck et al. 1996, NGC7714 by Brandl et al. 2004, M82 by Strickland et al. 2004, IRAS12112+0305 by Imanishi et al. 2007) with star formation rates (SFR) ranging from 6 to 600 M$_{\odot}$ yr$^{-1}$, two absorption line galaxy templates (Rowan-Robinson et al. 2008, Maraston 2005) and 4 composite templates that are known to harbour both an AGN and a starburst (NGC1068 by Matt et al. 1997, Mrk231 by Berta et al. 2006; 2007, IRAS19254-7245 by Berta et al. 2003, IRAS22491-1808 by Farrah et al. 2003). Except Mrk231 which is optically classified as a Broad Line QSO with a massive young nuclear starburst, the remainder three composite objects are all optically classified as Seyfert-2s with intense starbursts. We have adopted the model described in Ruiz et al. (2010) which fits all SEDs using a $\chi$$^{2}$ minimization technique within the fitting tool Sherpa (Freeman et al. 2001). Our fitting allows for two additive components, one associated with the AGN emission and the other associated with the starburst emission. The SEDs are built  and fitted in the rest-frame. We have chosen the fit with the lowest reduced $\chi$$^{2}$ as our best fit model. Fractions of AGN and starburst contributions are derived from the SED fitting normalisations  as these are derived from Ruiz et al. (2010) model,

\begin{equation}
 {F_{\nu}=F_{BOL}~(\alpha~u_{\nu}^{AGN}~+~(1-\alpha)~u_{\nu}^{SB})}
\end{equation}

\noindent where F$_{BOL}$ is the total bolometric flux, $\alpha$ is the relative contribution of the AGN to F$_{BOL}$ , F$_{\nu}$ is the total flux at frequency $\nu$, while $u_{\nu}^{AGN}$ and $u_{\nu}^{SB}$ are the normalized AGN and SB templates. \\ \\
\noindent Among our 758 broad emission line objects (FWHM $>$ 1000 km/sec) all are best fitted with one of our two available QSO SED templates, with 150 also requiring  starburst contributions of at least 5$\%$. Among the 252 narrow emission line objects (EW $>$ 5\AA; FWHM $<$ 1000 km/sec), 17 have been fitted with one of the composite templates which harbor a Seyfert-2, 208 have been fitted with a
Seyfert-2 template (65 having significant starburst contribution),
12 have been fitted with a starburst template, and 15 have been fitted
with a QSO template (10 of which require $>5\%$ starburst
contribution).  Among the 230 absorption line galaxies, 130 have been fitted with an
elliptical SED template, 91 with a mixture of Seyfert-2 and starburst templates and 9 objects
with a QSO template. Since the best fit model was determined only
using the lowest reduced $\chi$$^{2}$ value with no preselection based on
spectroscopic classification, these results indicate an excellent overall agreement between the SED fitting code and optical spectroscopic classification in the cases
of broad line and narrow emission line objects and fair agreement in
the ALGs. Figure 8 shows examples of sources fitted
with QSO, Seyfert-2 and composite templates. 
\begin{figure*} 
\begin{center}
\begin{minipage}[c]{8.1 cm}
        \includegraphics[width=1.0\textwidth]{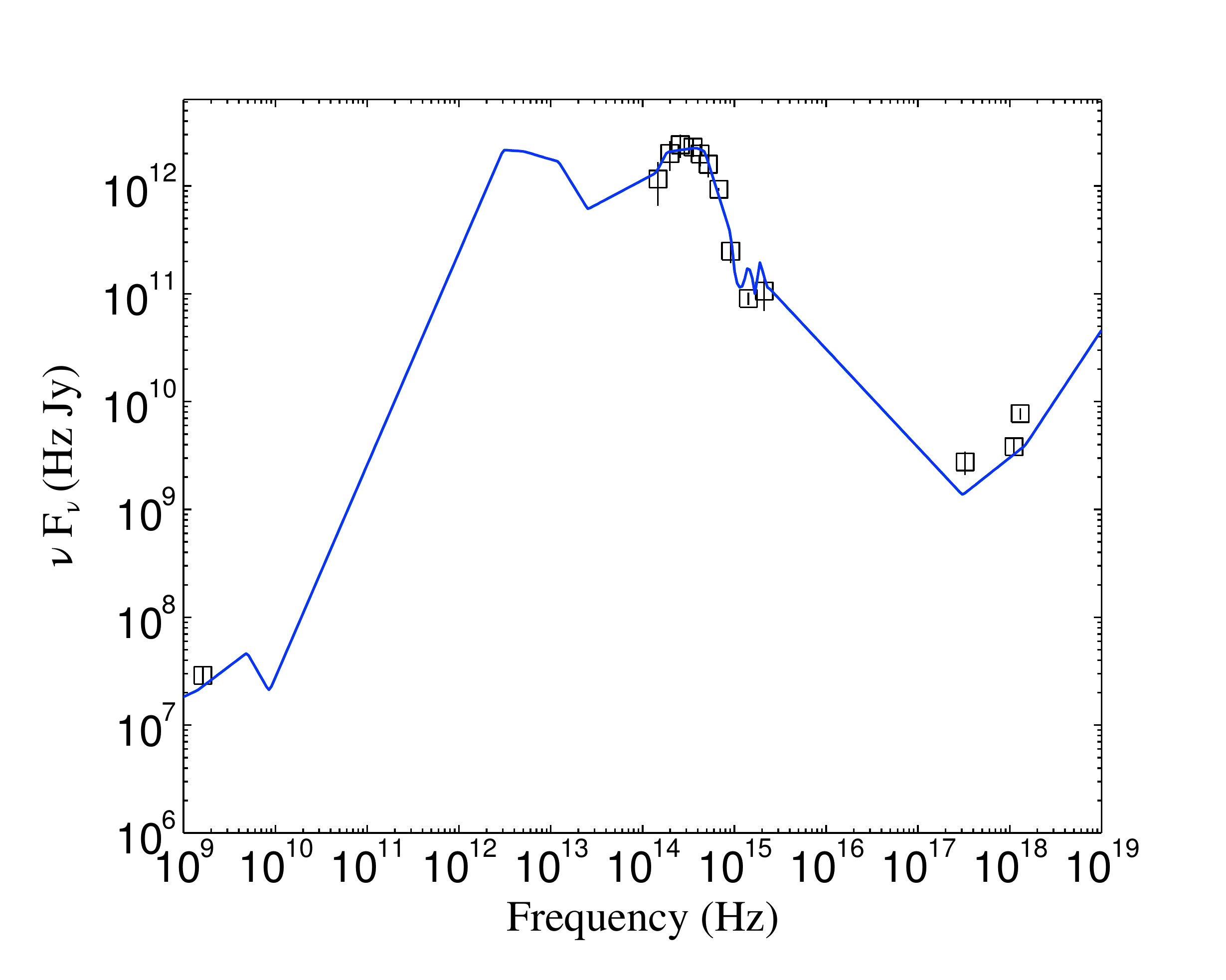}
\end{minipage}
\begin{minipage}[c]{8.1 cm}
        \includegraphics[width=1.0\textwidth]{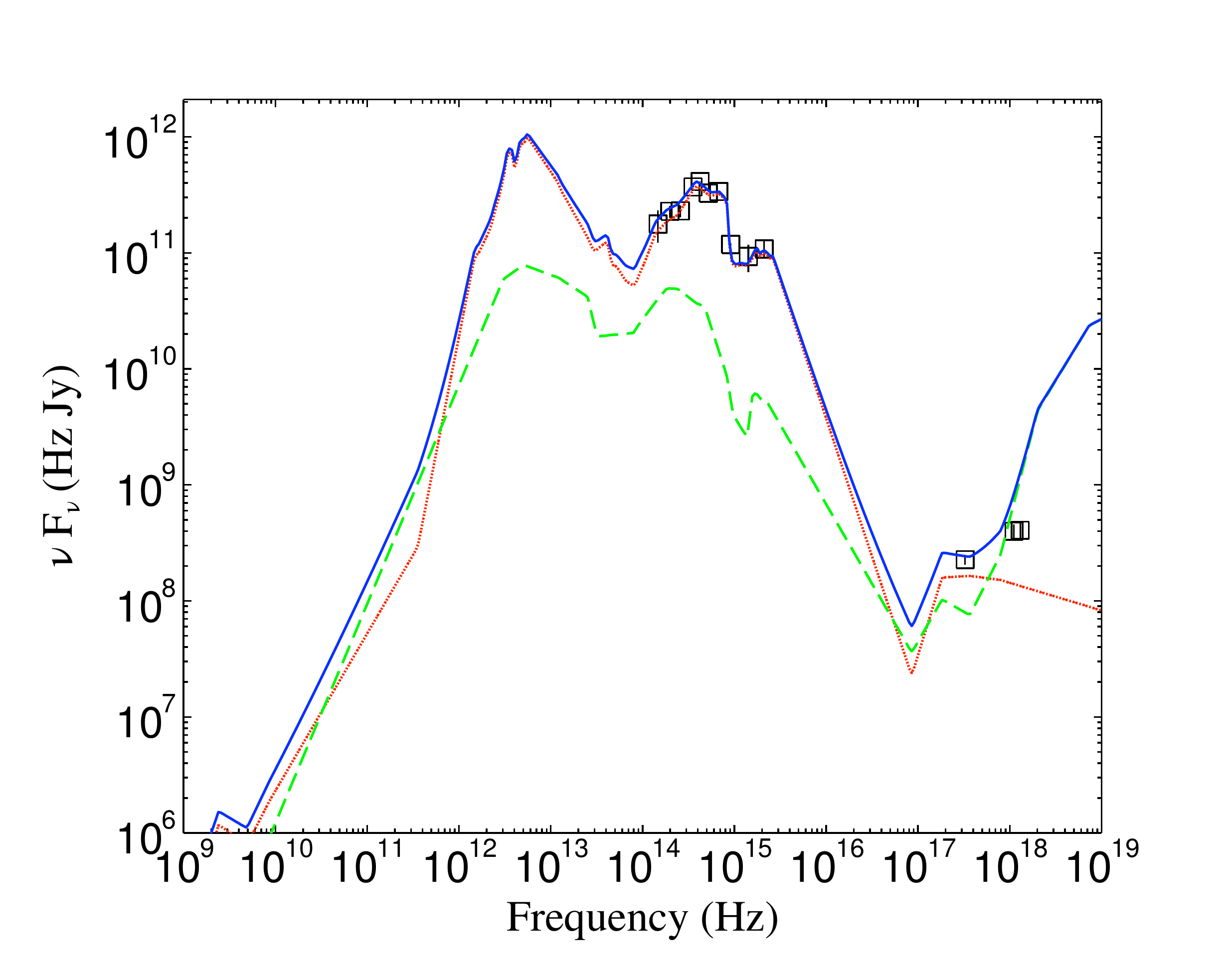}
\end{minipage}
\begin{minipage}[c]{8.1 cm}
        \includegraphics[width=1.0\textwidth]{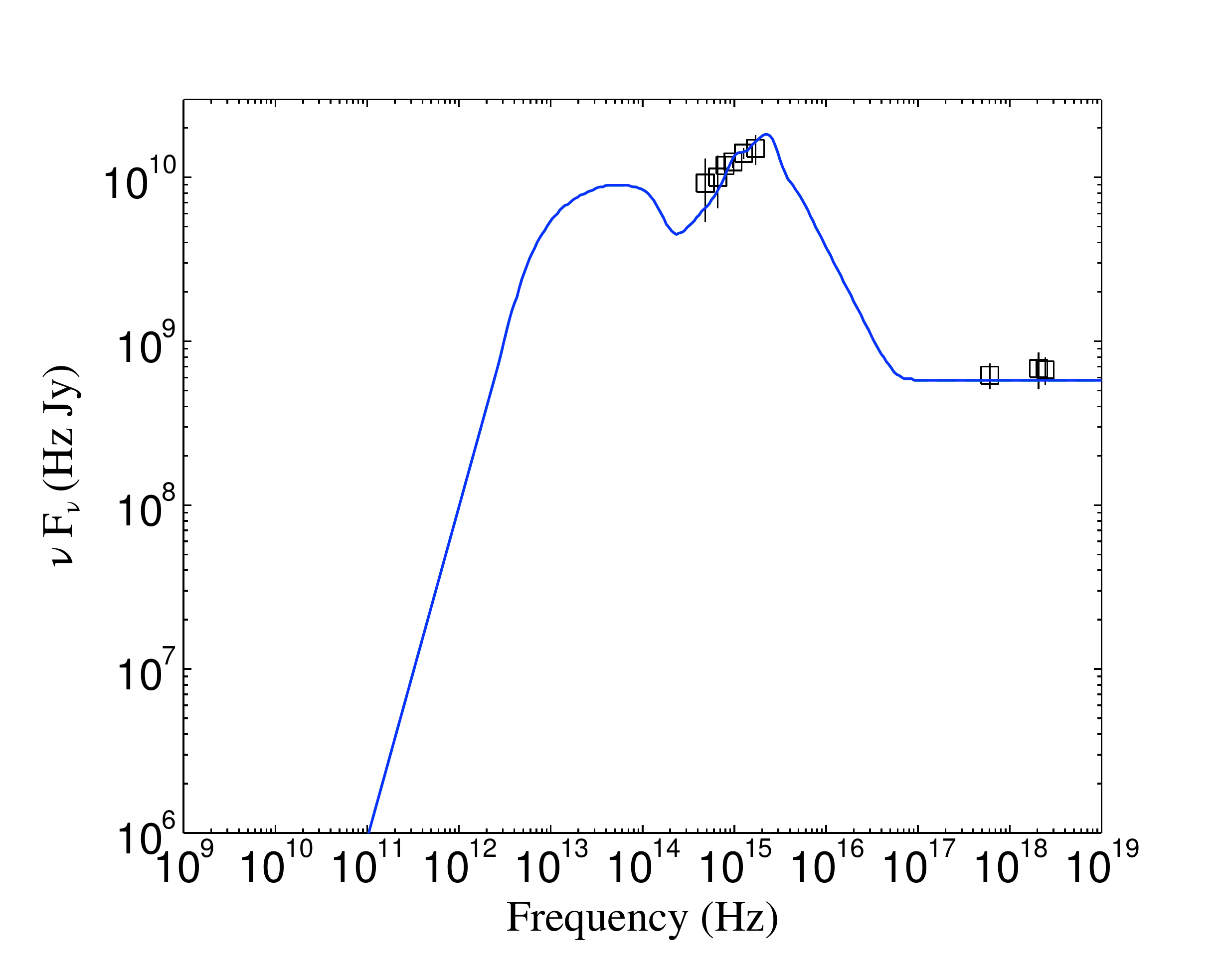}
\end{minipage}
\begin{minipage}[c]{8.1 cm}
        \includegraphics[width=1.0\textwidth]{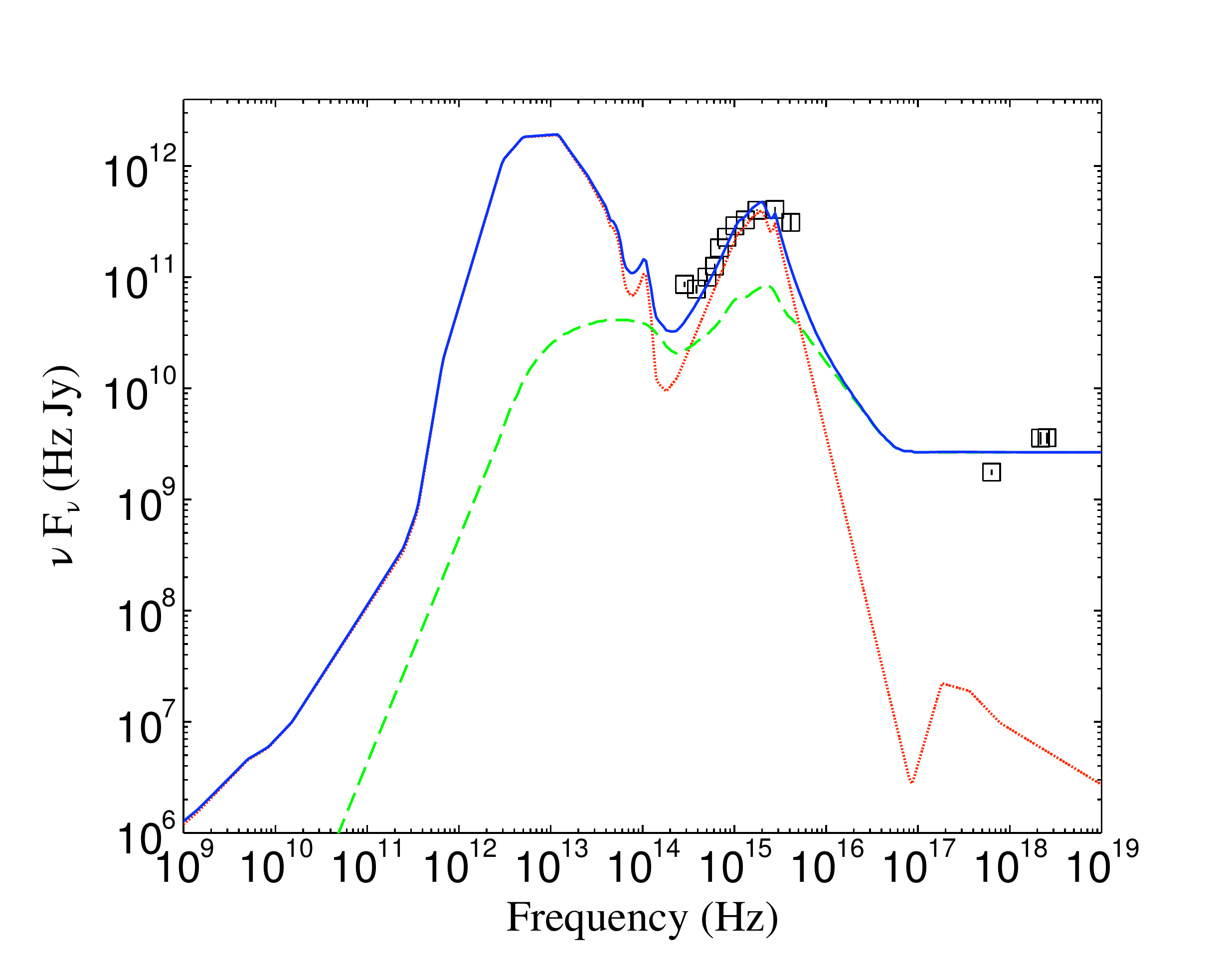}
\end{minipage}
\begin{minipage}[c]{8.1 cm}
        \includegraphics[width=1.0\textwidth]{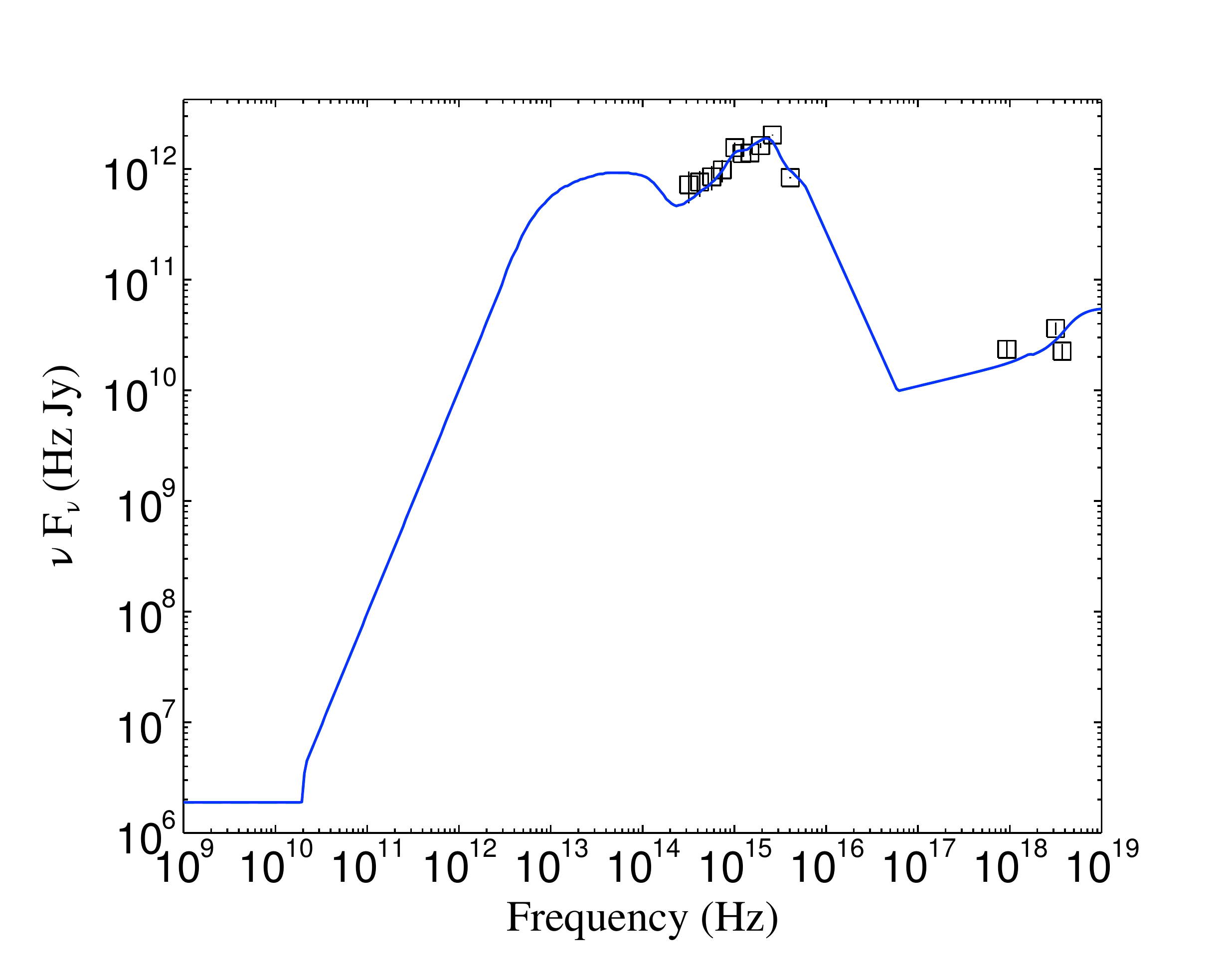}
\end{minipage}
\begin{minipage}[c]{8.1 cm}
        \includegraphics[width=1.0\textwidth]{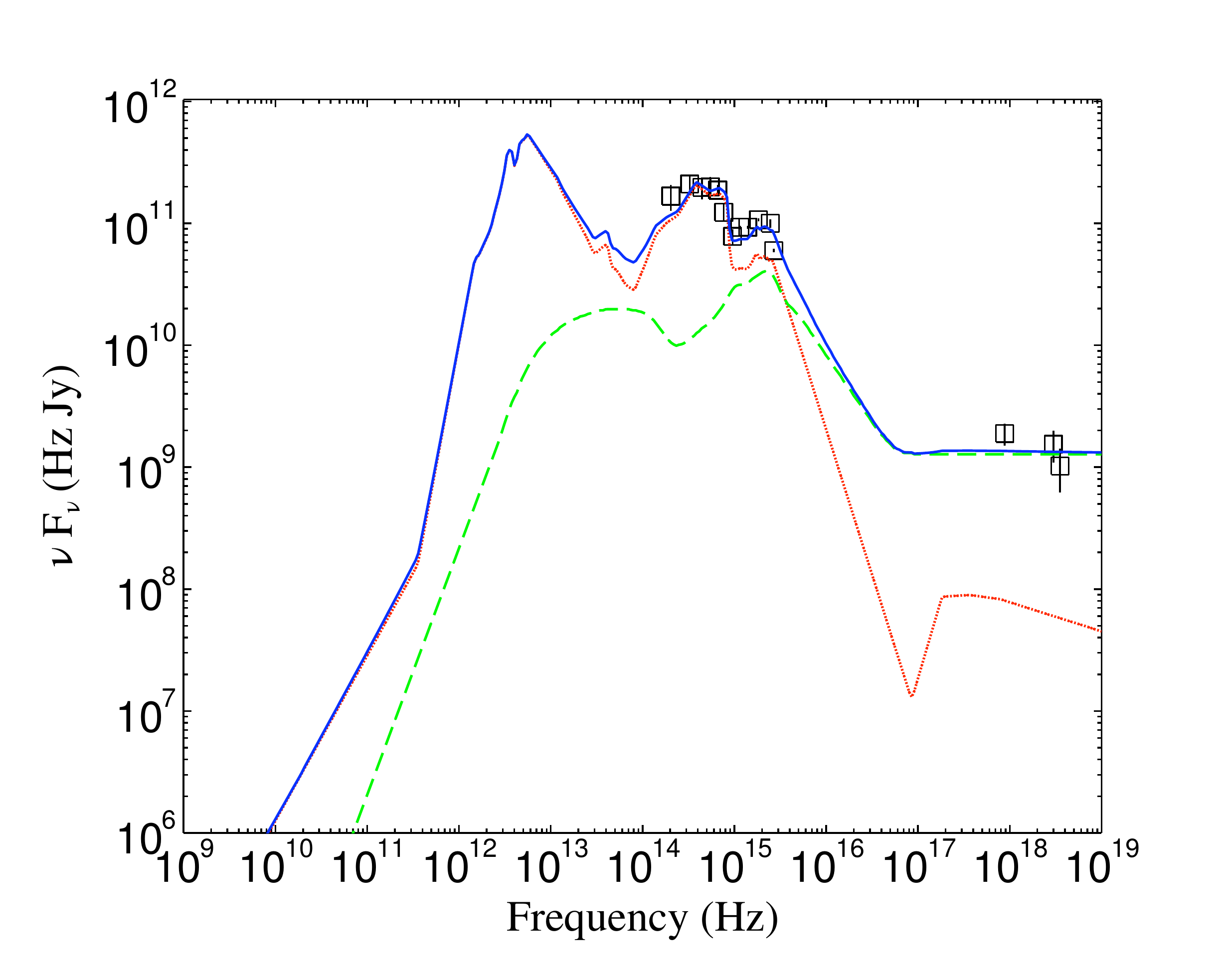}
\end{minipage}
\end{center}
    \caption
    {\scriptsize Examples of best-fit SEDs (blue line). Red solid and green dashed lines are the starburst and AGN templates respectively. Square boxes indicate available photometry. Vertical lines are associated $>$10$\%$ errors. $\bf{Top~Left:}$ XS0518B7007 with $L_{2-10~keV}$$=$1.1$\times$10$^{40}$ erg s$^{-1}$ at $z=0.08$. Best fitted with a Seyfert-2 NGC3393 template. $\bf{Top~Right:}$ XS05618B6006 with $L_{2-10~keV}$$=$3.1$\times$10$^{40}$ erg s$^{-1}$ at $z=0.08$. Best fitted with a Seyfert-2 NGC4507 and a starburst  NGC7714 templates, contributing 15$\%$ and 85$\%$ respectively to the $L_{BOL}$. $\bf{Middle~Left:}$ XS00546B2012,  with $L_{2-10~keV}$$=$9.1$\times$10$^{43}$ erg s$^{-1}$ at $z=1.01$. Best fitted with an Elvis et al QSO template. $\bf{Middle~Right:}$ XS02251B7002,  with $L_{2-10~keV}$$=$9.5$\times$10$^{43}$ erg s$^{-1}$ at $z=1.1$. Best fitted with an Elvis et al QSO and a starburst NGC5253 templates, contributing 15$\%$ and 85$\%$ respectively to the $L_{BOL}$.  $\bf{Bottom~Left:}$ XS0907B3001 with $L_{2-10~keV}$$=$9.6$\times$10$^{44}$ erg s$^{-1}$ at $z=2.08$. Best fitted with a Hopnkins et al (2007) QSO template. $\bf{Bottom~Right:}$ XS04151B6006 with $L_{2-10~keV}$$=$4.1$\times$10$^{44}$ erg s$^{-1}$ at $z=1.92$. Best fitted with a Hopkins et al. QSO and a starburst  NGC7714 templates, contributing 20$\%$ and 80$\%$ respectively to the $L_{BOL}$}
\end{figure*}
\section{X-ray Spectral Fitting}
\noindent For all ChaMP X-ray sources in our spectroscopic sample, we perform X-ray spectral fitting using the CIAO 
{\it Sherpa}\footnote{http://cxc.harvard.edu/sherpa} tool in an automated
script.  For each source we fit three power-law models.
While quasars are typically well-fit by a power-law, it is well-known
that AGN-dominated spectra are complex, including potentially a soft
excess and/or reflection component, Fe K$\alpha$ line emission, and
neutral, partially ionized and/or partially covering absorption (e.g. Reeves $\&$ Turner, 2000).  When active SMBH accretion is weak or
non-existent, emission related to the stellar component or ISM
give rise to the X-ray emission in galaxies.  X-ray spectra from lower
luminosity objects may have a significant power-law component arising
from X-ray binary populations (e.g. Persic $\&$ Rephaeli, 2002; Fragos et al. 2009).  Thermal components arising from hot
ISM or shocked gas may be present, which at high
signal-to-noise would be poorly fit with power-law models. However,
since our sample has a median of 45 net broadband counts, detailed
spectral fits are not warranted, so we remain content with power-law models.

\noindent The three X-ray spectral models we fit all contain an
appropriate neutral Galactic absorption component frozen at the 21\,cm
value:\footnote{Neutral Galactic column density $\nhgal$ taken from
Dickey et al. (1990) for the $Chandra$, aimpoint position on the sky.}
   (1) photon index $\Gamma$, with no intrinsic absorption component
(model ``{\tt PL}'')
   (2) an intrinsic absorber with neutral column $\nhintr$ at the
source redshift, with photon index frozen at $\Gamma=1.8$ (model
``{\tt PLfix}'').  Allowed fit ranges are $-1.5<\Gamma<3.5$ for {\tt
PL} and $10^{18}<\nhintr<10^{25}$ for {\tt PLfix}.
   (3) a two-parameter absorbed power-law where both $\Gamma$ and the
$\nhintr$ are free to vary within the above ranges while $\nhgal$ is fixed (model ``{\tt
PL\_abs}'').  All models are fit to the ungrouped data using Cash
statistics (Cash, 1979).  The latter model, {\tt PL\_abs},
is our default, for several reasons described below.

\noindent Overall, we find (Figure 9) that the best-fit
$\Gamma$ from our default 
model is not correlated with  $\nhintr$, which illustrates that these
parameters are fit with relative independence even in low count
sources.  The best-fit $\Gamma$ in the default  {\tt PL\_abs}
model correlates well with that from the {\tt PL} model for
the majority of sources; the median of the difference in 
fitted slopes for these two models$\Delta\Gamma$ is just 15\%
of the median uncertainty in slopes $\sigma_{\Gamma}$.  On the other
hand, 67 (5.4\%) of sources have their best-fit power-law slope
``pegged" at $\Gamma=5$.  Most (42 or 63\%) of these are optical ALG,
likely passive elliptical galaxies poorly fit by a power-law model. Only 10
(15\%) are QSOs.  Many of the sources with $\Delta \Gamma$ larger than
$\sigma_{\Gamma}$ also have detectable $\nhintr$ (which we define as
those fits where 90\% confidence lower limit of $\nhintr > 0$), which
justifies the softer $\Gamma$ result.

\begin{figure} 
\begin{center}
\begin{minipage}[c]{8.3cm}
        \includegraphics[width=1.0\textwidth]{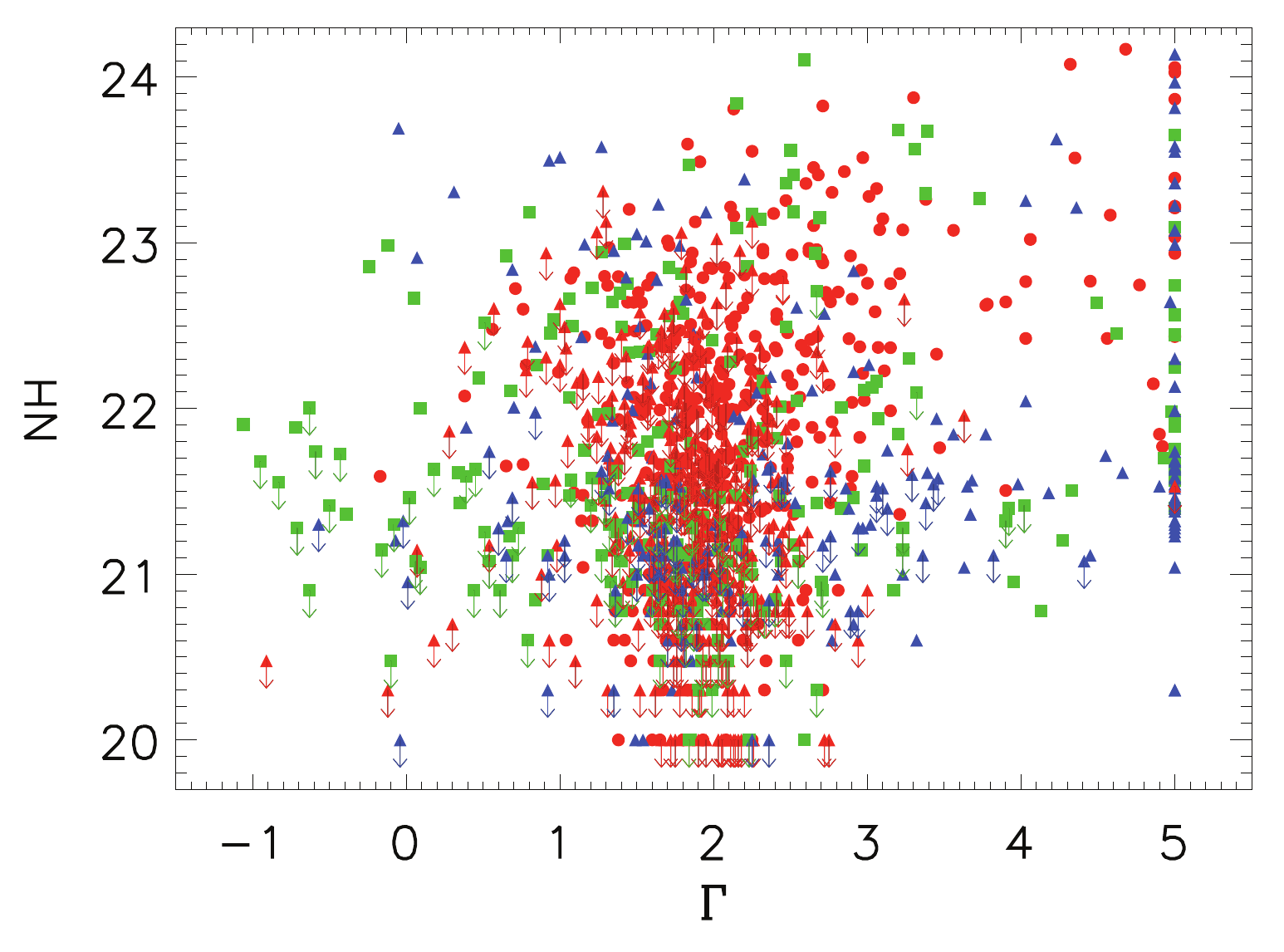}
\end{minipage}
\end{center}
\caption
           {\scriptsize Best fist X-ray spectral slope $\Gamma$ versus column density in logarithmic scale. Downward arrows represent the 90$\%$ upper limits to \nhintr,
whenever the 90$\%$ lower limit encompassed zero.  Red circles, green squares and blue triangles denote broad-line AGN, narrow emission line and absorption line galaxies respectively.}
\end{figure}

\section{Black Hole Masses $\&$ Star Formation Rates}
\noindent  Black hole masses are only available for sources with available SDSS spectra.  For our broad line objects, black hole masses  have been
retrieved from Shen et al (2011), who have compiled
virial black hole mass estimates of all SDSS DR7 QSOs using
Vestergaard $\&$ Peterson (2006) calibrations for H$\beta$ and C$\,IV$
and their own calibrations for Mg$\,II$. There are 329 broad-line objects within our sample with black hole mass estimates. For our narrow line objects, a total of 119 NELG for which high quality SDSS spectra are available, M$_{BH}$ values are calculated using the M-$\sigma$ relation of Graham et al. (2011).  A suite of optical and X-ray properties of these NELG are presented and discussed in Constantin et al. (2009).

\noindent Star formation rates have been estimated using the output of our SED fitting code. In the case that a starburst template is required in the fitting we extract the 8 - 1000 $\mu$m starburst luminosity which is a proxy for the far-infrared luminosity attributed to star formation. The L$_{SB}$ (8 -1000 $\mu$m) is then converted to a star formation rate using Kennicutt (1998) relation.
\begin{equation}
 {SFR~ (M_{\odot} ~yr^{-1})=4.5\times 10^{-44}\times L_{SB} ~(erg~ s^{-1})}
\end{equation}
\begin{figure}
\begin{center}
\begin{minipage}[c]{8.5cm}
       \includegraphics[width=1.0\textwidth]{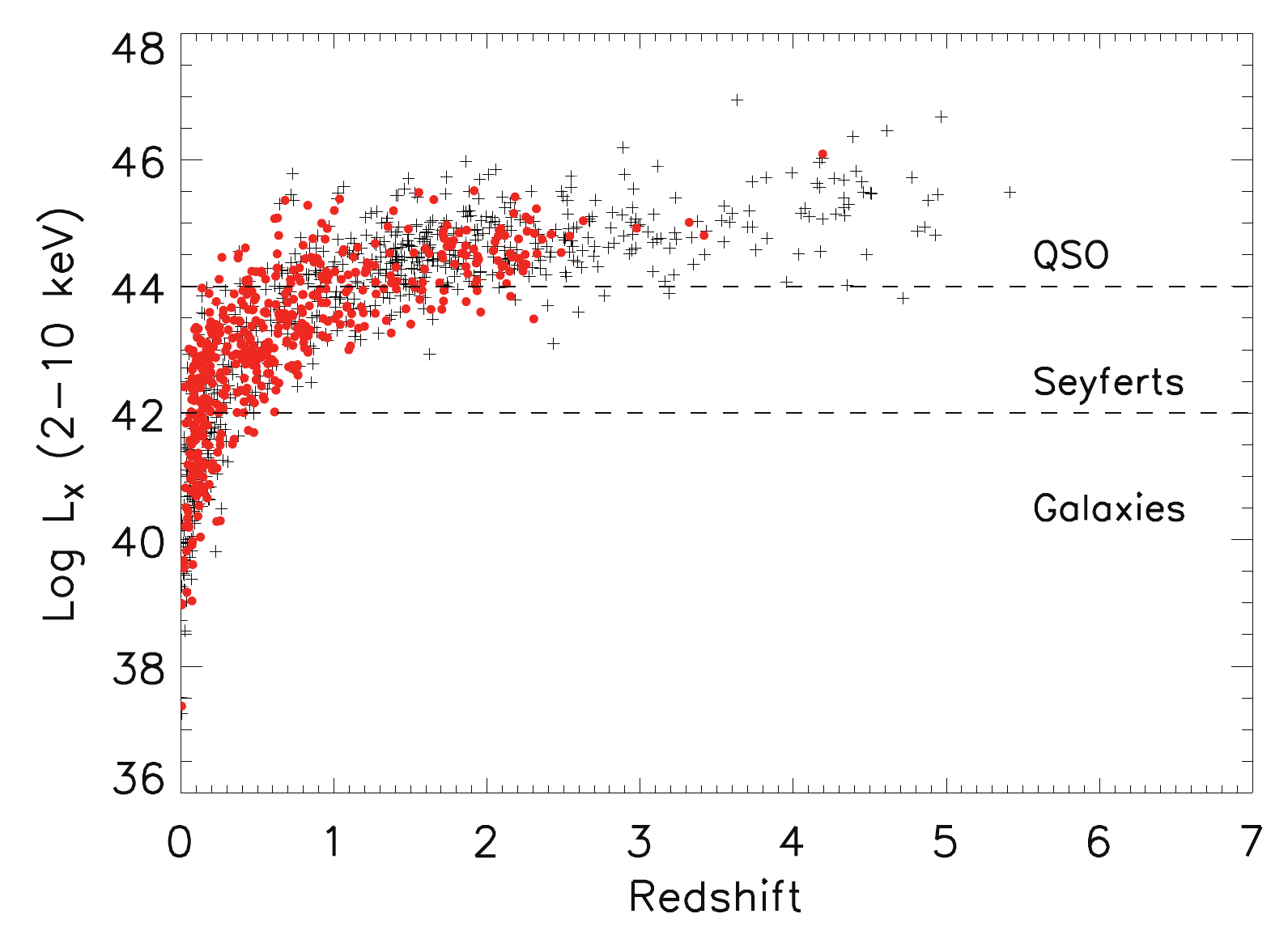}
\end{minipage}
\end{center}
    \caption{\scriptsize Distribution of rest-frame Logarithmic 2-10 keV X-ray luminosity and spectroscopic redshift for the 1202 hard-band X-ray detected sources from ChaMP. Red circles are sources that require a starburst component to fit available photometry. Black crosses are sources that their SED fits do not require any contribution from a starburst component. It is evident that as the X-ray luminosity increases, the occurrence of star formation activity decreases.}
\end{figure}

\section{X-ray Luminosity - Redshift plane}
\noindent X-ray luminosity is by itself an important discriminator of
the primary power source. The maximum achievable X-ray emission by
young stellar populations in an ultra-luminous starburst is
$L_{2-10~keV}$$\sim$1.5$\times$10$^{42}$ erg s$^{-1}$ (Persic et
al. 2004). Figure 10 shows the hard X-ray luminosity versus redshift
for all hard X-ray detected sources in our sample. By restricting the
sample to those sources detected in the most penetrating 2$-$10 keV
band we can minimize the influence of obscuration on our
results. However, among our 1242 extragalactic ChaMP sources with
available spectroscopy, 1202 have hard-band detections and as a result
selecting them does not bias our results. Here $L_{X}$ is the restframe 2-10 keV X-ray luminosity calculated with the method described in the Appendix of Green et al. (2011).  Red circles are those
sources that require a starburst component to fit the observed photometry. \\ \\
\noindent The $L_{X}$ - $z$ plane of
Figure 10 shows a striking trend.  Star formation occurrence increases from 43\% in sources with log\,$L_{2-10~keV}$ $<$ 42 to 58\%
amongst objects with 42 $<$ log\,$L_{2-10~keV}$ $<$ 44
erg\,s$^{-1}$ and drops sharply to 26\% when X-ray luminosity reaches QSO limits i.e. when log\,$L_{2-10~keV}$$>$44. However, the latter can be considered as a conservative upper limit to the occurrence of star formation in QSOs since the majority (69\%) of X-ray QSOs with starburst events lie at z$>$1 where contamination from torus emission to the mid-infrared can be significant (Rowan-Robinson et al. 2009; Mor $\&$ Netzer 2011). Since the longest infrared wavelength used for our SED fits is the 22$\mu$m band from WISE we expect that the number of apparent star forming QSOs at z$>$1 may be significantly lower, which would make the trend we see in Figure 10 even stronger. \\\\
\noindent Figure 11 shows the average fractional contribution of AGN and/or starburst components to the bolometric luminosity per 2-10 keV luminosity bin of 0.6. Again, there is a striking trend which seems to support the aforementioned indication of quenching of star formation in powerful QSOs. Although there seems to be a broad flat evolution of AGN and starburst contributions in low and moderate X-ray luminosities ($L_{X}$$<$10$^{41.5}$ erg s$^{-1}$), when powerful AGN activity is triggered ($L_{X}$$>$10$^{42}$ erg s$^{-1}$), star formation seems to pick up and reaches its maximum at log $L_{X}$$\sim$ 42.5. At this stage AGN and star formation appear to contribute the same fraction of the bolometric output but as the AGN becomes more powerful, as indicated both from the X-ray emission and AGN contribution, star formation decreases rapidly and eventually quenched when the AGN reaches extreme rates of accretion. At this stage of the evolution, AGN reaches its maximum emission (100\% contribution from an AGN component) without the presence of any identifiable starburst events. This finding is consistent with the picture drawn by Figure 10.  If the absence of star formation in QSOs, as depicted in Figure 10 were a detection bias, then we would expect that the QSOs with SFR to follow the relationship of Figure 11 and show enhanced starburst contribution compared to X-ray Seyferts. However, this is not the case, suggesting that intense SFR has stopped while accretion continues to rise in agreement with QSO-mode feedback models (e.g. Hopkins et al. 2005, Netzer 2009). 
\begin{figure} 
\begin{center}
\begin{minipage}[c]{8.5cm}
 \includegraphics[width=1.0\textwidth]{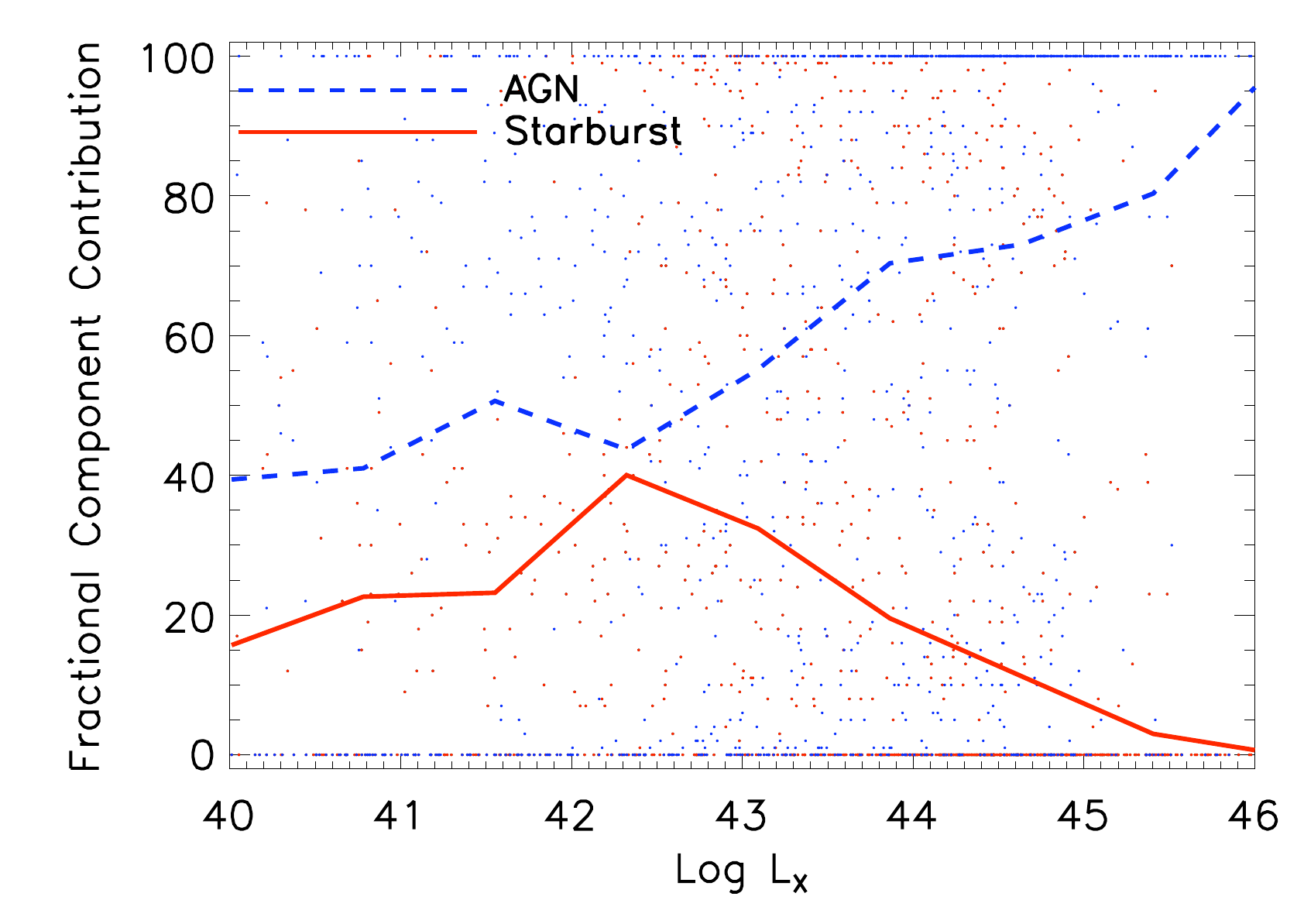}
 \end{minipage}
\end{center}
    \caption
           {\scriptsize Fractional contribution of
AGN (blue dots) and  starburst components (red dots) to bolometric luminosity versus $log$ $L_X$. Points
represent the fractional AGN and/or starburst contributions for all
objects with L$_{X (2-10 keV}$$>$$10^40$ erg s$^{-1}$. Solid red line represents the average value of starburst fractional
contribution per log L$_{X}$ bin of 0.6.  Dashed blue line represents the
average value of AGN fractional contribution per log L$_{X}$ bin of 0.6.}
\end{figure}
\section{Starburst Detectability}
\noindent Figures 10 and 11 suggest that star formation is either quenched (weak or absent compared to the L$_{Bol}$) in QSOs or simply not detectable by our SED fitting method using the currently available photometry.  The latter may occur because a given host ($L_{SB}$) falls below photometric survey detection limits towards higher redshift, and/or because a given host ($L_{SB}$) becomes more difficult to detect in contrast to a more luminous nuclear source. \\\\
\noindent To address the detectability issue we have performed a series of SED fits to a large set of simulated sources drawn from our original sample.  We have focused our tests to the population of 987 sources with L$_{X (2-10 keV)}$ $>$ 10$^{42}$ erg  s$^{-1}$. Of them, 456 have  10$^{42}$ $<$ L$_{X (2-10 keV)}$ $<$  10$^{44}$ erg s$^{-1}$ (XSEY hereafter), and 531 have L$_{X (2-10 keV)}$ $>$ 10$^{44}$ erg s$^{-1}$ (XQSO hereafter). To compare starbust detectability (i.e., completeness) between the predominantly lower redshift XSEY and the XQSOs, from the observed SEDs of sources that originally required at least 5$\%$ starburst contribution to their L$_{Bol}$ (the ``SF'' sample hereafter), we first artificially remove the appropriately normalized fluxes of the best-fit starburst template from the observed photometric bands.  The resulting "stripped" sample can then be treated identically in our detectability experiment as the "naked" sample - those sources whose observed SEDs were originally fitted with a pure AGN template. In this way, we can then add back in a starburst component at various levels to the observed SEDs, and refit the simulated SEDs, to test the sensitivity of our method.  384 of 531 XQSOs are naked, and 176 of 456 XSEY are naked.  Combining the ``stripped" and the ``naked" samples yields 987 pure AGN SEDs with no starburst contribution. \\ \\
\begin{figure} 
\begin{center}
\begin{minipage}[c]{7.9cm}
 \includegraphics[width=1.0\textwidth]{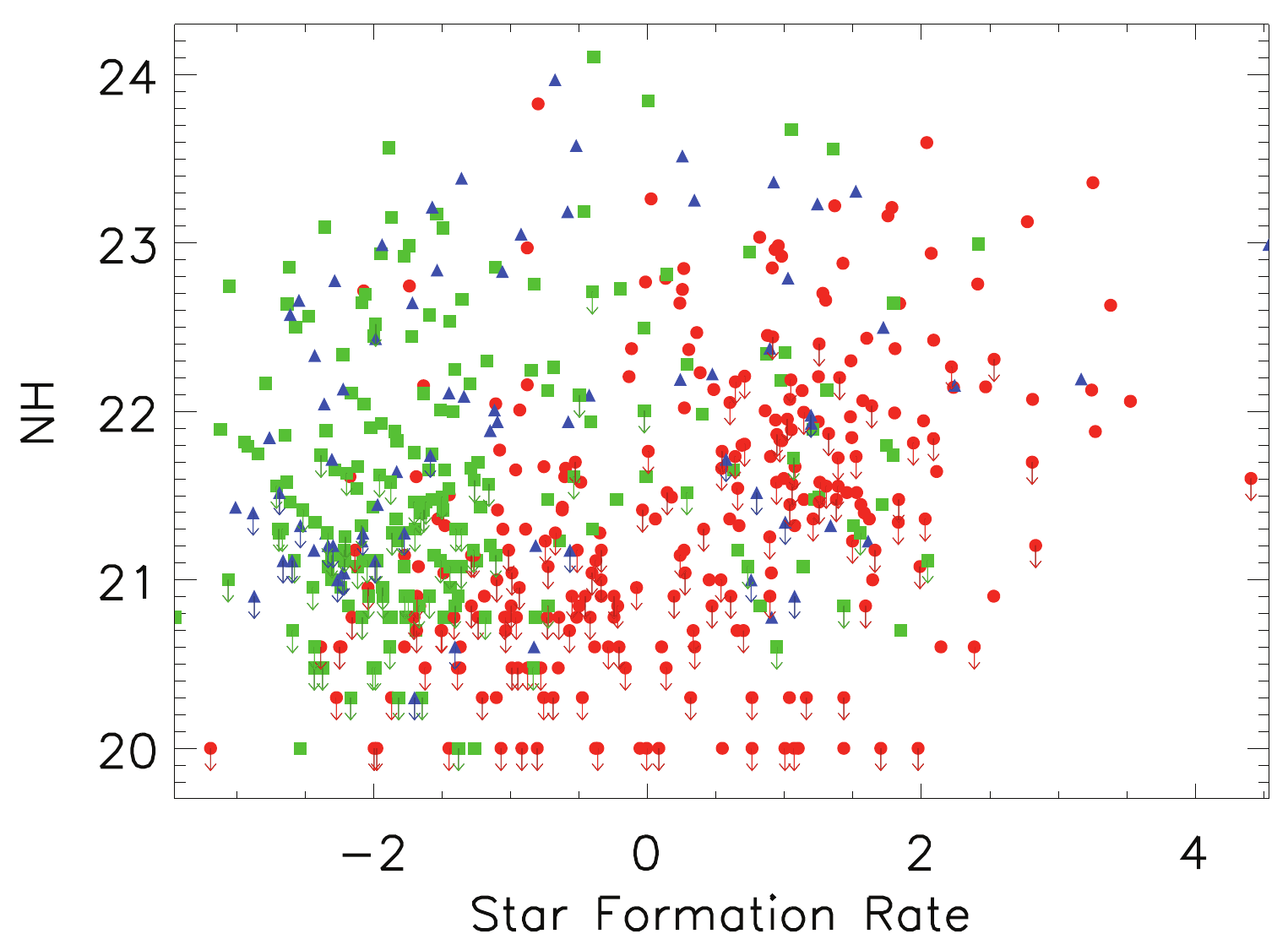}
  \end{minipage}
\end{center}
    \caption
           {\scriptsize Star formation rates versus column density in logarithmic scale. Downward arrows represent the 90$\%$ upper limits to \nhintr,
whenever the 90$\%$ lower limit encompassed zero. Red circles, green squares and blue triangles denote broad-line AGN, narrow emission line and absorption line galaxies respectively.}
\end{figure}
\noindent In the next step, for each of these 987 objects we fit a pure AGN template to estimate of its bolometric flux. Then, using each one of our 4 starburst templates, we compute an additional starburst component representing a fixed fraction of the total bolometric flux. The additional component ranges from 0$\%$ to 50$\%$ in intervals of 5$\%$. We then interpolate the new starburst component at the frequency points of original SED, thus obtaining starburst fluxes. Finally we add these simulated starburst fluxes to the "striped/naked" fluxes of the source. From each object/template pair, we thus create 11 new simulated SEDs, yielding $987 \times 4 \times 11 = 43428$  simulated sources.  We then run again our SED fitting code on all these simulated sources to constrain the  starburst contribution. Of the XQSOs simulated with $\ge5\%$ SB, our best-fit SED that includes $SB=5\%$ is $C_Q$=39\%.  By contrast, of all the XSEY simulated with $\ge5\%$ SB, our best-fit that includes $SB=5\%$  is  $C_S$=52\%   Thus, our completeness for XQSOs {\em relative} to XSEY is $R_C =C_Q/C_S$=39/52=0.75.  We can compensate crudely for this relative incompleteness by dividing our actual sample fraction of XQSO with $SB\ge5\%$ with $R_C$, $FSB_Q/R_C$ which yields 26$\%$/0.75=35$\%$ of XQSO.  Since that fraction is still significantly lower than the actual fraction of XSEY with $SB\ge5\%$, $FSB_S=58\%$, we can claim  that the star formation in XQSOs appears significantly weaker from what  we would expect if starburst luminosity increased with accretion luminosity. 
\section{The Relationship Between Absorption and Star Formation}
\noindent In previous studies a connection has been made between X-ray absorption and star formation in AGN (e.g. Page et al. 2004), though this remains controversial (e.g. Lutz et al. 2010; Shao et al. 2010). To test for whether any correlation exists between \nhintr\, and star
formation in our sample, we must use statistical analysis that can
account for upper limits.  We examine log  \nhintr\,  vs. log SFR and
also vs.  starburst fraction FS for the 524 of our 1242 objects for
which FS$>5\%$.  When the 90\% lower limit to \nhintr\, from our X-ray
spectral fits is consistent with zero, we assign its 90\% upper limit
value as the upper limit for the correlation tests.  We test for
significance using the Cox Proportional Hazard, Kendall's $\tau$ and
Spearman's $\rho$ tests, as implemented in the ASURV (Survival
Analysis for Astronomy) package (Lavalley et al. 1992).  Between
log\,\nhintr\, and the log\,SFR, we find that the correlation is
significant (i.e., the null hypothesis of no correlation is rejected)
at the 0.1\% level in the Spearman's $\rho$ test.  However, the Cox
and Kendall's $\tau$ test show $P$  19\% and 13\%, respectively,
which indicates no significant correlation.   The lack of a
significant relation with star formation rate SFR is perhaps not so
surprising; since SFR is essentially a luminosity, a strong distance
effect is encoded therein, which may mask an intrinsic physical
relationship.
\begin{figure} 
\begin{center}
\begin{minipage}[c]{8.5cm}
 \includegraphics[width=1.0\textwidth]{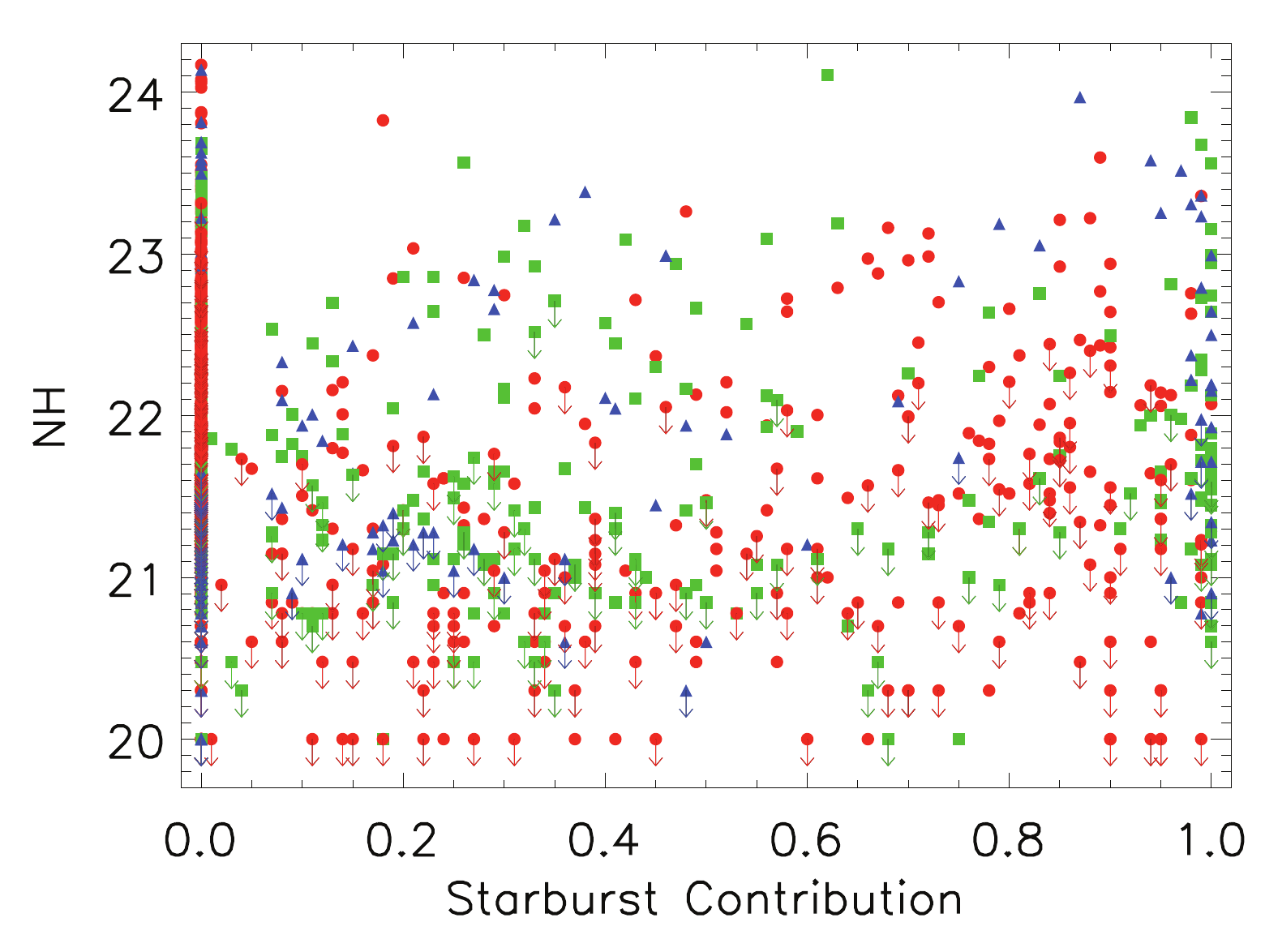}
  \end{minipage}
\end{center}
    \caption
           {\scriptsize Percentage of starburst contribution to the bolometric luminosity versus column density in logarithmic scale. Downward arrows represent the 90$\%$ upper limits to \nhintr,
whenever the 90$\%$ lower limit encompassed zero. Red circles, green squares and blue triangles denote broad-line AGN, narrow emission line and absorption line galaxies respectively.}
\end{figure}

\noindent When we test the correlation between log \nhintr  and the star
formation fraction FS (Figures 12 and 13), we find $P\leq 0.1$ for all 3 tests;
the correlation is highly significant.  The slope of the best-fit
linear regression, however, is rather flat: FS =
0.0122$\pm$0.0016\,log\nhintr + 19.84 with a standard deviation of
0.724 from the Buckley-James algorithm.

\noindent There are a number of reasons why this treatment may not be ideal.
First, we lump all source types together (BLAGN, NELG, ALG) which may
be inappropriate to the actual physics in question.  Second, there is
no {\em a priori} reason why we should consider the logarithm of
\nhintr\, vs. FS, except that the distribution is more regular.  Also,
we exclude objects with FS$<5\%$, which could be treated similarly as
upper limits.  However, the number of limits would thereby greatly
exceed the number of detections of a starburst component.  These would
be all piled up at FS=0, and would strongly bias our regression fits. Perhaps more importantly, we note that our X-ray spectral fits do
{\em not} include (nor in general have sufficient S/N to warrant) an
emission model component appropriate to strong starburst activity.
X-ray emission from star formation regions would generally be rather
soft due to thermal emission, making the soft X-ray absorption
features even more difficult to detect. 

A more credible test may be to examine only the X-ray Seyferts,
the subsample of objects with 10$^{42}$ $<$ $L_{2-10~keV}$ $<$ 10$^{44}$ erg
s$^{-1}$.  
Although the obscuration of some AGN is a consequence simply
of the geometry of the surrounding material and our line of sight to
the nucleus (Antonucci 1993), the prevalence of X-ray absorption in
star-forming AGN could imply an alternative source of absorbing
material, perhaps related to the gas that is fuelling the star
formation or to outflowing material from the early stages of AGN
feedback. In order to test this hypothesis we restrict our sample to
the 430 X-ray Seyferts among which 273 objects have only upper limits
to \nhintr.   First we perform two-sample tests for a difference in
log~\nhintr\, between the 263 objects with FS$\geq$5\% (167 \nhintr
limits), and the 167 with FS$<$5\% (106 \nhintr\, limits).  The mean
log~\nhintr\, values are indistinguishable for the two sub-samples at
21.0$\pm$0.1, and their distributions using Wilcoxon and LogRank tests 
are also indistinguishable ($P>88\%$).  We also searched for
correlations between star formation and absorption amongst only those
263 X-ray Seyferts with FS>5\%, to avoid being dominated entirely by
FS non-detections.  Among  this subsample, there are 96 detections of
\nhintr, but we find no evidence for a significant correlation either
between log~\nhintr\, and log SFR, or between log~\nhintr\, and FS. 

\begin{figure}
\begin{center}
\begin{minipage}[c]{8.5cm}
       \includegraphics[width=1.0\textwidth]{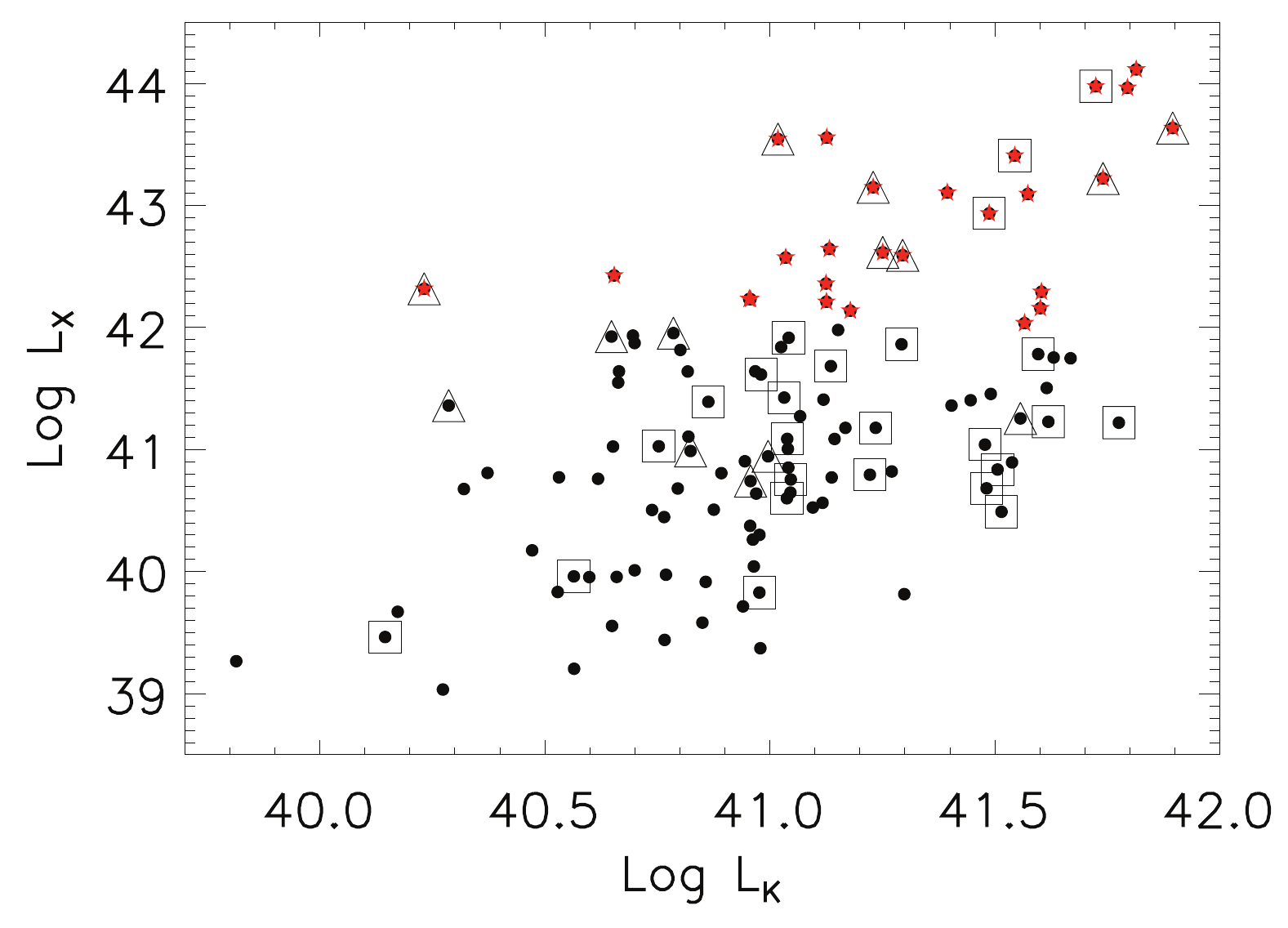}
\end{minipage}
       \end{center}
    \caption{\scriptsize Rest-frame 2-10 keV logarithmic X-ray luminosity versus logarithmic K-band luminosity for all ChaMP spectroscopically identified absorption line galaxies  with K-band detections. Red stars is the population of  25 XBONGs, ALGs with $L_{X (2-10~keV)}$$>$1.5$\times$10$^{42}$ erg s$^{-1}$. Open triangles are radio loud sources. Open squares are sources with log \nhintr $>$ 22.}
\end{figure}

\begin{figure*} 
\begin{center}
\begin{minipage}[c]{5.4cm}
        \includegraphics[width=1.0\textwidth]{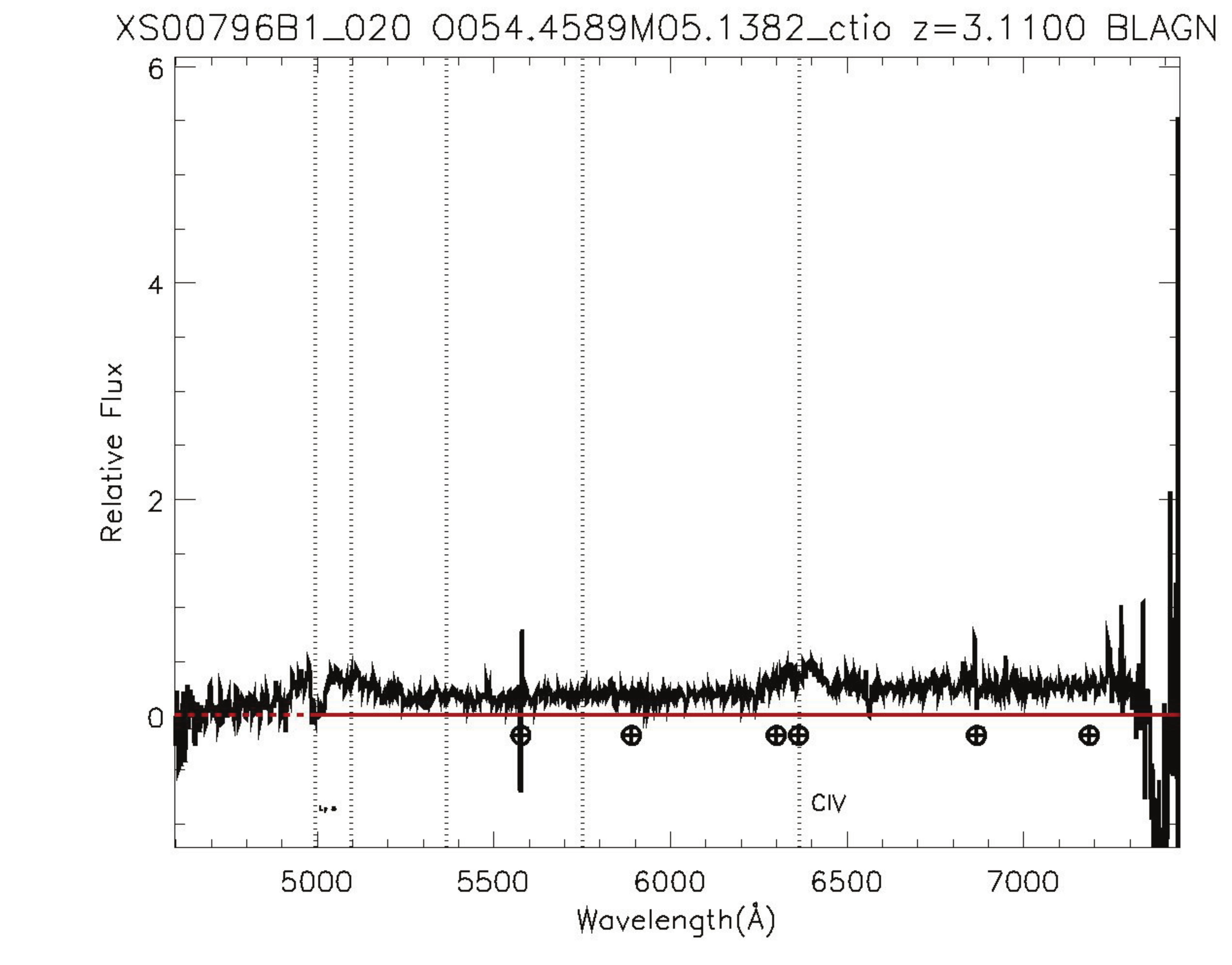}
\end{minipage}
\begin{minipage}[c]{5.4cm}
        \includegraphics[width=1.0\textwidth]{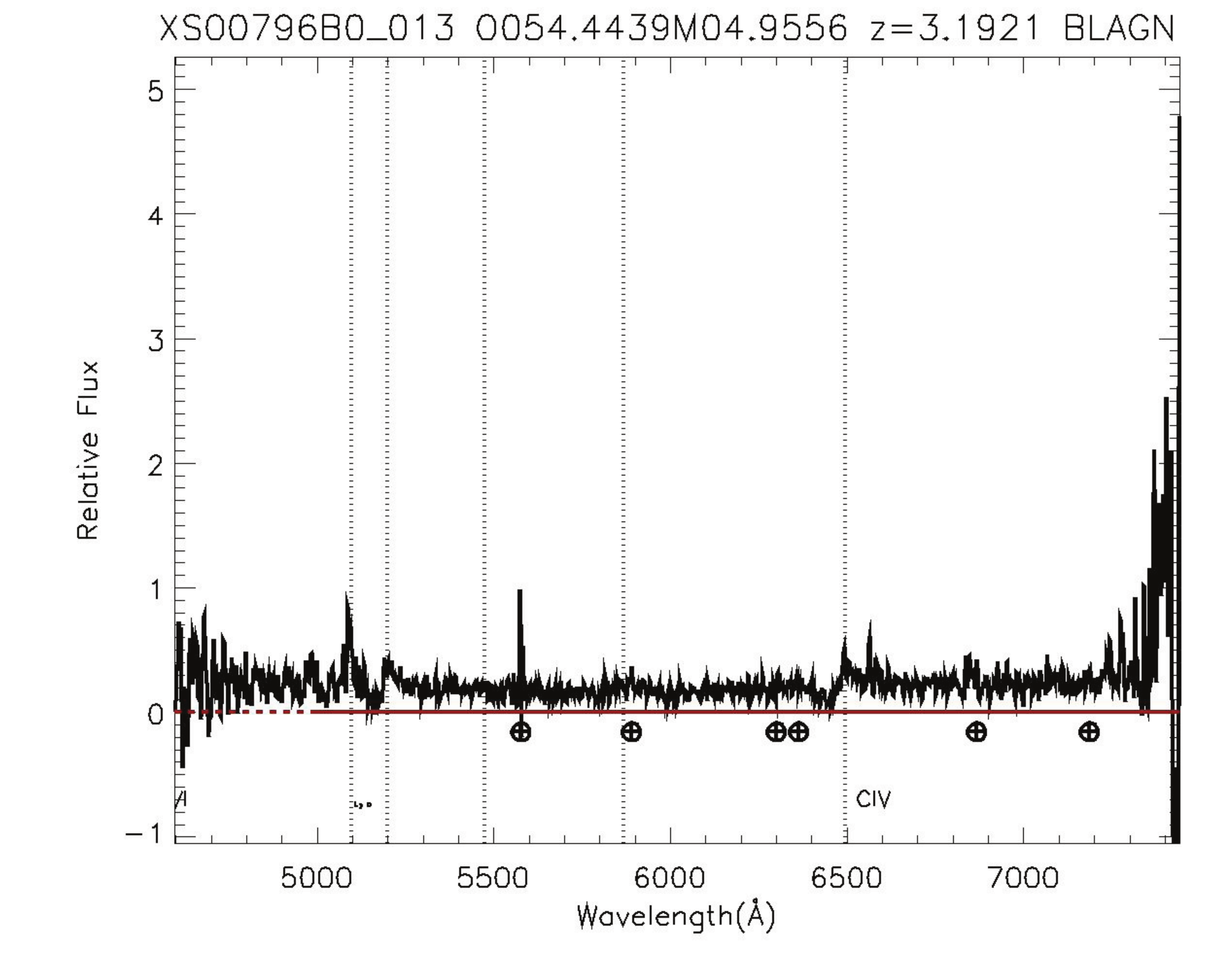}
\end{minipage}
\begin{minipage}[c]{5.4cm}
        \includegraphics[width=1.0\textwidth]{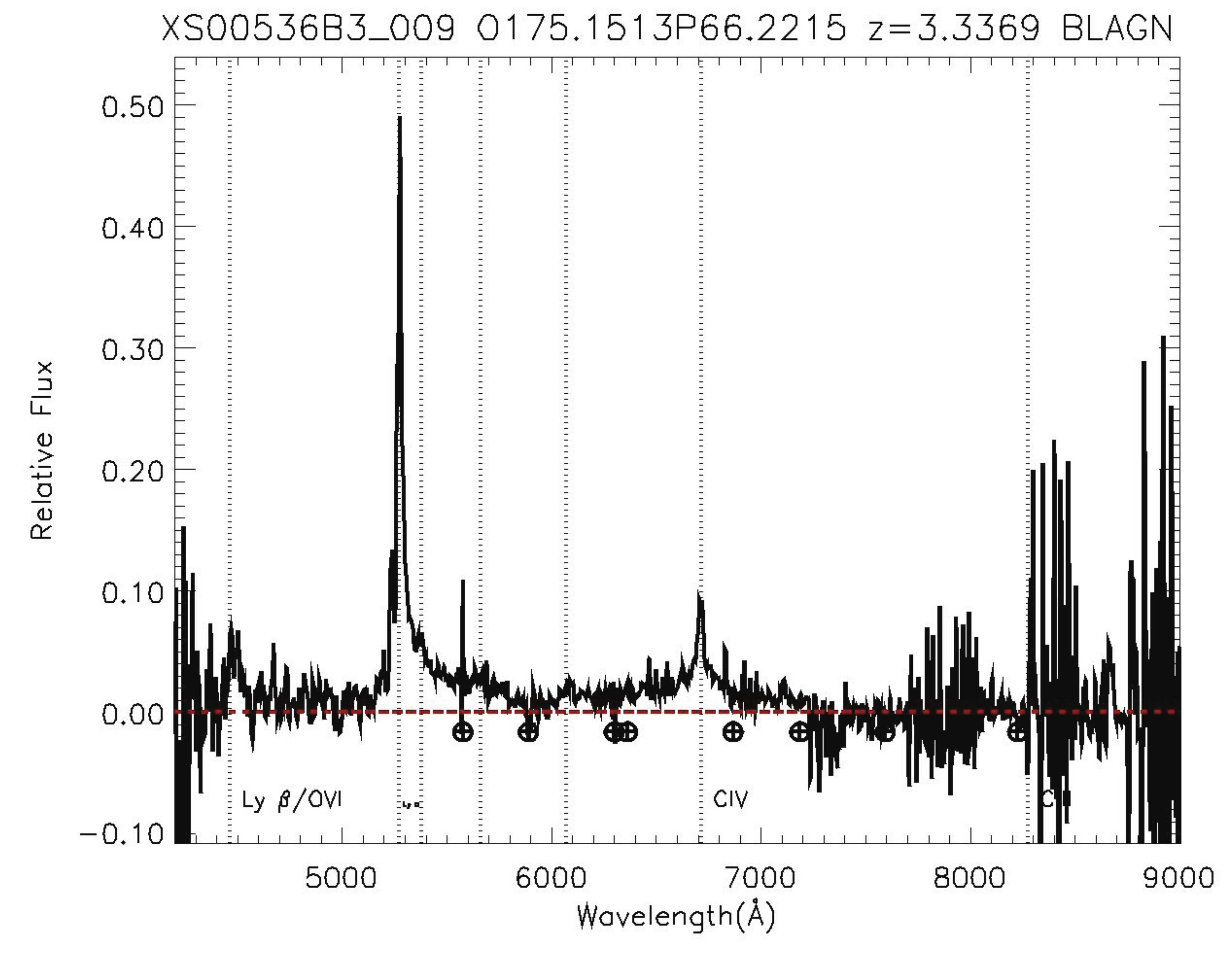}
\end{minipage}
\\
\begin{minipage}[c]{5.4cm}
        \includegraphics[width=1.0\textwidth]{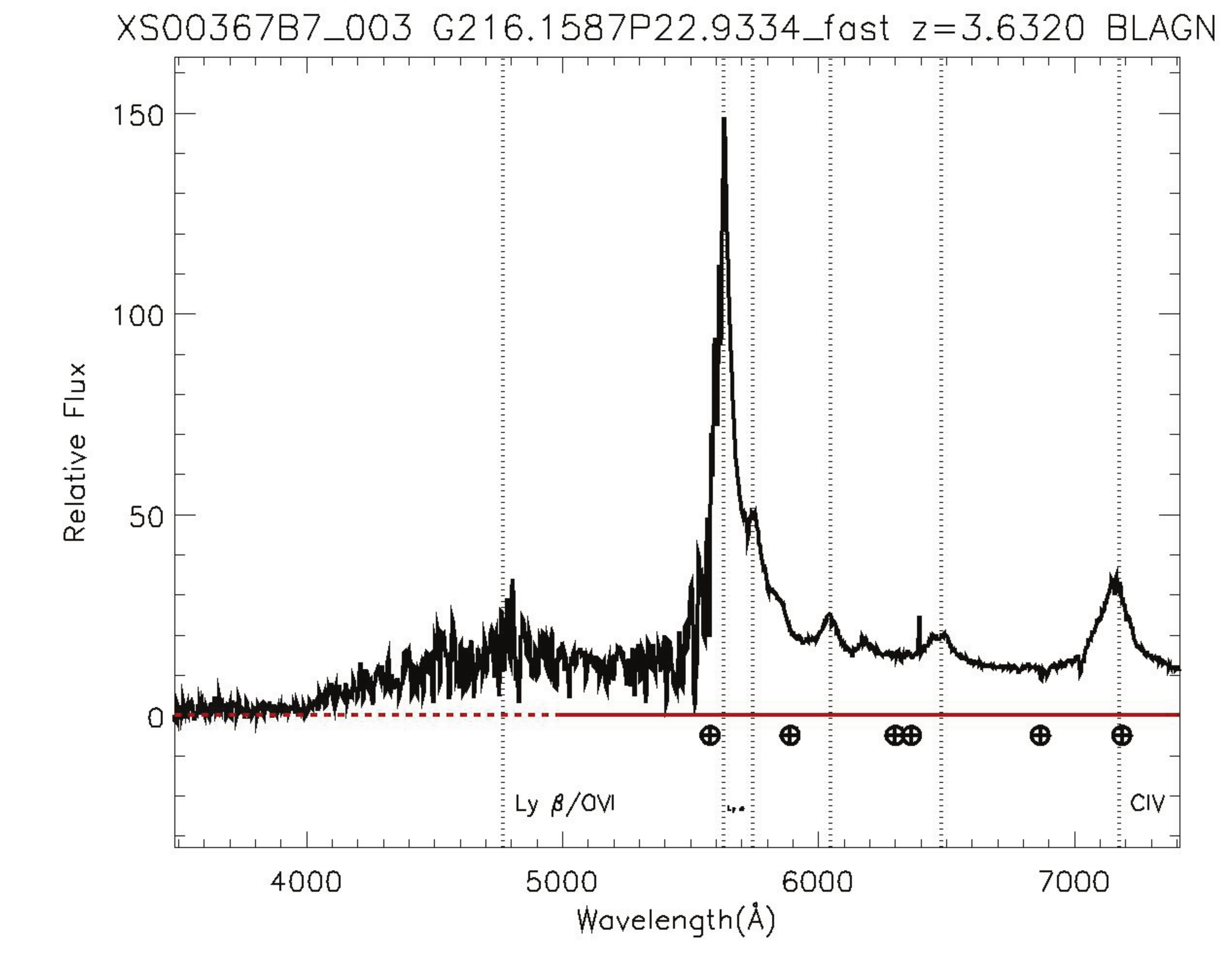}
\end{minipage}
\begin{minipage}[c]{5.4cm}
        \includegraphics[width=1.0\textwidth]{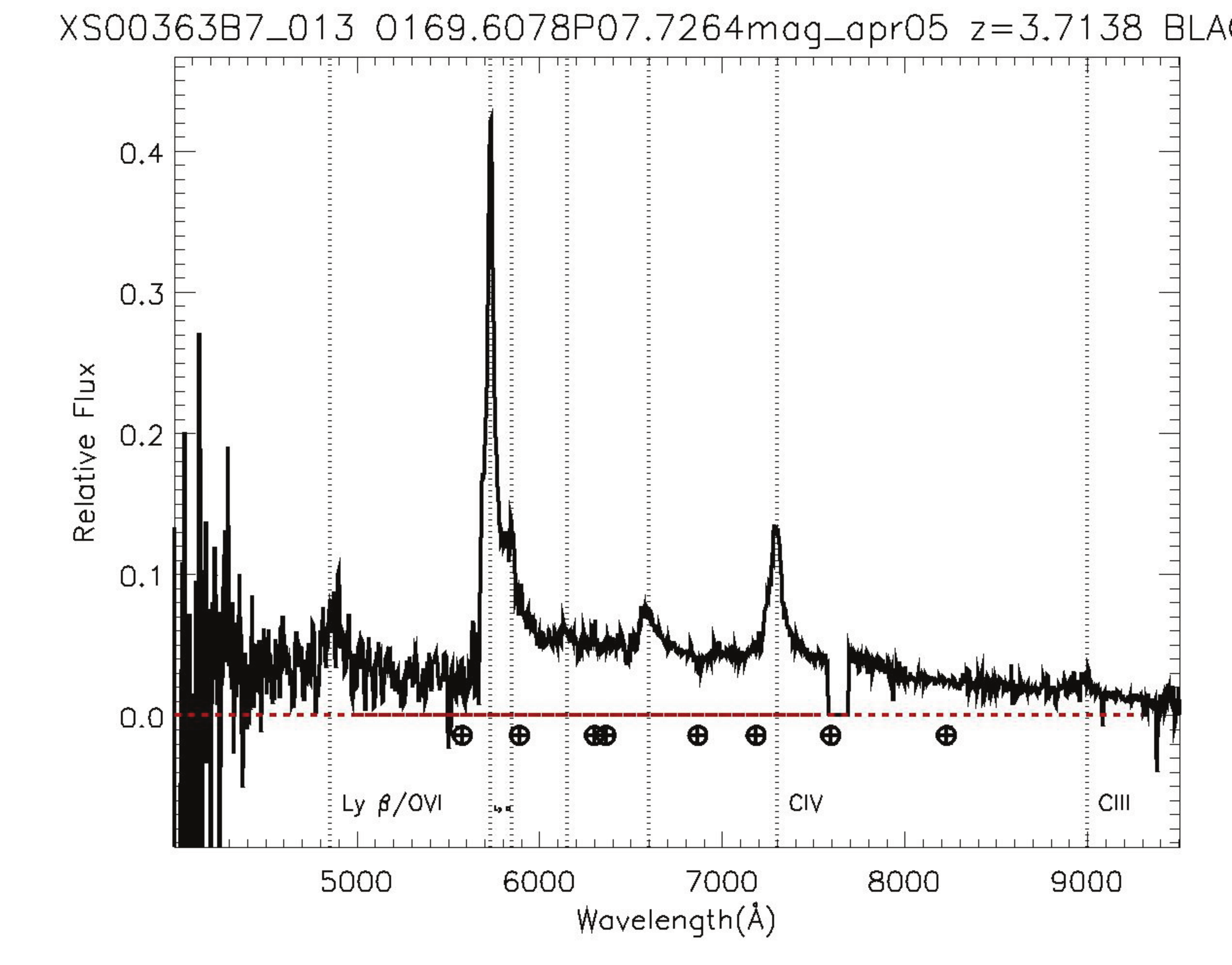}
\end{minipage}
\begin{minipage}[c]{5.4cm}
        \includegraphics[width=1.0\textwidth]{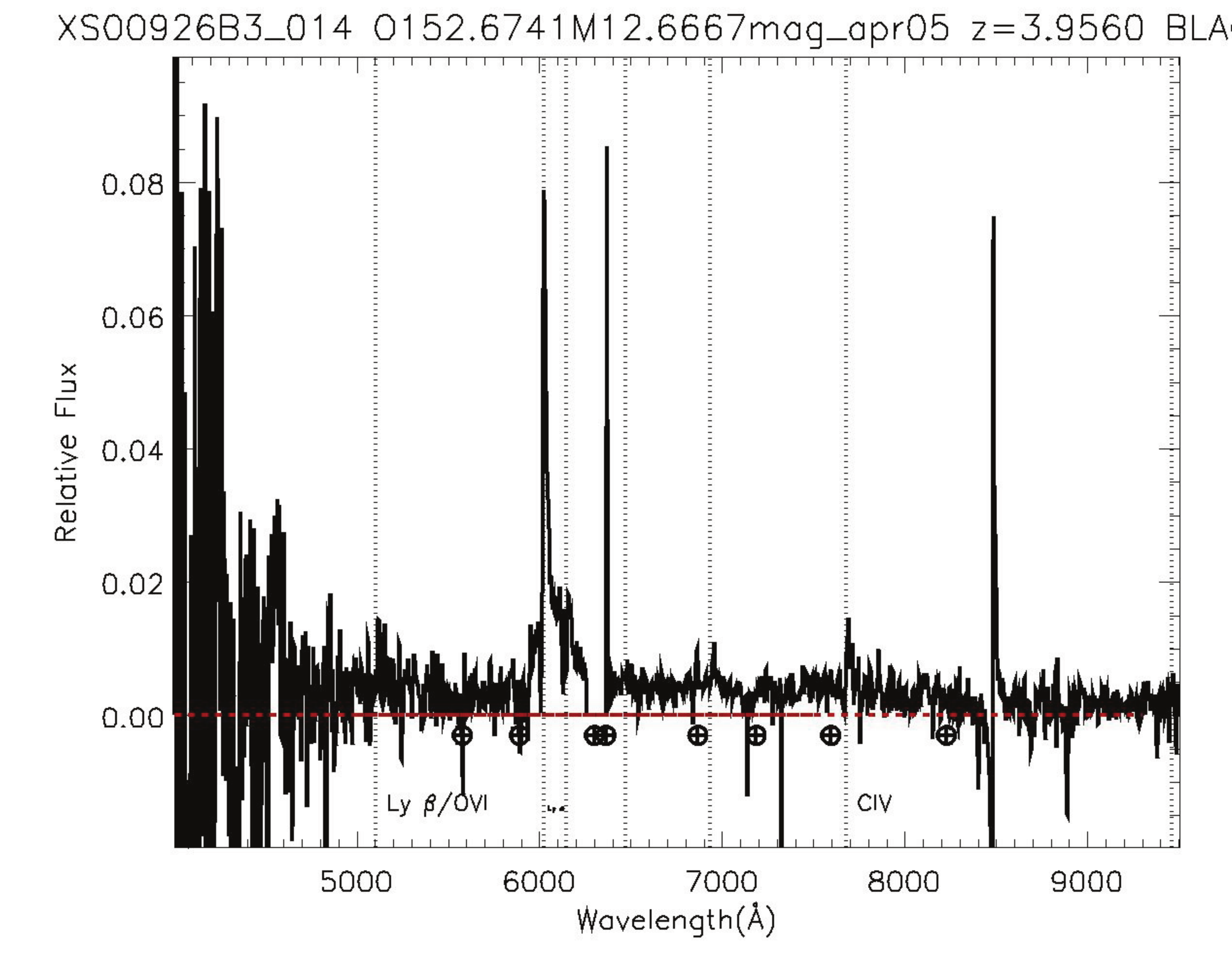}
\end{minipage}
\\
\begin{minipage}[c]{5.4cm}
        \includegraphics[width=1.0\textwidth]{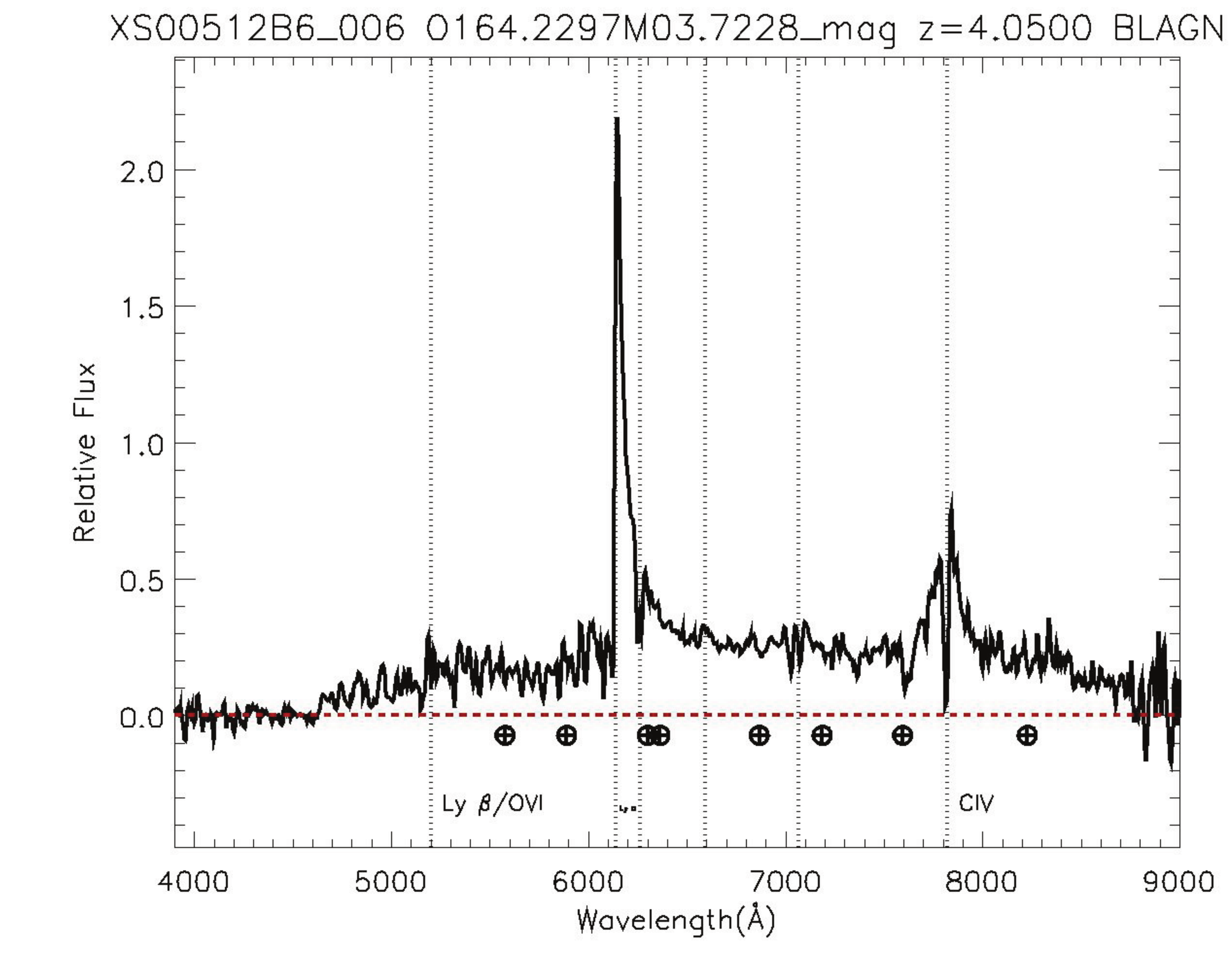}
\end{minipage}
\begin{minipage}[c]{5.4cm}
        \includegraphics[width=1.0\textwidth]{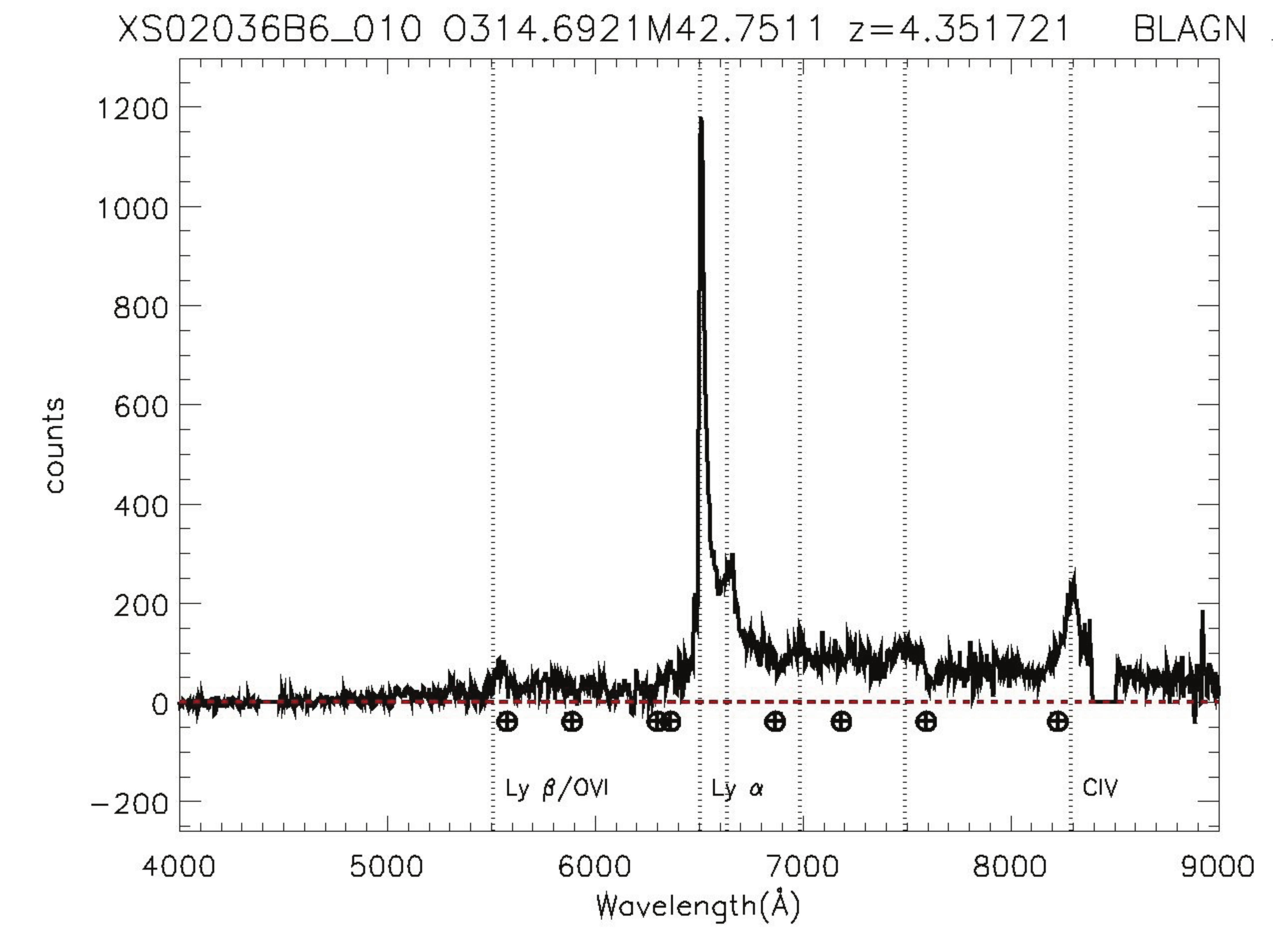}
\end{minipage}
\end{center}
    \caption {\scriptsize ChaMP $z>3$ objects obtained using CTIO, FAST, WIYN and Magellan telescopes. Spectra are not flux calibrated. CHANDRAOBSID, SPECOBJID, REDSHIFT and CLASS are given in the top of each plot.}
\end{figure*}
\section{X-ray Bright Optically Inactive Galaxies (XBONG)}
\noindent Following the definition of an XBONG, ALGs with $L_{X (2-10~keV)}$$>$1.5$\times$10$^{42}$ erg s$^{-1}$ (Comastri et al. 2002), we have identified a total of 81 XBONGs within our spectroscopic sample. However, as can be seen from Figure 7 a large number of ALGs appear to occupy the same parameter space as BLAGN in the f$_{X}$/ f$_{r}$ plane. This might indicate the presence of AGN activity that has been missed due to the shallow optical spectroscopy. In order to address this issue we further restrict our sample by looking for XBONGs in the K-band selected ALG sample. K-band detections come from UKIDSS DR4. We have identified a population of 25 XBONG within our K-band selected sample in the redshift range 0.035$<$z$<$0.948, 20$\%$ of the total K-band ALG selected sample. Eleven of the sources appear to be associated with AGN as their X-ray spectrum is described by a steep photon index that ranges between 1.4$<$$\Gamma$$<$1.9. We find evidence for significant X-ray absorbing columns in 7 of our sources - those that have $N_{H}$$>$$10^{22}$.  Figure 14 shows the X-ray vs. K-band luminosity of all the K-band detected ALG sources within the ChaMP spectroscopic sample.   Four possible explanations have been proposed for the nature of these objects (Green et al. 2004): a $''$buried$''$ AGN (Comastri et al. 2002), a low luminosity AGN (Severgnini et al. 2003), a BL Lac object (Yuan $\&$ Narayan 2004) and galactic scale obscuration (Rigby et al. 2006; Civano et al. 2007). \\
\begin{figure*} 
\begin{center}
\begin{minipage}[c]{6.5cm}
        \includegraphics[width=1.0\textwidth]{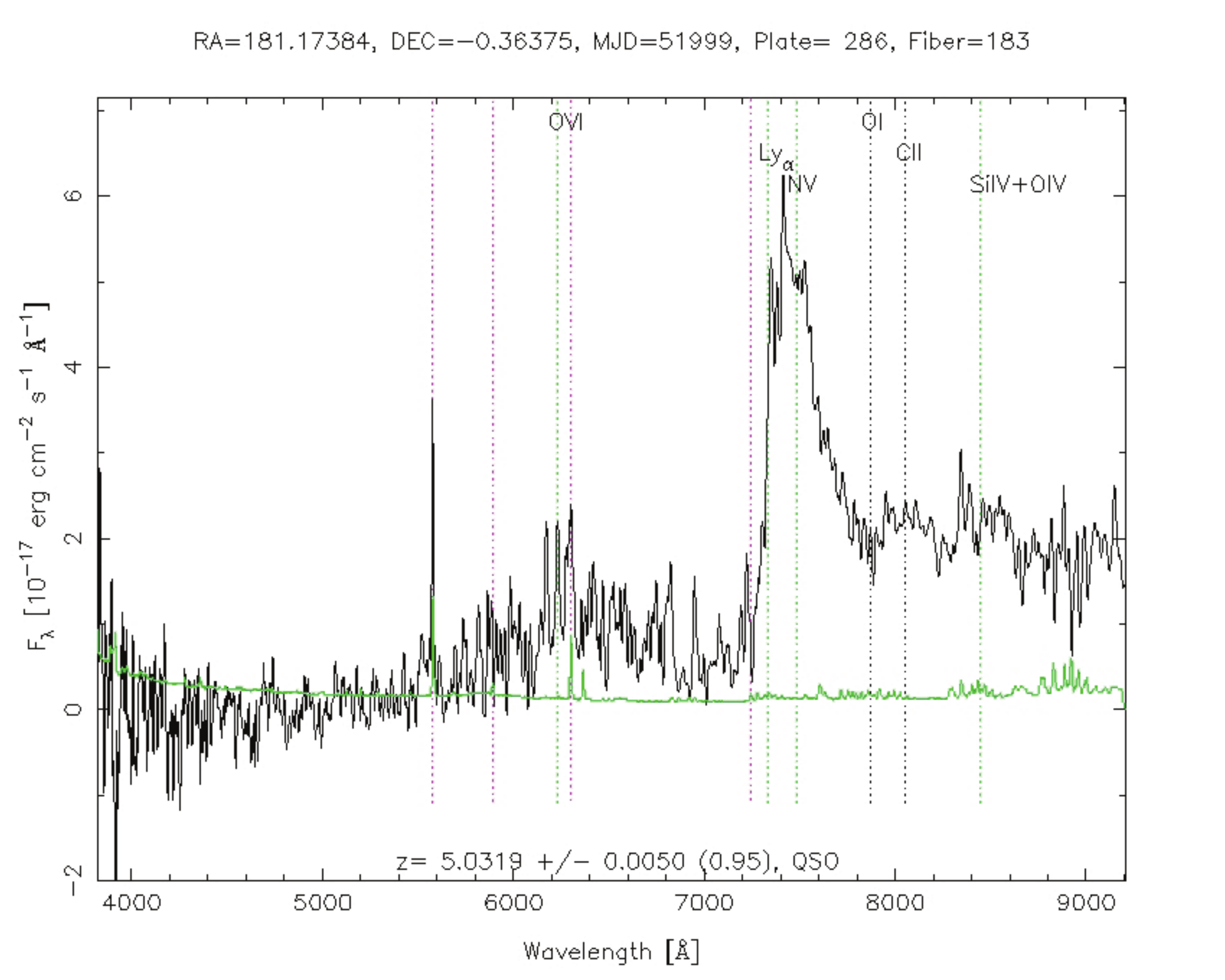}
\end{minipage}
\begin{minipage}[c]{6.5cm}
        \includegraphics[width=1.0\textwidth]{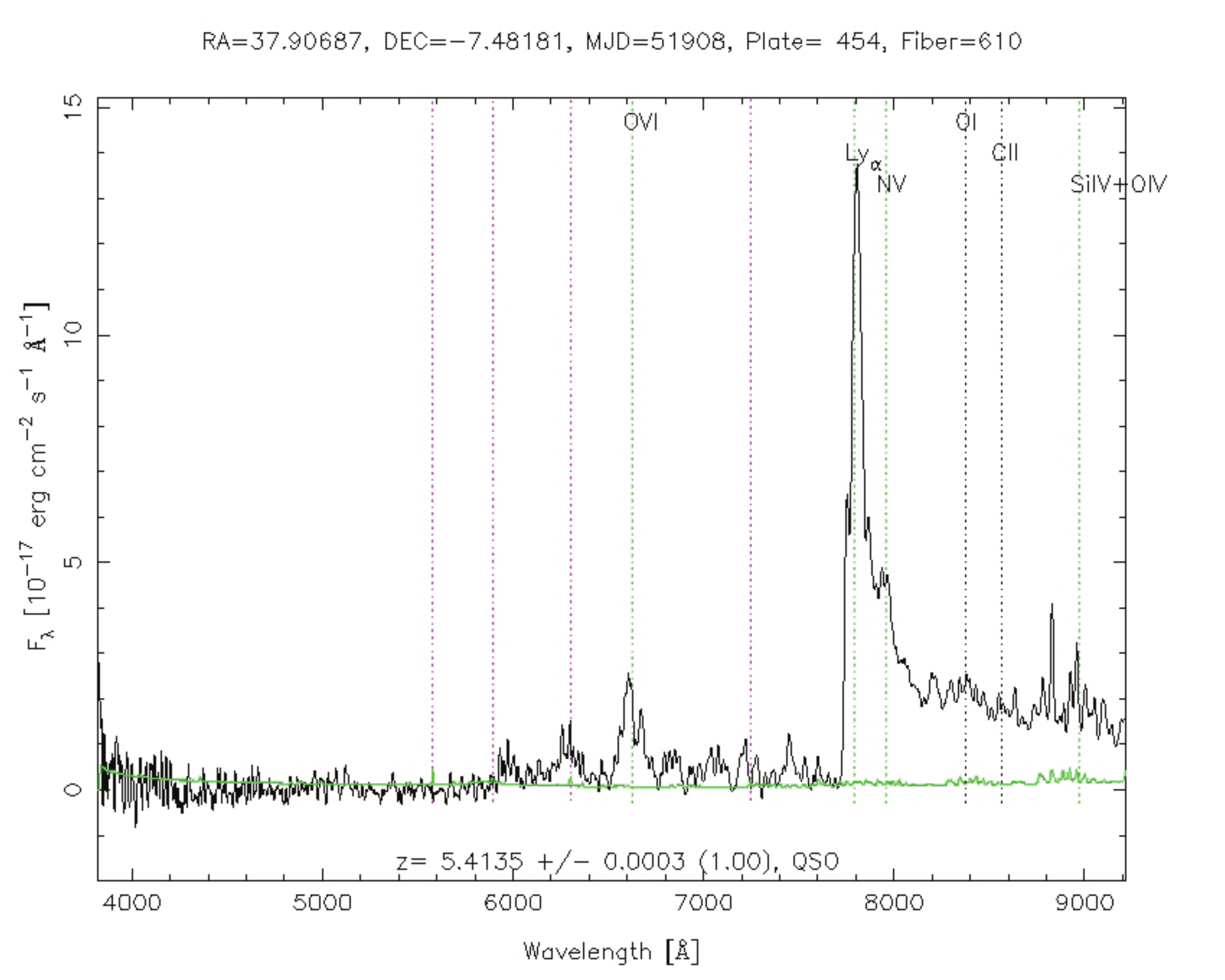}
\end{minipage}
\end{center}
    \caption
           {\scriptsize SDSS optical spectra of the two highest redshift objects found in our sample. They both are broad-line X-ray QSOs at $z>5$. }
\end{figure*}

\noindent Green et al. (2004) defined a $''$buried$''$ AGN as an object that has either no, or only narrow emission lines in its optical spectrum, strong evidence for $N_{H}$$>$$10^{22}$ in the rest frame and $L_{X (2-10~keV)}$$>$10$^{43}$ erg s$^{-1}$ without absorption correction. Among our 22 objects there are 4 objects that fulfill all of the above criteria and are probably Type-2 QSOs. Our SED fitting suggests that only one of the objects is consistent with an elliptical like broad-band spectrum. The remaining three are all fitted with a combination of a Type-2 Seyfert  and a starburst template consistent with a $''$buried$''$ AGN. In two of the latter, the starburst contribution dominates the bolometric emission at the 95$\%$ level. There are additional 6 XBONGs in our sample that have  $L_{X (2-10~keV)}$$>$10$^{43}$ erg s$^{-1}$, no signs of broad optical emission lines in their spectra and soft X-ray spectra which are most likely  $''$buried$''$ Type-2 AGN. Our SED fitting has identified half of them as ellipticals, 2 as Type-1 QSOs with some $<35\%$  star formation contribution and one Type-2 Seyfert with $10\%$ star formation contribution. There are only 2 XBONGs in our sample that exhibit strong radio emission. Both of them have $L_{X (2-10~keV)}$$>$10$^{42}$ with $N_{H}$$<$$10^{21}$ and are fitted with an elliptical template with no evidence of star formation. These can both be BL Lac candidates but the lack of high S/N optical spectra that would allow us to measure the 4000 $\AA$ break does not permit us to verify their BL Lac status. All the above suggest that the XBONGs found in our sample comprise a mixed bag of objects primarily including normal elliptical galaxies, Type-1 AGN and most importantly Type-2 QSOs that need further investigation.

\section{High-Redshift X-ray QSOs}
\noindent There are a total of 78 $z>3$ X-ray objects in our spectroscopic catalogue, 70 from SDSS and 8 from our own follow-up campaigns. The latter 8 spectra are presented in Figure 15. 76 are broad line AGN, one narrow emission line galaxy at $z=3.417$ and one absorption line galaxy at $z=3.32$. There are also two X-ray QSOs at $z>5$ whose SDSS spectra are shown in Figure 16. All of our high-z sources have $0.5-2$ $keV$ detections and 76 have also $2-8$ $keV$ band detections. 7 sources have UV, all of them have $ugriz$, 18 have $JHK$, 22 have WISE and 13 have radio detections. According to our SED fitting method, 76 are fitted with an Type-1 AGN template that in 22 cases requires a significant starburst contribution that ranges between 10$\%$ and 70$\%$ of the bolometric luminosity. The two non-broad line high redshift objects are both fitted with a starburst template, suggesting that these might be high-z Type-2 AGN shrouded in a powerful starburst. The high-z sources in our sample have a relatively high number of detected counts, 31 is the median value, with respect to the depth of the X-ray observations.  There are 15 sources that have more than 100 counts. In this count regime,  we can assume that the extracted spectral fit results are reliable enough to be used for column density estimations. 45 of our sources have $N_{H}$$>$22 suggesting that these sources are obscured. \\

\noindent Our sample represents the largest spectroscopically selected sample of z$>$3 X-ray sources and the second largest compared to samples with available photometric redshifts (Civano et al. 2011). Our sample has almost doubled the number of spectroscopically identified z$>$5 X-ray QSOs by adding two more sources (Figure 16) to the three previously known at z=5.19 (Barger et al. 2005), z=5.3 (Civano et al. 2011) and z=5.4 (Steffen et al. 2004). Here we report the identification of the highest redshift X-ray QSO with optical spectroscopy ever found at z=5.4135 (Figure 16).

\section{Conclusions}
\noindent We present the complete optical spectroscopic follow-up of ChaMP sources. We utilize a large suite of multi-slit and multi-fiber instruments on FLWO, SAAO, WIYN, CTIO, KPNO, MMT, Magellan and Gemini to identify both bright and faint serendipitous X-ray sources as well as archival SDSS optical spectra. These observations resulted in a total of 1569 spectroscopic identifications of X-ray sources. Among the latter, there are 1242 extragalactic sources, half of which are broad-line QSOs. For these sources we have collected an extensive library of ancillary multi-wavelength data including X-ray, UV, optical, near-IR, mid-IR and radio data from our own photometric follow-ups as well as various public catalogues including GALEX, 2MASS, UKIDSS, WISE, NVSS and FIRST. Multi-wavelength photometry in combination with available optical spectroscopy has allowed us to distinguish among different populations, study the X-ray-to-radio SEDs in order to estimate luminosities and assess the level of AGN and star formation contribution, estimate column densities via X-ray spectral fitting and estimate black hole masses and star formation rates. \\\\
\noindent Based on our observations, although X-ray Seyferts appear to be hosted in galaxies with powerful star formation events, when accretion onto the SMBH reaches its peak, as indicated from both hard X-ray luminosity and AGN contribution to the bolometric luminosity, star formation is quenched, resulting in only a small percentage of X-ray QSOs showing prevalent starburst events. According to Ebrero et al. (2009), objects with $L_{2-10~keV}$$\sim$10$^{44}$ erg s$^{-1}$ at 1$<$z$<$3 are at the peak  of their accretion rates. Therefore, the higher fraction of X-ray Seyferts with star formation compared to the X-ray QSOs with star formation, imply that 
the most prodigious episodes of star formation are common in the host galaxies of 1$<$ z $<$ 3 AGN, but avoid powerful AGN in which accretion is at its peak. This systematic separation of the peak periods of star formation and accretion implies a palpable interaction between the two processes, and provides a powerful discriminator for the form of AGN feedback which 
is responsible for terminating star formation in the host galaxy.\\ \\
\noindent  In ``QSO mode" feedback, a luminous 
AGN generates a powerful wind which terminates star formation by driving the interstellar medium from the 
surrounding host galaxy. In ``radio-mode" feedback, star formation is suppressed because collimated jets of relativistic particles emitted by a radiatively-inefficient AGN prevent gas in the surrounding hot halo from cooling, 
thereby starving the galaxy of cool gas from which to form stars. ``Radio-mode" feedback is commonly invoked in semi-analytical models to limit galaxy masses and luminosities (Croton et al. 2006). In these models, black holes grow through luminous accretion episodes and black hole mergers. The 
correlation between black hole and bulge mass comes from assuming that a fixed fraction of the gas is accreted 
by the nucleus during each star forming episode that results from a galaxy merger or disc instability, and hence 
star formation and accretion rate should be correlated over the full range of luminosity. \\ \\
\noindent Our observations are 
therefore globally inconsistent with models such as Croton et al. (2006) in which AGN influence their host galaxies only through 
radio mode feedback. In contrast, models of galaxy formation in which quasar-mode feedback is responsible 
for terminating the star formation (Hopkins et al. 2006) predict that the AGN luminosity peaks later than the 
star formation rate, and thus are consistent with our observations that show that star formation not only occurs less often but is also weaker in X-ray QSOs compared to X-ray Seyferts. These models also predict that residual 
star formation, at the level of a few tens of per cent of the peak, will continue during the period in which the 
AGN luminosity is at its maximum, consistent with our  results which show that some of the X-ray QSOs are still forming stars. Finally, while the observations presented 
here provide strong evidence for the violent quenching of star formation as AGN reach peak luminosity, they 
do not rule out radio-mode feedback as the agent by which galaxy growth is subsequently suppressed. Further observations and/or studies with far-infrared/submm data are essential in order to verify our last finding and to reduce any ambiguity based on template fitting results.\\ \\
\noindent Our findings regarding the relationship between X-ray column density and starburst contribution and star formation rate further support  the prediction that AGN obscuration is a consequence 
simply of the geometry of the surrounding material and our line of sight to the nucleus (Antonucci 1993) rather than a common material feeding both mechanisms. In this work, we have shown that obscuration does not seem to be associated to star formation  either in the general population or the population of star forming X-ray AGNs. \\ \\
\noindent Finally we report the identification of 25 K-band selected XBONGs. Among the latter, 10 appear to be Type-2 QSOs with an AGN buried in active starburst events. We have also identified a significant population (78) of $z>3$ objects. There are two non-broad line objects in this sample that are quite probably high-z Type-2 AGN. 45 of the high-z objects in our sample appear to be highly obscured. We finally report here the identification of two $z>5$ X-ray QSOs who are among the highest spectroscopic redshift X-ray selected QSOs ever observed.

\noindent{\bf Acknowledgements}: The authors would like to thank Francesca Civano and Hagai Netzer for their useful comments. Support for this work was provided by
the National Aeronautics and Space Administration through {\em
  Chandra} Award Numbers AR9-0020X and AR1-12016X, issued by the {\em
Chandra} X-ray Observatory Center, which is operated by the
Smithsonian Astrophysical Observatory for and on behalf of the
National Aeronautics Space Administration under contract NAS8-03060.

\end{document}